\documentclass[trackchanges,twocolumn]{aastex7}

\usepackage{array}
\usepackage{rotating}
\usepackage{makecell}
\usepackage{graphicx}
\usepackage{color}
\usepackage{amsmath}
\usepackage{amssymb}
\usepackage{txfonts}
\usepackage{tikz}
\usepackage{stackengine}
\usepackage{pifont}

\usepackage{float}
\usepackage{multirow}
\usepackage[most]{tcolorbox}
\usepackage{ragged2e}
\usepackage{longtable}
\usepackage{booktabs}
\usepackage{upgreek}
\newcolumntype{P}[1]{>{\centering\arraybackslash}p{#1}}
\usepackage{soul}
\usepackage[table,dvipsnames]{xcolor}
\usepackage{stfloats}
\usepackage{wrapfig}

\usepackage{todonotes}

\usepackage{silence}
\newcommand{\tablesize}{\normalsize}

\newcommand{\tblrulevskip}{\noalign{\vskip 0.6pt}}

\newcommand{\cmark}{\ding{51}}

\newcommand{\vsepblack}{\kern0pt{\color{black}\vrule width 0.6pt}\kern0pt}   
\newcommand{\vsepgray}{\kern0pt{\color{gray!50}\vrule width 0.6pt}\kern0pt}  

\DeclareRobustCommand{\CIRCLE}{%
  \ensuremath{%
    \mbox{
      \tikz[baseline=-0.6ex]{%
        \fill (0,0) circle[radius=2.75pt];
      }%
    }
  }%
}

\DeclareRobustCommand{\mycircle}{%
  \ensuremath{%
    \mbox{
      \tikz[baseline=-0.6ex]{%
        \draw (0,0) circle[radius=2.75pt];
        \fill (0,0) circle[radius=1.75pt];
      }%
    }
  }%
}

\DeclareRobustCommand{\Bigcircle}{%
  \ensuremath{%
    \mbox{%
      \tikz[baseline=-0.6ex]{%
        \draw[very thick] (0,0) circle[radius=2.5pt];
      }%
    }%
  }%
}

\newcommand\sbullet[1][.5]{\mathbin{\vcenter{\hbox{\scalebox{#1}{$\bullet$}}}}}

\newtcbox{\myboxi}[1][]{%
  nobeforeafter,
  tcbox raise base,
  colframe=black!15!white,
  colback=black!15!white,
  height=8pt,
  width=8pt, 
  square,
  circular arc,
  halign=center,
  valign=center,
  raster valign=center,
  box align=base,
  top=0pt,
  bottom=0pt,
  left=0pt,
  right=2pt,
  boxrule=0pt,
  boxsep=2.5pt,
  before upper=\strut,
  #1
}
\newcommand{\mybox}[2][1.1ex]{\raisebox{#1}{\myboxi{#2}}}

\newtcbox{\myboxxi}[1][]{%
  nobeforeafter,
  tcbox raise base,
  colframe=orange!80!white,
  colback=orange!80!white,
  height=8pt,
  width=8pt,
  square,
  circular arc,
  halign=center,
  valign=center,
  raster valign=center,
  box align=base,
  top=0pt,
  bottom=0pt,
  left=0pt,
  right=2pt,
  boxrule=0pt,
  boxsep=2.5pt,
  before upper=\strut,
  #1
}
\newcommand{\myboxx}[2][1.1ex]{\raisebox{#1}{\myboxxi{#2}}}
        
\newtcbox{\myboxxxi}[1][]{%
  nobeforeafter,
  tcbox raise base,
  colframe=green!75!black,
  colback=green!75!black,
  height=8pt,
  width=8pt,
  square,
  circular arc,
  halign=center,
  valign=center,
  raster valign=center,
  box align=base,
  top=0pt,
  bottom=0pt,
  left=0pt,
  right=2pt,
  boxrule=0pt,
  boxsep=2.5pt,
  before upper=\strut,
  #1
}
\newcommand{\myboxxx}[2][1.1ex]{\raisebox{#1}{\myboxxxi{#2}}}

\begin{document}


\title{Chemical Divergence and Water Depletion: \\
Gas Properties of Evolved Upper Scorpius Disks Revealed by JWST/MIRI}

\author[0009-0002-2380-6683]{Eshan Raul}
\affil{Department of Astronomy, University of Wisconsin--Madison, Madison, WI 53706, USA}
\email[show]{eraul@wisc.edu}

\author[0000-0002-0661-7517]{Ke Zhang}
\affil{Department of Astronomy, University of Wisconsin--Madison, Madison, WI 53706, USA}
\email{ke.zhang@wisc.edu}

\author[0000-0002-1566-389X]{Abygail Waggoner}
\affil{Department of Astronomy, University of Wisconsin--Madison, Madison, WI 53706, USA}
\email{awaggoner2@wisc.edu}

\author[0000-0001-8184-5547]{Chengyan Xie}
\affil{Lunar and Planetary Laboratory, The University of Arizona, Tucson, AZ 85721, USA}
\email{cyxie@arizona.edu}

\author[0009-0005-2896-2150]{Nicholas Tallon}
\affil{Department of Astronomy, University of Wisconsin--Madison, Madison, WI 53706, USA}
\email{ntallon@wisc.edu}

\author[0000-0003-4335-0900]{Andrea Banzatti}
\affil{Department of Physics, Texas State University, 749 N Comanche St, San Marcos, TX 78666, USA}
\email{banzatti@txstate.edu}

\author[0000-0003-3682-6632]{Colette Salyk}
\affiliation{Department of Physics and Astronomy, Vassar College, 124 Raymond Avenue, Poughkeepsie, NY 12604, USA}
\email{cosalyk@vassar.edu}

\author[0000-0001-7552-1562]{Klaus Pontoppidan}
\affil{Jet Propulsion Laboratory, California Institute of Technology, 4800 Oak Grove Drive, Pasadena, CA 91109, USA}
\email{klaus.m.pontoppidan@jpl.nasa.gov}

\author[0000-0001-7962-1683]{Ilaria Pascucci}
\affil{Lunar and Planetary Laboratory, The University of Arizona, Tucson, AZ 85721, USA}
\email{pascucci@arizona.edu}

\author[0000-0003-2631-5265]{Nicole Arulanantham}
\affil{Space Telescope Science Institute, 3700 San Martin Drive, Baltimore, MD 21218, USA}
\email{narulanantham@schmidtsciences.org}

\author[0000-0002-4147-3846]{Miguel Vioque}
\affiliation{European Southern Observatory, Karl-Schwarzschild-Str.\ 2, 85748 Garching bei M\"{u}nchen, Germany}
\email{miguel.vioque@eso.org}

\author[0000-0001-6448-7178]{Aaron Empey}
\affil{Department of Physics, University College Dublin, Belfield, Dublin 4, Ireland}
\email{aaron.empey@ucdconnect.ie}

\author[0000-0003-3562-262X]{Carlo Manara}
\affil{European Southern Observatory, Karl-Schwarzschild-Str.\ 2, 85748 Garching bei M\"{u}nchen, Germany}
\email{cmanara@eso.org}

\author[0000-0003-0787-1610]{Geoffrey A. Blake}
\affiliation{Division of Geological \& Planetary Sciences, MC 150-21, California Institute of Technology, Pasadena, CA 91125, USA}
\email{gab@caltech.edu}

\author[0000-0001-8764-1780]{Paola Pinilla}
\affiliation{Mullard Space Science Laboratory, University College London, Holmbury St Mary, Dorking, Surrey RH5 6NT, UK}
\email{p.pinilla@ucl.ac.uk}

\author[0000-0002-7607-719X]{Feng Long}
\affil{Kavli Institute for Astronomy and Astrophysics, Peking University, Beijing 100871, China}
\email{long.feng@pku.edu.cn}

\author[0009-0002-0525-8222]{Jinghuai Yao}
\affil{Department of Astronomy, University of Wisconsin--Madison, Madison, WI 53706, USA}
\email{jyao224@wisc.edu}

\author[0000-0003-0386-2178]{Jayatee Kanwar}
\affiliation{Department of Astronomy, University of Michigan, 1085 South University Avenue, Ann Arbor, MI 48109, USA}
\email{jkanwar@umich.edu}

\author[0000-0003-3401-1704]{Naman S. Bajaj}
\affiliation{Lunar and Planetary Laboratory, The University of Arizona, Tucson, AZ 85721, USA}
\email{namanbajaj@arizona.edu}

\author[0000-0002-5296-6232]{Mar\'ia Jos\'e Colmenares}
\affiliation{Department of Astronomy, University of Michigan, 1085 South University Avenue, Ann Arbor, MI 48109, USA}
\email{mjcolmen@umich.edu}

\author[0000-0001-8240-978X]{Till Kaeufer}
\affil{Department of Physics and Astronomy, University of Exeter, Exeter, EX4 4QL, UK}
\email{T.Kaeufer@exeter.ac.uk}

\author[0000-0002-1103-3225]{Benoit Tabone}
\affil{Université Paris-Saclay, CNRS, Institut d’Astrophysique Spatiale, 91405 Orsay, France}
\email{benoit.tabone@cnrs.fr}

\author[0000-0003-4179-6394]{Edwin Bergin}
\affiliation{Department of Astronomy, University of Michigan, 1085 South University Avenue, Ann Arbor, MI 48109, USA}
\email{ebergin@umich.edu}

\author[0000-0002-2828-1153]{Lucas A. Cieza}
\affiliation{Instituto de Estudios Astrof\'isicos, Universidad Diego Portales, Avenida Ejercito 441, Santiago, Chile}
\email{lucas.cieza@mail.udp.cl}

\author[0000-0002-0554-1151]{Mayank Narang}
\affil{Jet Propulsion Laboratory, California Institute of Technology, 4800 Oak Grove Drive, Pasadena, CA 91109, USA}
\email{mayankn1154@gmail.com}

\author[0000-0002-1575-680X]{James Miley}
\affil{Joint ALMA Observatory, Alonso de Córdova 3107, Vitacura, Santiago, Chile}
\affil{European Southern Observatory, Alonso de Córdova 3107, Vitacura, Santiago, Chile}
\affil{Millennium Nucleus on Young Exoplanets and their Moons (YEMS), Chile}
\email{james.miley@alma.cl}

\author[0000-0002-3291-6887]{Sebastiaan Krijt}
\affiliation{Department of Physics and Astronomy, University of Exeter, Exeter, EX4 4QL, UK}
\email{s.krijt@exeter.ac.uk} 

\author[0000-0003-4853-5736]{Giovanni Rosotti}
\affil{Dipartimento di Fisica, Università degli Studi di Milano, via Giovanni Celoria 16, 20133, Milano, Italy}
\email{giovanni.rosotti@unimi.it}

\author[0000-0001-8407-4020]{Aditya Arabhavi}
\affil{Jet Propulsion Laboratory, California Institute of Technology, 4800 Oak Grove Drive, Pasadena, CA 91109, USA}
\email{adityamarabhavi.jpl@gmail.com}

\begin{abstract}

Tracing the chemical evolution of protoplanetary disks over time requires observations of disks at different ages.
However, most JWST/MIRI surveys published to date have targeted younger ($\sim$1-3 Myr) rather than older systems.
We present the results of a JWST/MIRI MRS survey of the inner regions of 10 protoplanetary disks (ages $\sim$2-6~Myr, spectral types M0-M4.5) in the Upper Scorpius region previously characterized by the ALMA AGE-PRO large program. Using MCMC slab modeling, we fit to a wide variety of detected molecules, including H$_2$O, CO, C$_2$H$_2$, $^{13}$CCH$_2$, HCN, HC$_3$N, CO$_2$, $^{13}$CO$_2$, C$_2$H$_6$, C$_4$H$_2$, and OH, as well as C$_6$H$_6$, CH$_3$, and H$_2$ visually.
We classify each disk along two independent axes—a Water Classification based on H$_2$O line luminosity (Water-Rich, Water-Poor, or Water-Absent) and a Chemotype based on the dominant non-water chemistry (Organic-Rich, CO$_2$-Dominated, or Molecule-Absent)—and find an unexpectedly high diversity of distinct chemical compositions within our population.
We leverage the heterogeneity of detected molecules in our sample to present new characteristic ``diagnostic" wavelength regions for most species.
We find that carbon-based molecules consistently exhibit markedly lower excitation temperatures ($\lesssim$300 K) compared to younger ($\sim$1–3~Myr) star-forming regions ($\sim$600-1000 K), hinting at relatively colder molecular reservoirs.
We also determine that Upper Scorpius disks show systematically lower water luminosities by factors of 10-1000.
In particular, disks with strong carbon-based molecular features but no observed H$_2$O defy expectations of an inner-disk dust cavity or a low ($\lesssim3$) $R_{\rm gas}/R_{\rm dust}$ ratio, instead suggesting that the presence of a strong outer-disk dust trap largely controls the chemical outcome of the terrestrial planet-forming region.

\end{abstract}


\section{Introduction}\label{sec:intro}
Understanding the volatile (e.g., CO, H$_2$O, C$_2$H$_2$, HCN, CO$_2$) content inside the innermost few astronomical units (au) of protoplanetary disks is crucial in predicting the initial chemical composition of close-in terrestrial planets and ultimately their atmospheres. The relative abundances of common carriers of carbon and oxygen reflect the local gas-phase carbon-to-oxygen ratio (C/O) and therefore influence which solids and gases are available during planet assembly; small changes in this chemical inventory can lead to vastly different planetary and atmospheric compositions \citep{oberg11,Madhusudhan12}.
Disks can also re-distribute ices and refractories, which can alter the inner-disk volatile budget on million-year (Myr) timescales \citep[see][]{Morfill1984,stevenson88,Cyr1998,Ciesla06,najita13,Morbidelli16,booth17,Banzatti20}. This makes empirical constraints on gas-phase abundances across different disk stages an essential input to the chemistry of planet-formation, as this allows us to posit which evolutionary mechanisms become most relevant for planet formation.  

Observationally, mid-infrared spectroscopy is the most direct probe of warm molecular gas in the terrestrial planet-forming region. 
Spitzer/IRS first established that simple molecules (H$_2$O, OH, HCN, C$_2$H$_2$, CO$_2$) are common in inner disks around T~Tauri stars \citep{pontoppidan10a,salyk11,carr11,Pascucci2013,pontoppidan14}, though limited resolution and line blending left many ambiguities in the interpretation of spectral features \citep{Banzatti2013}.
The improved spectral resolution/sensitivity of the James Webb Space Telescope’s Mid-Infrared Instrument \citep{Rieke2015,Wright2015} Medium Resolution Spectrometer \citep{Wells2015,Argyriou20} (JWST MIRI/MRS) has now enabled us to (1) resolve many of the ambiguities in Spitzer spectra by separating overlapping lines and (2) reveal a variety of new molecules and isotopologues in the inner few au of multiple disks \citep[see][]{Grant23,Gasman23,Tabone23,Pontoppidan2024,Xie2023,Long25,Colmenares24,Ramirez-Tannus2025,Kanwar2024b,vanDishoeck2025,Arabhavi24,Arabhavi2025,Arulanantham25,Banzatti25}, therefore enabling analysis at the population level.

In order to build a time-sequence of inner-disk chemistry,
it is essential to observe disks with JWST MIRI at distinctly different evolutionary stages. While previously targeted by Spitzer (e.g., \citealt{Pascucci2013}), 
MIRI/MRS surveys have so far not systematically targeted the inner chemistry of an ensemble of older disks, representative of the tail end of their gas-rich phase ($\gtrsim$2~Myr).
This age range is especially consequential because it coincides with the epoch when planet formation and disk dispersal jointly shape the final volatile inventories of planets \citep{Fedele10,Pfalzner2022}.

Upper Scorpius (hereafter Upper Sco) is a complex association composed of sub-regions with mean ages spanning $\sim$2–14~Myr \citep{Blaauw1978,Ratzenbock2023a,Ratzenbock2023b}; by selecting disks in the $\sim$2–6~Myr range, we target systems undergoing the critical late stages of planet formation and disk evolution.
Filling this observational gap with JWST/MIRI MRS spectroscopy observations of Upper Sco thereby provides a direct view of the inner $\lesssim$1~au chemistry in a population that is both nearby (mean distance $\sim$145~pc; \citealt{BailerJones2021}) and representative of older disks. 

In this paper, we present the design and results of our Upper Sco MIRI/MRS survey. We fit molecular emission from slab models to our MIRI/MRS spectra in order to quantify emitting temperatures, column densities, and projected emitting areas for a suite of molecules, which we then use to calculate characteristic emission luminosities for each molecule in each disk.
Section~\ref{sec:method} describes the sample selection/properties, JWST observations/data reduction, continuum subtraction, slab‐modeling, and line luminosity calculations. Section~\ref{sec:results} details the results of our molecular retrievals, our new definitions of the wavelength ranges of various molecular diagnostic lines, descriptions of each new observational class of disk seen in our sample, and comparisons to younger (1-3~Myr) star-forming regions. In Section~\ref{sec:discussion}, we present several possible theoretical mechanisms that could explain the water depletion, chemical diversity, and cold organics of our sample, as well as comment on the implications for planet formation. Finally, Section~\ref{sec:conclusion} concludes with a summary and prospects for future age‐dependent studies of disk chemistry.

\section{Observations and Methods} \label{sec:method}

In the following section, we formally introduce the Upper Sco disk sample, and subsequently outline the data reduction, continuum subtraction, slab modeling, and line luminosity calculation methodology used in this paper.

\subsection{Selected Sample} \label{subsec:sample}

\begin{splitdeluxetable*}{llcccccccccccBccccccccccccc}
\tablewidth{\textwidth}
\tabletypesize{\scriptsize}
\tablecaption{
Summary of physical source properties for the 10 USco~disks, sorted primarily by Water Classification Type (see Table~\ref{tab:classes} and Sec.~\ref{subsec:classes}). Notable values that we reference in Sec.~\ref{sec:discussion} are labeled in bold. d2g refers to the dust-to-gas mass ratio. Spectral index n$_{13-26}$ (description in Sec.~\ref{subsec:cavities}) values are measured in MIRI using line-free regions as defined in \citet{Banzatti25}. For the sub-mm substructure taken from \citet{Vioque2025}, the \CIRCLE\ denotes a centrally-concentrated radial profile, \mycircle\ one with centrally-concentrated + ``shouldered" radial profile, \Bigcircle\ one with a dust ring, and - denotes an only marginally-resolved disk with ALMA. Confirmed dust and gas inner-disk cavities from \citet{Vioque2025} are labeled with \textcolor{green}{\cmark}, while tentative inner-disk cavities are labeled with \textcolor{orange}{\cmark}. While \citet{Vioque2025} additionally report USco~6, 7, and 8 as having inner-disk dust cavities, we note that we simply refer to these as dust traps (see Sec.~\ref{subsec:cavities} for more).
\label{tab:sources}
}
\tablehead{
\multicolumn{1}{l}{\textbf{Water}} &
\multicolumn{1}{l}{} &
\colhead{USco} &
\colhead{2MASS} &
\colhead{d$^a$} &
\colhead{Age$^b$} &
\colhead{Spectral$^{c,d}$} &
\colhead{$M_{\star}$$^b$} &
\colhead{$L_{\star}$$^b$} &
\colhead{$\dot{M}$$^d$} &
\colhead{log($M_{\rm gas}$)$^e$} &
\colhead{$M_{\rm dust}$$^b$} &
\colhead{i$^f$} &
\colhead{\textbf{Water}} &
\colhead{Chemo-} &
\colhead{USco} &
\colhead{log($L_{acc}$)$^g$} &
\colhead{$R_{\rm gas,90\%}$$^h$} &
\colhead{$R_{\rm dust,90\%}$$^f$} &
\colhead{$R_{\rm gas,90\%}$} &
\colhead{d2g} &
\colhead{n$_{13-26}$} &
\colhead{Sub-mm$^f$} &
\colhead{$R_{\rm dust\,ring}$$^f$} &
\colhead{Inner Dust$^{f,i}$} &
\colhead{Inner Gas$^{f,i}$} \\
\multicolumn{1}{l}{\textbf{Classification}} &
\multicolumn{1}{l}{Chemotype} &
\colhead{\#} &
\colhead{Name} &
\colhead{(pc)} &
\colhead{(Myr)} &
\colhead{Type} &
\colhead{$(M_{\odot})$} &
\colhead{$(L_{\odot})$} &
\colhead{$(M_{\odot}\ \mathrm{yr}^{-1})$} &
\colhead{(log $M_{\odot}$)} &
\colhead{$(M_{\Earth})$} &
\colhead{$^{\circ}$} &
\colhead{\textbf{Class}} &
\colhead{Type} &
\colhead{\#} &
\colhead{(log(L$_\odot$))} &
\colhead{(au)} &
\colhead{(au)} &
\colhead{($R_{\rm dust,90\%}$)} &
\colhead{(\%)} &
\colhead{Index} &
\colhead{Substructure} &
\colhead{(au)} &
\colhead{Cavity} &
\colhead{Cavity}
}
\startdata
    \noalign{\vskip 2.5pt}
    \multicolumn{1}{l}{Water-Rich (WR)} & - & 5 & J16145026-2332397 & 144.0 & 3.8 & M3 & 0.29 & 0.11 & - & -5.4 & 0.4 & 15
    & WR & - & 5 & -1.31 & 30.8 & $<$19 & $>$2.6 & \textbf{26.1} & -0.02 & - & - & - & - \\
    \arrayrulecolor{gray!50}\noalign{\vskip 0.6pt}\hline\arrayrulecolor{black}\noalign{\vskip 0.6pt}
    \multicolumn{1}{l}{Water-Poor (WP)} & - & 10 & J16090075-1908526 & 136.9 & 1.5 & M0 & 0.53 & 0.35 & 1.7$\times10^{-9}$ & -2.1 & 13.4 & 49
    & WP & - & 10 & -2.20 & 74.6 & 42.0 & 1.8 & 0.6 & 0.36 & \mycircle\ & 35 & - & - \\
    \multicolumn{1}{l}{Water-Poor (WP)} & \multicolumn{1}{l}{CO$_2$-Dominated (CD)} & 3 & J16020757-2257467 & 139.6 & 3.6 & M2 & 0.37 & 0.15 & 1.1$\times10^{-11}$ & -3.2 & 1.0 & 58
    & WP & CD & 3 & -4.15 & 34.1 & 27.0 & 1.3 & 0.5 & -0.12 & \CIRCLE\ & - & - & \color{orange}\cmark \\
    \multicolumn{1}{l}{Water-Poor (WP)} & \multicolumn{1}{l}{Organic-Rich (OR)} & 9 & J16082324-1930009 & 137.0 & 6.6 & M0 & 0.56 & 0.24 & 7.9$\times10^{-10}$ & -1.3 & 9.8 & 75
    & WP & OR & 9 & -2.62 & 180.8 & 47.8 & \textbf{3.8} & 0.1 & -0.56 & \mycircle\ & 40 & \color{orange}\cmark & \color{green}\cmark \\
    \multicolumn{1}{l}{Water-Poor (WP)} & \multicolumn{1}{l}{Organic-Rich (OR)} & 2 & J16054540-2023088 & 137.6 & 2.0 & M4.5 & 0.13 & 0.07 & 3.6$\times10^{-10}$ & -3.9 & 2.6 & 56
    & WP & OR & 2 & -3.22 & 50.6 & $<$17 & \textbf{$>$4.0} & 6.2 & -0.51 & - & - & - & - \\
    \arrayrulecolor{gray!50}\noalign{\vskip 0.6pt}\hline\arrayrulecolor{black}\noalign{\vskip 0.6pt}
    \multicolumn{1}{l}{Water-Absent (WA)} & \multicolumn{1}{l}{CO$_2$-Dominated (CD)} & 7 & J16202863-2442087 & 152.7 & 1.8 & M2 & 0.34 & 0.23 & - & -3.4 & 1.2 & 32
    & WA & CD & 7 & -2.30 & 160.5 & 47 & 3.4 & 0.9 & -0.79 & \Bigcircle\ & 30 & \color{orange}\cmark & - \\
    \multicolumn{1}{l}{Water-Absent (WA)} & \multicolumn{1}{l}{Organic-Rich (OR)} & 8 & J16221532-2511349 & 139.0 & 2.1 & M3 & 0.29 & 0.14 & - & -2.4 & 7.8 & 56
    & WA & OR & 8 & -3.68 & 144.1 & 27.19 & \textbf{5.3} & 0.6 & -1.11 & \Bigcircle\ & 17 & \color{orange}\cmark & - \\
    \multicolumn{1}{l}{Water-Absent (WA)} & \multicolumn{1}{l}{Molecule-Absent (MA)} & 1 & J16120668-3010270 & 131.9 & 4.5 & M0.5 & 0.51 & 0.25 & 4.0$\times10^{-10}$ & -2.5 & 9.0 & 37
    & WA & MA & 1 & -1.69 & 167.0 & 87.64 & 1.9 & 0.8 & \textbf{1.37} & \Bigcircle\ & 78 & \color{green}\cmark & \color{orange}\cmark \\
    \multicolumn{1}{l}{Water-Absent (WA)} & \multicolumn{1}{l}{Molecule-Absent (MA)} & 6 & J16163345-2521505 & 158.4 & 5.8 & M0.5 & 0.52 & 0.18 & 1.2$\times10^{-11}$ & -4.2 & 0.9 & 63
    & WA & MA & 6 & -3.35 & 159.7 & 63 & 2.5 & 3.7 & \textbf{1.10} & \Bigcircle\ & 35 & \color{green}\cmark & \color{green}\cmark \\
    \multicolumn{1}{l}{Water-Absent (WA)} & \multicolumn{1}{l}{Molecule-Absent (MA)} & 4 & J16111742-1918285 & 136.9 & 2.1 & M0.25 & 0.50 & 0.35 & - & -3.7 & 0.1 & \textbf{79}
    & WA & MA & 4 & - & 46.3 & $<$56 & $>$1.3 & 0.1 & -0.08 & - & - & - & - \\
\enddata
\tablerefs{ \\
a: \citet{BailerJones2021}; \\
b: \citet{AGEPRO_IV_UpperSco}; \\
c: \citet{Manara2020}, \citet{Luhman2022_uppstars}; \\
d: \citet{Manara_PPVII}; \\
e: \citet{Trapman2025b}; \\
f: \citet{Vioque2025}; \\
g: A. Empey et al. (accepted); \\
h: \citet{Trapman2025a}; \\
i: \citet{Sierra2024}. \\
}
\end{splitdeluxetable*}

The JWST data in this paper are part of a JWST Cycle 2 Program (PID: 3034, PI: Zhang). The sources in this program are adopted from ALMA Cycle 8 Large Program (PID: 2021.1.00128.L, PI: Zhang), entitled the ALMA survey of Gas Evolution in PROtoplanetary disks (AGE-PRO) (details, including moment-zero and 1.3 mm emission maps, can be found in \citealt{Zhang25}).
The goal of this survey is to observe protoplanetary disks in three different star-forming regions, roughly representative of three different evolutionary stages: SED Class I and flat spectrum sources in the Ophiuchus region (embedded disk phase), Lupus (the middle age), and Upper Sco (the end of the disk's lifetime). Previous studies have established typical disk dust masses of $\sim$~$10^{-1}$–$10^{1}\,M_\earth$ \citep{carpenter14, Vioque2025} and median gas masses of $\sim$~$7\cdot10^{-5}\,M_\odot$ (down from median $\sim$~$7\cdot10^{-4}\,M_\odot$ for Ophiuchus) \citep{Trapman2025b} in Upper Sco, indeed indicating substantial gas depletion by 10~Myr (see Table~\ref{tab:sources} for dust-to-gas ratios).

This paper focuses on protoplanetary disks in the Upper Sco association. 
Within the Upper Sco region, the AGE-PRO ALMA survey specifically targeted sources with (i) prior detections in both mm continuum and CO line emission from previous shallow ALMA surveys and (ii) no known close companions within 600 au \citep{Zhang25}. Ten M0–M4.5 Upper Sco disks were selected that span the range of continuum luminosities within Upper Sco \citep{Zhang25,Agurto-Gangas25}, henceforth referred to as USco~\#1-10. These sources are listed in Table~\ref{tab:sources}, which summarizes the physical properties of the systems as derived from GAIA DR3 \citep{BailerJones2021,Luhman2022_uppstars}, AGE-PRO studies \citep{Agurto-Gangas25,Vioque2025,Trapman2025a,Trapman2025b}, and optical surveys (\citealt{Manara2020,Manara_PPVII}, A. Empey et al., accepted) of Upper Sco.

Our survey consists of disks of ages that span $\sim$2–6~Myr and stellar masses of $\sim0.1-0.6M_{\odot}$, which we note are derived from isochrone fitting (see \citealt{Zhang25,Agurto-Gangas25}), which results in the error bars reported in \citet{Agurto-Gangas25}. We also note that the 10 sources presented in this paper can approximately be further sub-categorized into two age bins: five sources with approximate ages 2-2.5~Myr, and five sources with approximate ages of 3.5-6~Myr (see Fig.1 in \citealt{Zhang25} and Xie et al., in press for isochrones). Every source in this sample falls into the category of relatively low mass T~Tauri star (M$_{\star}$$\lesssim 0.6M_{\odot}$); we note that USco~2 firmly qualifies as a very low-mass star (VLMS, M$_{\star}$$<0.3M_{\odot}$), whereas USco~5, 8, 7, and 3 lie increasingly near the boundary (see Table~\ref{tab:sources}). 
Parallel analyses of the embedded Ophiuchus and intermediate-age Lupus samples will be presented in a forthcoming paper (Waggoner et al., in prep).

\subsection{Observations and Data Reduction} \label{subsec:reduction}

MIRI-MRS provides integral-field spectroscopy across four channels covering 4.9–28.0 $\upmu$m with spectral resolving power $R\sim1500$–3500 \citep{Wells2015,Argyriou20,Pontoppidan2024,Banzatti25}. The observations, as listed in Table~\ref{tab:observations}, were taken during fall of 2024 and employed a four-point dither pattern optimized for all four channels. The data were reduced using the same general procedures described in \citet{Pontoppidan2024} but reprocessed with the updated JWST Disk Infrared Spectral Chemistry Survey (JDISCS) pipeline (internal v9.1). The JDISCS pipeline builds on the JWST Calibration Reference Data System reductions (stage 2b; JWST pipeline v1.18.0, CRDS 12.1.5 \citealp{Bushouse2025}) and applies dedicated background subtraction and fringe correction using both stellar and asteroid calibrators \citep{Pontoppidan2024,Arulanantham25};  these improvements yield up to a $\sim$6× signal-to-noise gain in MIRI channel 4. The fringe calibration has been refined \citep{Humes2024} and the spectrophotometric calibration has been benchmarked against CALSPEC standards \citep{Bohlin2020}.

The wavelength-dependent spectral extraction apertures are tied to the diffraction limit so that the physical extraction radius increases with wavelength; specifically, for channels 1–4 the aperture radii are 1.4, 1.3, 1.2, and 1.1 times 1.22$\uplambda$/D, respectively. To improve spectrophotometric fidelity the identical aperture definition is applied to each science target and its calibrator so that wavelength-dependent structure in the point-spread function is effectively removed during calibration.
Updates to the internal JDISCS reduction pipeline since \citet{Pontoppidan2024} (now reflected in v9.1) lead to differences between the reductions presented here and those in \citet{Arulanantham25} using pipeline v8.4, resulting in $\sim$10–20\% higher fluxes in the 18–27 $\upmu$m range.

\subsection{Continuum Subtraction} \label{subsec:csub}

\begin{figure*}[ht!]
    \centering
    \includegraphics[width=\textwidth]{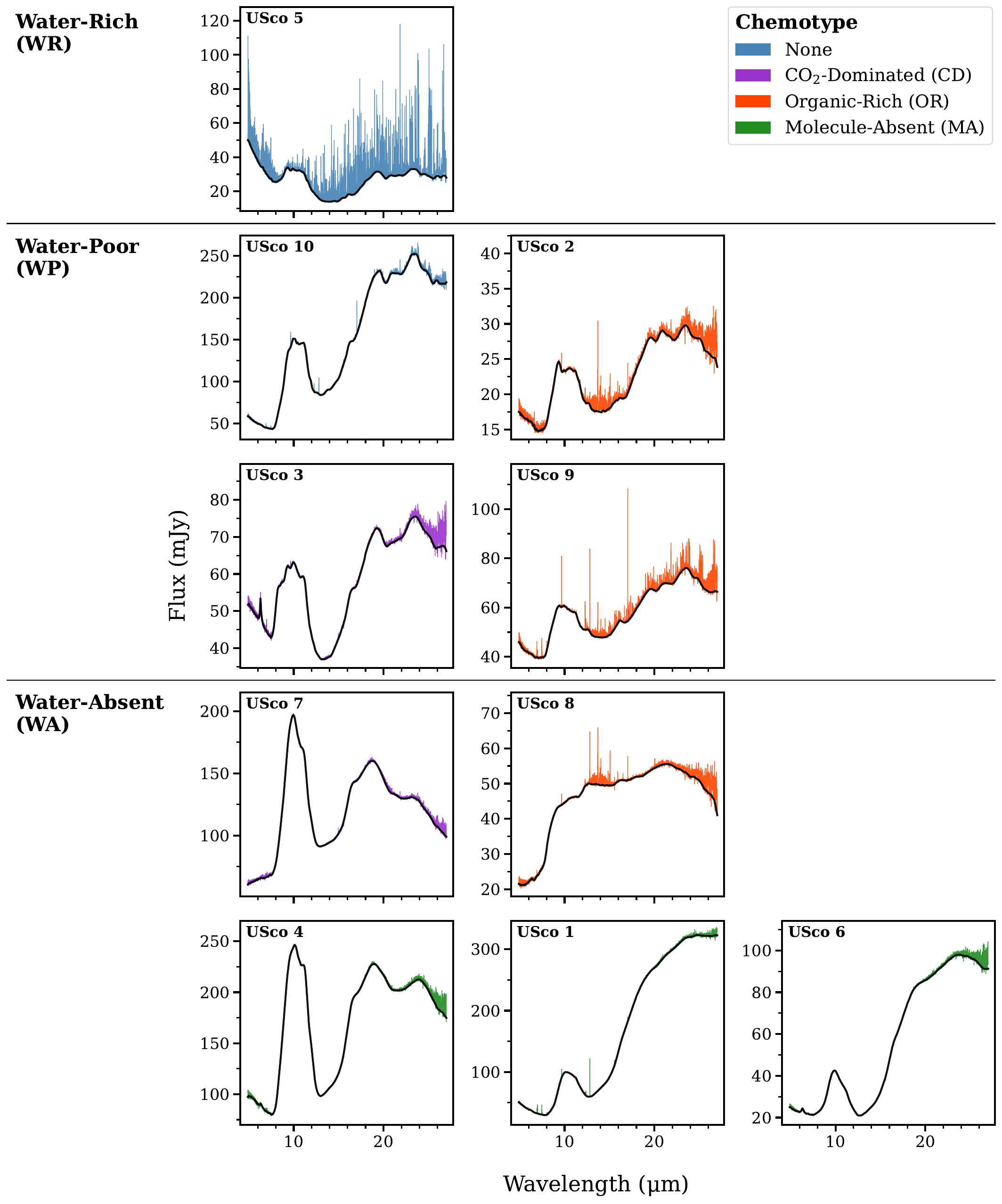}
    \caption{Full MIRI/MRS of the Upper Sco sample over 4.9-28 $\upmu$m, grouped primarily by Water Classification Type (as defined in Table~\ref{tab:classes}). The black lines within each panel denote the continuum, whereas colored lines represent the total flux density of the spectra, grouped by Chemotype. We note that while several patterns within each group are perceptible in the silicate features presented here, these are not discussed in this work and will be analyzed in a separate forthcoming publication.
    \label{fig:continuum}}
\end{figure*}

We subtract the continuum of MIRI spectra using \texttt{ctool} \citep{Pontoppidan_2026_ctool} as described in \citet{Pontoppidan2024} and updated in \citet{Banzatti25}, where Fig.~\ref{fig:continuum} shows the fit to the continuum for each source. This method determines the continuum using an iterative smoothing and rejection routine: the spectrum is smoothed with a broad filter, data points that lie above a given threshold are rejected and replaced by linear interpolation, and this process is then repeated. We use 5 iterations and a box size of 55.
This algorithm is designed to estimate the underlying dust continuum across all four channels of MIRI in cases where narrow molecular lines are superimposed on broad silicate or ice emission.

While this method provides a reasonable continuum baseline across all four channels when observing narrow gas emission on top of the broad dust emission, we note that estimations become more uncertain when dense clustering of organic line emission produce a so-called ``pseudo-continuum" \citep{carr11, Liu19b, Tabone23, Kanwar2024b, Arabhavi24, Kaeufer2024b}. Consequently, in the disks in our sample that exhibit high SNR organic features (USco~2, 8, and 9), we exclude (interpolate over) the organic region from $\sim$ 13.4-14.1 $\mathrm{\upmu m}$ from our continuum fitting procedure in order to avoid oversubtracting organic features of this region, similar to the approach taken by \citet{Arabhavi24, Banzatti25}. This window encompasses the overlapping wings of HCN and C$_2$H$_2$ emission complexes, where line blending is most severe and the pseudo-continuum dominates; however, this approach can still neglect that the true continuum may be hidden beneath a forest of blended transitions \citep{Tabone23}.

\subsection{Slab Modeling} \label{subsec:model}

In order to accurately assess the multitude of overlapping molecular emission in the JWST/MIRI spectra, we employ slab models using the medium/high-resolution IR molecular spectra analysis package \texttt{spectools\_ir} \citep{Salyk22}. This framework reproduces the primary molecular features in the observed spectrum of each disk by assuming molecular emission arises from a slab of gas in local thermodynamic equilibrium (LTE).
The best-fitting slab models are used to constrain the physical parameters of temperature (T) in K, column density (log(N)) in cm$^{-2}$, and projected emitting area (log(A)) in au$^2$ for each molecule. In addition, we use these best-fitting models, as opposed to the spectra, to measure line fluxes of most species (with the exception of H$_2$O, CH$_3$, and C$_6$H$_6$; see Sec.~\ref{subsec:integrations}) to separate the overlapping emission of many species.

Detection significance for tentatively detected lines are assessed with the spectral visualization tool iSLAT \citep{Jellison24, iSLAT_code}, which reports the S/N of features relative to the surrounding noise; lines with S/N$\geq$3 in the original continuum-subtracted spectrum are considered real detections, and lines with 1$<$S/N$<$3 are considered tentative. For non‐detections and tentative detections, we derive upper limits by fixing the slab temperature to the average T and log(N) found for that species in other disks in our sample and determining the maximum log(A) that remains consistent with the noise. 

Using the HITRAN 2024 emission line database \citep{Gordon2026}, we generate model spectra across the full 11.75 to 27 $\upmu$m range. The modeled spectra are split into two independently fitted windows from 11.75-18 $\upmu$m and 18-27 $\upmu$m, allowing us to select different molecules to fit to within each region, as different molecular features dominate different parts of the spectra. These windows are largely chosen due to the presence of non-LTE effects of H$_2$O and CO
below 12 $\upmu$m and extremely noisy data from 27-28 $\upmu$m \citep{Pontoppidan2024}. Our model spectra are convolved with a Gaussian line profile with a full width at half maximum of 110 km/s from 11.75-18 $\upmu$m and 130 km/s for 18-27 $\upmu$m \citep{Pontoppidan2024,Banzatti25}. Our setup using \texttt{spectools\_ir} also includes a local velocity distribution term that incorporates both the thermal and turbulent velocity, for which we use a width (standard deviation) of 1 km/s (compared to 2 km/s as used in \citet{Tabone23,Long25}, as thermal broadening is typically small around low-mass stars).

We retrieve slab parameters (T, log(N), and log(A)) for each molecule using a Markov chain Monte Carlo (MCMC) ensemble sampler \citep{Foreman-Mackey2013}. We note that for some species, such as C$_2$H$_2$, HCN, and CO$_2$, the strong degeneracy between log(N), and log(A) dominates the formal uncertainties \citep{Arulanantham25}. Hence, we additionally compute and report the corresponding masses as a derived parameter for each molecule \citep{Arulanantham25}.
We iterate over the MCMC sample space using 5,000 steps for our initial models, with an additional 10,000 added to our final models to constrain the model parameters and their errors. We use 8 times as many walkers as there are free parameters in our model (3 for each molecular component included in the iteration), and we set the first 500 steps as burn-in (discarded in best-fitting value and error calculation).

The initial priors for most species in our models are as follows, which uniformly sample both linear and log space: T (K) = $\mathcal{U}$ (200, 1400), log(N) ($\mathrm{cm^{-2}}$) = $\mathcal{U}$ (14, 22), and log(A) ($\mathrm{au^2}$) = $\mathcal{U}$ (-3.0, 3.0). As our models improve over multiple iterations and we obtain better constraints on the free parameters of each molecular component for each source, we slightly tighten the ranges of our priors to speed up the computation while assuring that enough width is left for the sampler to adequately compute errors. Alternatively, if any best-fitting parameters approach the edge of one of the priors (e.g., in the case of cold C$_4$H$_2$), we expand the prior range for that singular molecular component and run a new iteration with the widened priors.

We initially include the molecules that typically exhibit the most prominent spectral features in disks within the MIRI/MRS wavelength range (C$_2$H$_2$, HCN, CO$_2$, 
H$_2$O, and OH) in a single retrieval for every disk over the primary LTE emitting regions of these species ($\sim$11.75-27 $\upmu$m). For H$_2$O, we initially simultaneously include 3 distinct temperature components, corresponding to both ``hot" and ``warm" H$_2$O ($\mathrm{H_2O_{Hot}}$ and $\mathrm{H_2O_{Warm}}$) in the 11.75-18 $\upmu$m independently-fitted window, and both ``warm" and ``cold" H$_2$O ($\mathrm{H_2O_{Cold}}$) from 18-27 $\upmu$m. This results in $\mathrm{H_2O_{Warm}}$ being fit to the water emission across the full 11.75-27 $\upmu$m model range, while $\mathrm{H_2O_{Hot}}$ and $\mathrm{H_2O_{Cold}}$ are exclusively fit to the water emission in each wavelength window, as different temperatures of H$_2$O tend to dominate different parts of the spectrum \citep{Temmink2024,Banzatti25}. These correspond to adjusted initial priors of T$_\mathrm{H_2O_{Hot}}$ (K) = $\mathcal{U}$ (600, 1400), T$_\mathrm{H_2O_{Warm}}$ (K) = $\mathcal{U}$ (300, 600), and T$_\mathrm{H_2O_{Cold}}$ (K) = $\mathcal{U}$ (100, 300). We also include two distinct temperature components of OH, with corresponding priors of T$_\mathrm{OH_{Hot}}$ (K) = $\mathcal{U}$ (500, 2500), and T$_\mathrm{OH_{Cold}}$ (K) = $\mathcal{U}$ (100, 500).

This setup allows us to visually determine the molecules most likely to contribute to blending of other features and subtract these contributions, if present, from the continuum-subtracted spectra for identifications of more complex species in the residuals (e.g., C$_2$H$_6$, $^{13}$CO$_2$).
If any of these initial species are visually determined to definitively not be detected in the spectrum, it is removed from our retrieval and is not subtracted from the spectrum, while tentative molecular detections are kept for subsequent iterations of the model but are not subtracted from the spectra at this stage.

We then use iSLAT to identify any remaining features in the subtracted spectra against known HITRAN database lines \citep{Gordon2026}; when a candidate species is found, we rerun the MCMC sampler on the original continuum-subtracted spectrum with that molecule added to the existing suite. This iterative inclusion ensured comprehensive coverage of detectable species without presupposing their presence in the spectrum.
We note that HITRAN is missing line data for several molecules previously detected in disks (e.g., C$_6$H$_6$, C$_3$H$_4$), so we instead elect to simply identify these detections (as well as H$_2$ due to its typically isolated features) by eye based on the consistent presence of features at known wavelengths from other spectroscopic databases (e.g., \citealt{Delahaye2021}).

\subsection{Line Luminosity Calculations} \label{subsec:integrations}

Our slab modeling methodology, which isolates respective molecular contributions to the spectrum, enables us to quantify the discrete feature strength of each individual molecule in each source. We calculate line luminosities via a simple trapezoidal integration of the flux over a selected diagnostic region for every detected molecule in our sample, as are defined in Table~\ref{tab:line_luminosities}, which is then converted to luminosity using the distances listed in Table~\ref{tab:sources}. This is with the exception of CO due to its complex non-LTE emission within the MIRI wavelength range (see Fig.~\ref{fig:upsco5_contribution2}), H$_2$ as we do not fit to these lines, and $^{13}$CCH$_2$ due to blending with C$_2$H$_2$. The C$_2$H$_2$, HCN, and CO$_2$ integrated wavelength ranges are the same as used in \citet{Gasman2025} and \citet{Arabhavi2026}. Similarly, hot, warm, and cold water vapor each have their own corresponding diagnostic lines (two in the case of cold: H$_2$O$_{\mathrm{Cold,a}}$ and H$_2$O$_{\mathrm{Cold,b}}$), all of which are taken from \citet{Banzatti25}; these correspond to upper-level energies (E$_u$) of $\sim$6000 K, 3600 K, and 1500 K. Due to blending of different H$_2$O temperature components, we elect to integrate over the spectrum data for each diagnostic line rather than an individual H$_2$O temperature component. As CH$_3$ and C$_6$H$_6$ emission is not included in our models, we also integrate these features over the data after having subtracted the best-fitting slab model from the spectrum.

Errors and upper limits are computed in a similar manner as Xie et al. (in press), in which the noise level is determined by selecting a feature-free region near each molecular feature of each disk in which no molecular or atomic lines are present. This characteristic noise is then used as the standard deviation for our integration, which is then propagated to compute 3-$\sigma$ upper-limits for each tentative detection, as well as for every non-detection of C$_2$H$_2$, HCN, CO$_2$, and H$_2$O temperature component.

We additionally compute a characteristic derived luminosity value L$_\mathrm{organic}$, which is defined as the sum of all individual integrated luminosities or upper limits of C$_2$H$_2$ and HCN, as well as any \textit{definitively detected} organic molecules in the spectrum.
This allows us to compute ratios between several luminosities as derived quantities, which are subsequently used in Sec.~\ref{subsec:classes}.
Due to severe line blending of organic features, we calculate additional line luminosities for C$_2$H$_2$, HCN, and CO$_2$ (C$_2$H$_{2,\, \mathrm{tot}}$, HCN$_{\mathrm{tot}}$, and CO$_{2,\, \mathrm{tot}}$, respectively) integrated over the entire 12-16 $\upmu$m region (not included in L$_\mathrm{organic}$ calculation) in the same manner as, and allowing for a better comparison to \citet{Arulanantham25}.

\section{Results} \label{sec:results}

Using the methodology described in Sec.~\ref{subsec:model}, we report the slab model fits to the nine disks (USco~1, 2, 3, 5, 6, 7, 8,
9, and 10) with definitive or tentative molecular features (see Sec.~\ref{subsec:ma} for notes on the faint-emission source USco~4). The best-fitting temperatures, column densities, projected emitting areas, and
derived masses for each molecule and source are reported in
Table~\ref{tab:bestfit_params}, with the corresponding best-fit model spectra
shown in Figures~\ref{fig:contributions_org}, \ref{fig:contributions_standard},
and \ref{fig:upsco5_contribution1}.

\begin{figure*}[ht!]
    \centering
    \includegraphics[width=1.0\textwidth]{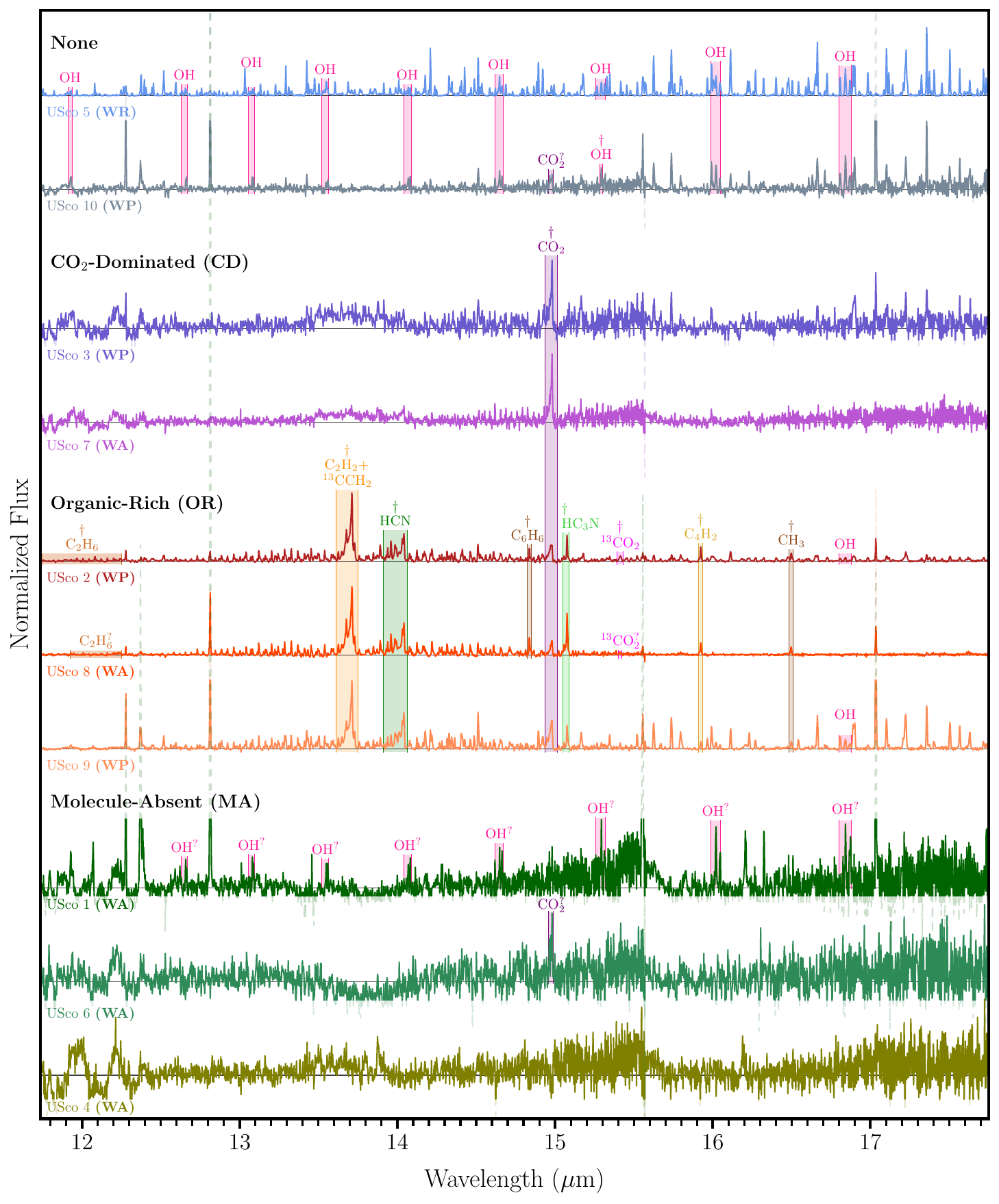}
    \caption{The full sample of the normalized continuum subtracted spectra of the ten Upper Sco disks presented in this paper, grouped primarily by Chemotype (as defined in Table~\ref{tab:classes}). Boxes highlighting the main features of each molecule are included to demonstrate the common features between each source.
    $\mathit{\dagger}$ symbols denote boxes that correspond to the diagnostic line regions defined in Table~\ref{tab:line_luminosities}, while $\mathit{?}$ symbols denote tentative features/detections.
    Water lines are not classified in this plot as they fill most of the displayed wavelength range (see Fig.~\ref{fig:upsco5_contribution1}). Each spectrum is normalized to the highest non-atomic line feature in the 11.75-18 $\upmu$m region. Extremely high or low (below-continuum) values in the continuum-subtracted spectra (typically from these atomic lines) are displayed as dashed lines for visual clarity; we note the [Ne II] atomic line of USco~1 takes up most of the graph. 
    \label{fig:identifications}}
\end{figure*}

We detect a rich diversity of molecules in our sample; Table~\ref{tab:detections} summarizes our molecular detections. Our definitive molecular detections are as follows: molecular hydrogen (H$_2$) is definitively detected in six sources, water vapor (H$_2$O) in five sources; CO$_2$ in five sources, OH in four; simple organics (C$_2$H$_2$, HCN, $^{13}$C$_2$H$_2$, HC$_3$N, and CH$_3$) and complex hydrocarbons (C$_2$H$_6$, C$_4$H$_2$, and C$_6$H$_6$) in three ``Organic-Rich" sources (USco~2, 8, and 9, with C$_6$H$_6$ only in USco~2 and 8 and C$_2$H$_6$ only in USco~2), $^{13}$CO$_2$ in one source, and CO in one source. Fig.~\ref{fig:identifications} shows the locations of each of these detections (barring H$_2$ and CO) in the 11.75-18 $\upmu$m region of the continuum-subtracted spectra.

\clearpage
\begin{deluxetable}{l l c c c c c c c c c c c c c c c c c c c}
\tabletypesize{\scriptsize}
\tablewidth{0pt}
\tablecaption{Summary of molecular detections in the MIRI spectra of the 10 Upper Scorpius disks, sorted primarily by Chemotype (as defined in Table~\ref{tab:classes}). Definitive detections are marked in green, tentative detections are marked in orange, and non-detections are marked in gray, as defined in Sec.~\ref{subsec:model}.
\label{tab:detections}}
\tablehead{
  \multicolumn{1}{l}{\textbf{Chemotype}} &
  \multicolumn{1}{l}{Water Classification} &
  \colhead{USco~\#} &
  \colhead{\rotatebox{90}{CO}}  & \colhead{\rotatebox{90}{H$_2$O$_\mathrm{H}$}} & \colhead{\rotatebox{90}{H$_2$O$_\mathrm{W}$}}
  & \colhead{\rotatebox{90}{H$_2$O$_\mathrm{C}$}} & \colhead{\rotatebox{90}{CO$_2$}}   & \colhead{\rotatebox{90}{OH$_\mathrm{H}$}} & \colhead{\rotatebox{90}{OH$_\mathrm{C}$}}  & \colhead{\rotatebox{90}{C$_2$H$_2$}} & \colhead{\rotatebox{90}{HCN}} &
  \colhead{\rotatebox{90}{$^{13}$CO$_2$}} & \colhead{\rotatebox{90}{$^{13}$CCH$_2$}}  & \colhead{\rotatebox{90}{HC$_3$N}}
  & \colhead{\rotatebox{90}{C$_4$H$_2$}} & \colhead{\rotatebox{90}{C$_2$H$_6$}} & \colhead{\rotatebox{90}{C$_6$H$_6$}}
  & \colhead{\rotatebox{90}{CH$_3$}} & \colhead{\rotatebox{90}{CH$_4$}} & \colhead{\rotatebox{90}{H$_2$}}
}
\startdata
- & Water-Rich (WR) & 5 & \myboxxx{} & \myboxxx{}  & \myboxxx{}  & \myboxxx{} & \mybox{} & \myboxxx{} & \myboxxx{} & \mybox{} & \mybox{} & \mybox{} & \mybox{} & \mybox{} & \mybox{} & \mybox{} & \mybox{} & \mybox{} & \mybox{} & \mybox{} \\
- & Water-Poor (WP) & 10 & \mybox{} & \mybox{}  & \myboxxx{}  & \myboxxx{} & \myboxx{}& \myboxxx{} & \myboxxx{} & \mybox{} & \mybox{} & \mybox{} & \mybox{} & \mybox{} & \mybox{} & \mybox{} & \mybox{} & \mybox{} & \mybox{} & \myboxxx{} \\
\arrayrulecolor{gray!50}\noalign{\vskip 0.6pt}\hline\arrayrulecolor{black}\noalign{\vskip 0.6pt}
CO$_2$-Dominated (CD) & Water-Poor (WP) & 3 & \mybox{} & \mybox{} & \myboxxx{} & \myboxx{} & \myboxxx{} & \mybox{} & \mybox{} & \mybox{} & \mybox{} & \mybox{} & \mybox{} & \mybox{} & \mybox{} & \mybox{} & \mybox{} & \mybox{} & \mybox{} & \myboxxx{} \\
CO$_2$-Dominated (CD) & Water-Absent (WA) & 7 & \mybox{} & \mybox{} & \mybox{} & \myboxx{} & \myboxxx{} & \mybox{} & \mybox{} & \mybox{} & \mybox{} & \mybox{} & \mybox{} & \mybox{} & \mybox{} & \mybox{} & \mybox{} & \mybox{} & \mybox{} & \mybox{} \\
\arrayrulecolor{gray!50}\noalign{\vskip 0.6pt}\hline\arrayrulecolor{black}\noalign{\vskip 0.6pt}
Organic-Rich (OR) & Water-Poor (WP) & 2 & \mybox{} & \myboxxx{}  & \myboxxx{} & \myboxxx{} & \myboxxx{} & \myboxxx{}  & \mybox{} & \myboxxx{} & \myboxxx{} & \myboxxx{} & \myboxxx{}  & \myboxxx{} & \myboxxx{} & \myboxxx{} & \myboxxx{}  & \myboxxx{} & \myboxx{} & \myboxxx{} \\
Organic-Rich (OR) & Water-Absent (WA) & 8 & \mybox{} & \mybox{} & \mybox{} & \mybox{} & \myboxxx{} & \mybox{} & \mybox{} & \myboxxx{} & \myboxxx{} & \myboxx{} & \myboxxx{} & \myboxxx{} & \myboxxx{} & \myboxx{} & \myboxxx{} & \myboxxx{} & \mybox{} & \myboxxx{} \\
Organic-Rich (OR) & Water-Poor (WP) & 9 & \mybox{} & \mybox{} & \myboxxx{} & \myboxxx{} & \myboxxx{} & \myboxxx{}  & \mybox{} & \myboxxx{} & \myboxxx{} & \mybox{} & \myboxxx{} & \myboxxx{} & \myboxxx{} & \mybox{}  & \mybox{} & \myboxxx{}  & \mybox{}  & \myboxxx{} \\
\arrayrulecolor{gray!50}\noalign{\vskip 0.6pt}\hline\arrayrulecolor{black}\noalign{\vskip 0.6pt}
Molecule-Absent (MA) & Water-Absent (WA) & 1 & \mybox{} & \mybox{} & \mybox{} & \mybox{} & \mybox{} & \myboxx{} & \mybox{} & \mybox{} & \mybox{} & \mybox{} & \mybox{} & \mybox{} & \mybox{} & \mybox{} & \mybox{} & \mybox{} & \mybox{} & \myboxxx{} \\
Molecule-Absent (MA) & Water-Absent (WA) & 6 & \mybox{} & \mybox{} & \mybox{} & \mybox{} & \myboxx{} & \mybox{} & \mybox{} & \mybox{} & \mybox{} & \mybox{} & \mybox{} & \mybox{} & \mybox{} & \mybox{} & \mybox{} & \mybox{} & \mybox{} & \mybox{} \\
Molecule-Absent (MA) & Water-Absent (WA) & 4 & \mybox{} & \mybox{} & \mybox{} & \mybox{} & \mybox{} & \mybox{} & \mybox{} & \mybox{} & \mybox{} & \mybox{} & \mybox{} & \mybox{} & \mybox{} & \mybox{} & \mybox{} & \mybox{} & \mybox{} & \mybox{} \\
\enddata
\end{deluxetable}

\subsection{New Diagnostic Line Luminosity Regions} \label{subsec:luminosities}
Because our sample spans Organic-Rich, Water-Rich, CO$_2$-dominant and Molecule-Absent disks, we are often able to identify spectral regions where a given molecule appears in isolation or with minimal blending. We take advantage of this spectral diversity and use these clean features as a ``diagnostic line" wavelength range for that molecule \citep[e.g.,]{Banzatti25,Arabhavi2026}.
This results in a consistent set of luminosities across species that exploits the complementary strengths of different sources in the sample. Table~\ref{tab:line_luminosities} summarizes the integrated line luminosities and their errors for each, as well as multiple derived quantities, using the methodology outlined in Sec.~\ref{subsec:integrations}. Our integrated line luminosities span $\sim10^{-9}$–$10^{-5}\,L_\odot$, with the strongest emitters in the Organic‐Rich and Water-Rich subsets (see Table~\ref{tab:line_luminosities}).

\subsection{Chemical Classes} \label{subsec:classes}

One of the main results of this survey is the large diversity of spectra among 10 disks drawn from a single star-forming region. Examination of the detection patterns (Table~\ref{tab:detections}) and line luminosities (Table~\ref{tab:line_luminosities}) reveals a natural division of the sample into empirically distinct chemical groups. We classify each disk along two independent axes: a \textit{Water Classification} based on the strength of H$_2$O emission, and an optional Special Classification (or, hereafter \textit{Chemotype}) based on the dominant non-water chemistry, if present. The definitions are presented in (Table~\ref{tab:classes}), based on the integrated line luminosities and their ratios. Every disk receives exactly one label on the Water Classification axis, and an additional one on the Chemotype axis, if applicable. The two axes are complementary, not mutually exclusive.

\begin{deluxetable}{l|l|l}
    \tablesize
    \tabletypesize{\tablesize}
    \tablewidth{0pt}
    \tablecaption{Criteria for each defined chemical class of disks. Luminosities are calculated using the methodology defined in Sec.~\ref{subsec:integrations} and are reported in Sec.~\ref{subsec:luminosities}. We use the definition of a complex organic molecule (COM) from \citet{Herbst-vanDishoeck2009}: any carbon-based molecule with at least 6 constituent atoms.
    \label{tab:classes}}
    \tablehead{
      \multicolumn{3}{@{}l@{}}{
        \makebox[4.6cm][l]{\large{Chemical Class}}
        \makebox[5.85cm][l]{\large{Criteria}}
        \makebox[4.4cm][l]{\large{Constituents}}
      }
    }
    \startdata
        \tblrulevskip\tableline\tblrulevskip
        \multicolumn{3}{@{}l@{}}{\bfseries Water Classification Types} \\
        \tblrulevskip\tableline\tblrulevskip
        $\textcolor{blue}{\hookrightarrow}$ Water-Rich \textbf{(WR)} &
        $\sbullet$ $\mathrm{L_{H_2O_{Warm}}}\ \gtrsim\ -6\ \log(L_\odot)$ &
        USco~5 \\[3pt]
        \arrayrulecolor{gray!50}\tblrulevskip\hline\arrayrulecolor{black}\tblrulevskip
        $\textcolor{cyan}{\hookrightarrow}$ Water-Poor \textbf{(WP)} &
        \makecell[l]{$\sbullet$ $\mathrm{L_{H_2O_{Warm}}}\ <\ -6\ \log(L_\odot)$ \\[3pt]
        $\sbullet$ H$_2$O detected} &
        USco~2, 3, 9, 10 \\[3pt]
        \arrayrulecolor{gray!50}\tblrulevskip\hline\arrayrulecolor{black}\tblrulevskip
        $\textcolor{CadetBlue}{\hookrightarrow}$ Water-Absent \textbf{(WA)} &
        $\sbullet$ No H$_2$O detected &
        USco~1, 4, 6, 7, 8 \\
        \tblrulevskip\tableline\tblrulevskip
        \tblrulevskip\tableline\tblrulevskip
        \multicolumn{3}{@{}l@{}}{\bfseries Chemotype} \\
        \tblrulevskip\tableline\tblrulevskip
        $\textcolor{violet}{\hookrightarrow}$ CO$_2$-Dominated \textbf{(CD)} &
        \makecell[l]{$\sbullet$ $\mathrm{L_{CO_2}}/\mathrm{L_{H_2O_{Warm}}}\ \gtrsim\ 5$ \\[3pt]
        $\sbullet$ $\mathrm{L_{CO_2}}/\mathrm{L_{organic}}\ \gtrsim\ 0.5$} &
        USco~3, 7 \\
        \arrayrulecolor{gray!50}\tblrulevskip\hline\arrayrulecolor{black}\tblrulevskip
        $\textcolor{orange}{\hookrightarrow}$ Organic-Rich \textbf{(OR)} &
        \makecell[l]{$\sbullet$ $\mathrm{L_{organic}}\ \gtrsim\ -6\ \log(L_\odot)$ \\[3pt]
        $\sbullet$ $\geq$1 COM detected} &
        USco~2, 8, 9 \\
        \arrayrulecolor{gray!50}\tblrulevskip\hline\arrayrulecolor{black}\tblrulevskip
        $\textcolor{SpringGreen}{\hookrightarrow}$ Molecule-Absent \textbf{(MA)} &
        $\sbullet$ No non-H$_2$ molecules detected &
        USco~1, 4, 6 \\
    \enddata
\end{deluxetable}

\subsubsection{Water Classification Types} \label{subsubsec:water_class}

We assign each disk to one of three water categories based on the detection of H$_2$O and
warm H$_2$O line luminosity
($L_\mathrm{H_2O,warm}$) as  described in Sec.~\ref{subsec:integrations} and listed in Table~\ref{tab:line_luminosities}:
\begin{itemize}
    \item \textbf{Water-Rich (WR):}
    $\log(L_\mathrm{H_2O_{Warm}}/L_\odot) \gtrsim -6$. This is the threshold above which most young ($\lesssim$ 3 Myr) disks emit (see Fig.~\ref{fig:lacc}). Only USco~5 meets this
    criterion. With robust detections of hot ($\sim$910~K), warm ($\sim$500~K),
    and cold ($\sim$210~K) H$_2$O as well, two OH components, and the only detection of CO in our sample, USco~5
    has the richest oxygen chemistry in the sample while notably lacking any
    hydrocarbon or CO$_2$ features. Physical explanations are offered in Sec.~\ref{subsec:d2g}.

    \item \textbf{Water-Poor (WP):}
    $\log(L_\mathrm{H_2O_{Warm}}/L_\odot) < -6$, with at least one H$_2$O component
    detected. This includes USco~2, 3, 9, and 10. These disks show H$_2$O at
    luminosities 0.5-2~dex below USco~5 and typically lack a hot H$_2$O
    component (with the exception of the $\sim$640~K component in USco~2).
    Physical explanations are offered in Sec.~\ref{subsec:evolution}.

    \item \textbf{Water-Absent (WA):}
    No H$_2$O features detected at S/N~$>3$. This includes USco~1, 4, 6, 7, and 8. We emphasize that USco~8 falls in this category despite otherwise being one of the most molecule-rich disks in the sample, implying that water absence does not require a globally molecule-depleted inner disk. Physical explanations are offered in Sec.~\ref{subsec:dust_trapping}.
\end{itemize}

We choose to only use $L_\mathrm{H_2O,warm}$ for this criteria, as this is the temperature component most commonly detected in our sample.
Tables~\ref{tab:sources},~\ref{tab:line_luminosities}, and Fig.~\ref{fig:continuum} group the sample by these classifications, showing the line luminosity, system property, and spectral differences among the three types.

\subsubsection{Chemotypes} \label{subsubsec:special_class}

\begin{figure*}[ht!]
    \centering
    \includegraphics[width=.9125\textwidth]{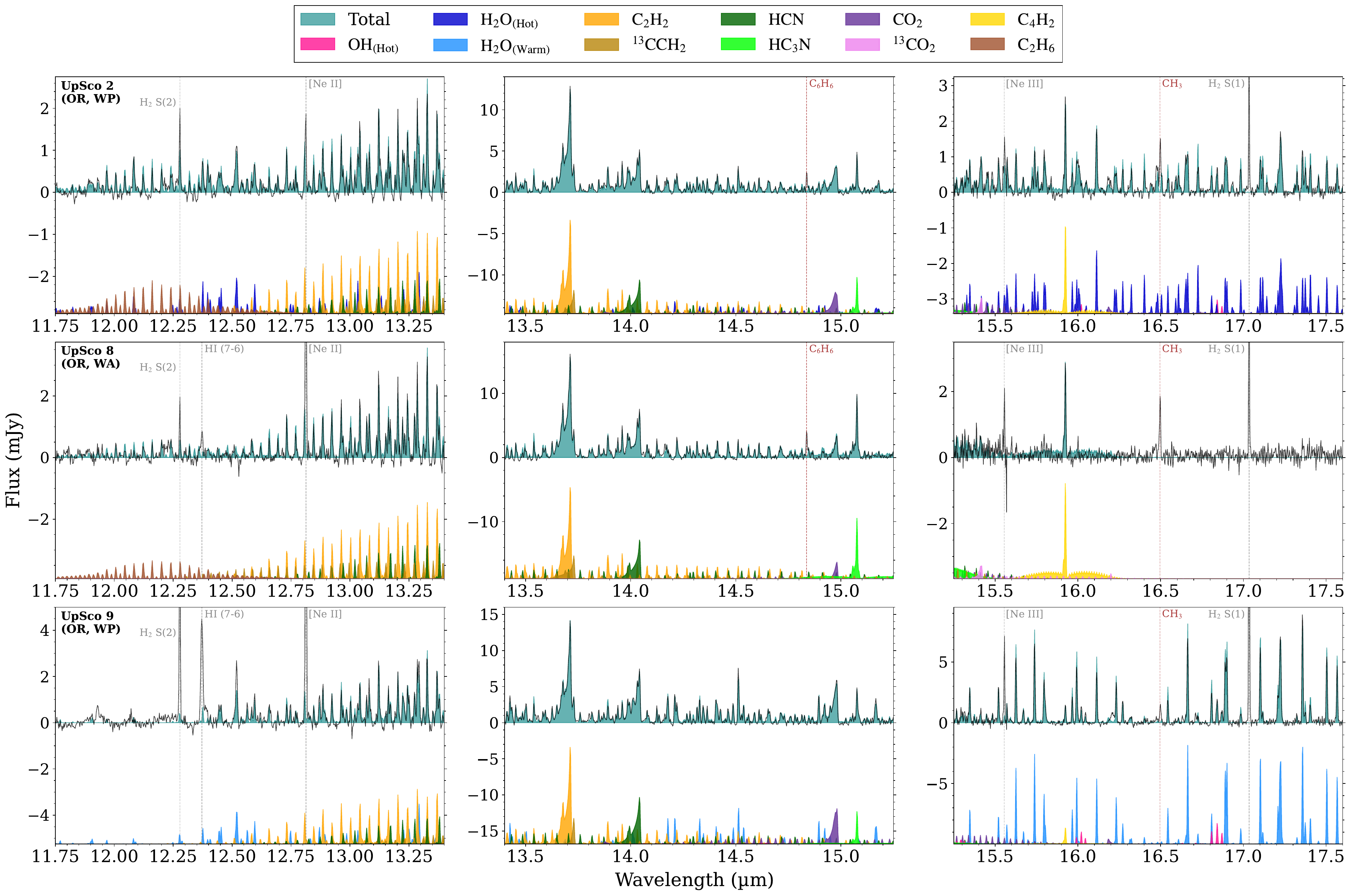}
    \caption{Total best-fitting slab models for all Organic-Rich (OR) disks in the AGE-PRO Upper Scorpius sample, where the bottom row of each panel shows the individual molecular contributions, while the top row shows the total model fit to the data (black line). This plot is split into three wavelength windows for each disk; each window possesses a different y-axis flux scale in order to highlight relatively smaller features compared to C$_2$H$_2$ in the other windows (e.g., tentative C$_2$H$_6$ and $\rm^{13}CO_2$ in USco 8). This shows the high S/N organic and CO$_2$ emission for molecules in the roughly-defined organic region ($\sim$11.75-18 $\upmu$m), in which USco~2, 8, and 9 most prominently display features in our sample.
    \label{fig:contributions_org}}
\end{figure*}

Orthogonal to the water axis, we identify three chemical types based on the
non-water molecular inventory:
\begin{itemize}
        \item \textbf{CO$_2$-Dominated (CD):}
    $L_\mathrm{CO_2}/L_\mathrm{H_2O_{Warm}} \gtrsim 5$ and
    $L_\mathrm{CO_2}$/$L_\mathrm{organic} \gtrsim 0.5$, indicating that CO$_2$ is the spectrally dominant species. USco~3 and 7 meet both criteria, each showing a prominent CO$_2$ feature with weak or absent H$_2$O and no hydrocarbon detections as shown in Fig.~\ref{fig:identifications} and \ref{fig:contributions_org}. These disks also have the lowest CO$_2$ excitation temperatures in the sample ($\sim$200~K; Table~\ref{tab:bestfit_params}). On the water axis, USco~3 is WP and USco~7 is WA. Physical explanations are offered in Sec.~\ref{subsec:evolution} and further comments are found in Sec.~\ref{subsec:cd}.

    \item \textbf{Organic-Rich (OR):}
    $\log(L_\mathrm{organic}/L_\odot) \gtrsim -6$ and at least one complex organic molecule (COM: a carbon-based molecules made up of at least 6 atoms \citealt{Herbst-vanDishoeck2009}) detected, where L$_\mathrm{organic}$ is the integrated line luminosity of all organic components (see Sec.~\ref{subsec:integrations}). USco~2, 8, and 9 satisfy this criterion, each showing detections of C$_2$H$_2$, HCN, $^{13}$CCH$_2$, HC$_3$N, C$_4$H$_2$, and CH$_3$. USco~2 (and tentatively USco~8) additionally exhibits C$_2$H$_6$, while both USco~2 and 8 showing C$_6$H$_6$ emission, as shown in Fig.~\ref{fig:identifications} and \ref{fig:contributions_org}. On the water axis, USco~2 and 9 are WP while USco~8 is WA. Physical explanations are offered in Sec.~\ref{subsec:evolution} and further comments are found in Sec.~\ref{subsec:or}.

    \item \textbf{Molecule-Absent (MA):}
    No confirmed detections of any non-H$_2$ molecular features at S/N~$>3$.
    USco~1, 4, and 6 fall in this category; all three are also WA.
    USco~1 shows tentative OH emission and is the only disk in our sample with a tentative protoplanet detection \citep{Sierra2024}.
    USco~6 shows a tentative CO$_2$ feature, but at the lowest line luminosities in the sample we do not consider this a robust detection.
    USco~4 has no molecular or atomic line detections, likely in part due to its faint emission.
    Physical explanations are offered in Sec.~\ref{subsec:cavities} and further comments are found in Sec.~\ref{subsec:ma}.
    
\end{itemize}

Table~\ref{tab:detections} and Fig.~\ref{fig:identifications} group the sample by these classifications, illustrating the Chemotype dependence upon molecular detections and features.
More specific comments about individual sources are contained in Sec.~\ref{sec:comments}, which is also grouped by Chemotype.

\begin{figure*}[ht!]
    \centering
    \includegraphics[width=\textwidth]{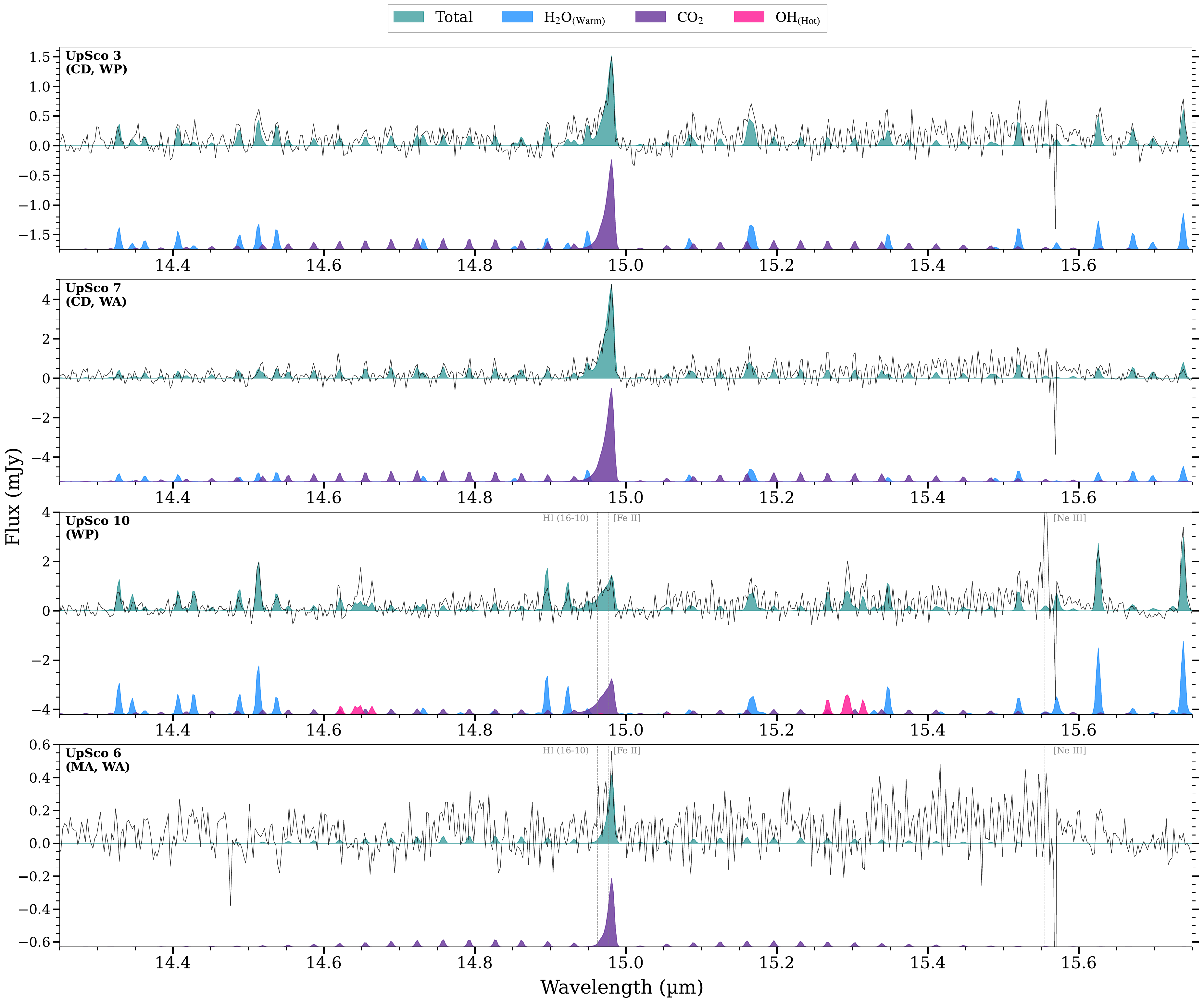}
    \caption{Total best-fitting slab models for our CO$_2$-Dominated (CD) disks (USco~3 and 7),
    as well as the two disks with ``tentative" CO$_2$ features (USco~10 and 6),
    done in the same manner as Fig.~\ref{fig:contributions_org}.
    Atomic lines are shown around the ``tentative" CO$_2$ features in USco~6 and 10 to demonstrate that feature cannot be solely replicated by the included atomic lines in this region. 
    \label{fig:contributions_standard}}
\end{figure*}

\subsection{Chemical Divergence from Younger Sources} \label{subsec:comparison}

\begin{figure*}[b!]
    \centering
    \includegraphics[height=0.4325\textheight]{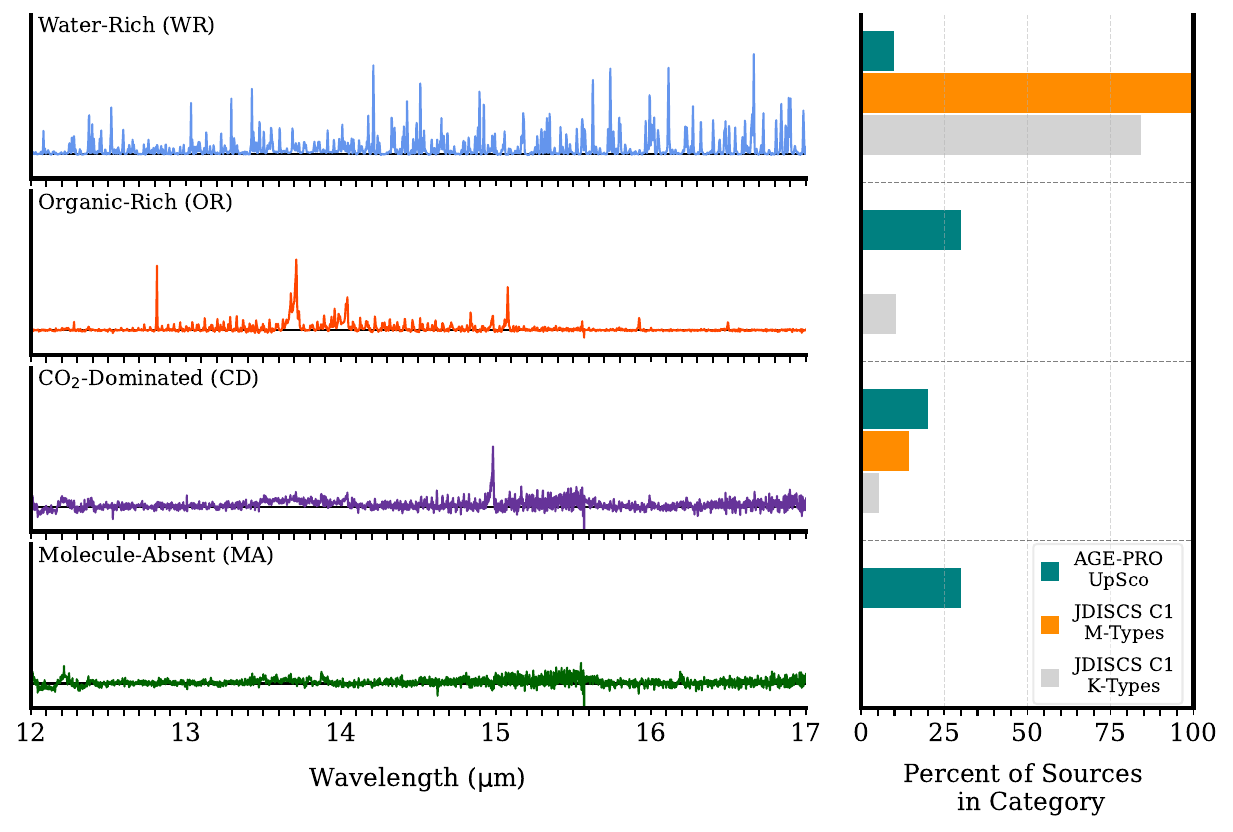}
    \caption{Plot showing the detection rate of different chemical classes of disks as defined in Sec.~\ref{subsec:classes} for our AGE-PRO Upper Scorpius sample vs. the 19 K- and 7 M-type stars from the JDISCS~C1 sample. All Upper Sco disks presented here are M-dwarfs. This shows the greatly increased chemical diversity of our more evolved Upper Scorpius sample, suggesting a divergence in chemistry at older ages.
    Within JDISCS~C1, the two Water-Poor (WP) disks are GO~Tau and MY~Lup \citep{Salyk2025}, the two Organic-Rich (OR) disks are GO~Tau and DoAr~33 \citep{Colmenares24}, and the two CO$_2$-Dominated (CD) disks are Sz~114 \citep{Xie2023} and MY~Lup.
    \label{fig:histogram}}
\end{figure*}

The AGE-PRO Upper Sco sample occupies a unique parameter space among recent JWST disk surveys, complementing efforts such as JDISCS~Cycle~1 \citep{Arulanantham25} and MINDS \citep{Henning24} by probing the end‐stage chemistry of terrestrial planet–forming zones at ages of 2-–6~Myr. In order to underscore the importance of age in observed disk chemistry, we primarily make comparisons to the JDISCS~Cycle~1 (hereafter referred to as JDISCS~C1) sample of $\sim$1-3~Myr sources \citep{Arulanantham25}. This sample includes 31 disks in total from multiple observing programs (PID: 1584, PI: C. Salyk; PID: 2025, PI: K. Oberg; PID: 1549, PI: K. Pontoppidan; PID: 1640, PI: A. Banzatti). However, we omit 4 A- and G-type stars in the same manner as \citet{Arulanantham25} for a more similar comparison to our entirely M-type sample. Additionally, we remove comparisons to AS205 S, a combined K- and M-type spectroscopic binary \citep{salyk14}, thus resulting in a total sample of 26 sources (19 K- and 7 M-types, which we compare separately) in JDISCS~C1.
As we reiterate in Sec.~\ref{subsubsec:trends}, our sample of only 10 sources in AGE-PRO Upper Sco is too small to establish conclusive patterns in comparison to 26 sources in JDISCS~C1 (of which only 7 are M-types), and a much larger homogeneous population will be needed to determine out real population-level trends (e.g., Waggoner et al., in prep).

In Fig.~11 of \citet{Arulanantham25}, JDISCS~C1 reports H$_2$O in 100\% of both K‐ and M‐type disks, whereas Upper Sco shows H$_2$O in only 50\%. Similarly, JDISCS~C1 find C$_2$H$_2$ and HCN in $\sim$75\% of K-types, and $\sim$85\% C$_2$H$_2$ and 100\% HCN detection rates in M-types, much greater than the 30\% of both we find in Upper Sco.
While $\gtrsim$90\% of the JDISCS~C1 sample also shows robust detections of CO across both spectral types, only one source in our sample does the same (10\%), which follows the same trend of water detection in both samples \citep{Arulanantham25}. However, we note that this discrepancy between samples may be due to stellar contamination or low accretion, as we did not perform any photospheric corrections in our data reduction (see \citet{Mallaney2026} for example of stellar photosphere hiding MIRI CO emission and see \citet{Dickson-Vandervelde2025} for discussion of relation of CO to accretion rate).
Lastly, while OH occurs at $\sim$90\% in both spectral types of JDISCS~C1, we find OH in only 40\% of our sample; however, as OH detections are strongly correlated with water vapor detections due to being a product of H$_2$O photodissociation, this result is unsurprising.

We compare detection rates of several of our defined (Table~\ref{tab:classes}) chemical classes of disks to the younger sources in JDISCS~C1 in Fig.~\ref{fig:histogram}, which emphasizes the chemical diversity we observe in our more evolved sample. For example, we find two CO$_2$-Dominated (CD) disks within our sample of 10 disks, compared to the two CD disks, Sz~114 (K-type, \citealt{Xie2023}) and MY~Lup (K-type, \citealt{Salyk2025}) within JDISCS~C1 (see Sec.~\ref{subsec:cd} for comments on HT~Lup). Similarly, we find that MA disks make up 30\% of our sample (see Sec.~\ref{subsec:ma} for more), but are absent from both spectral-type sub-group of JDISCS~C1 due to the ubiquitous presence of water lines in the JDISCS~C1 sample. 
Comparisons of the number of WR and OR sources in each sample are discussed below as major results of our analysis in Sec.~\ref{subsubsec:depletion} and \ref{subsubsec:complexity}, respectively.

\subsubsection{Strong Water Depletion} \label{subsubsec:depletion}

\begin{figure*}[ht!]
    \centering
    \includegraphics[width=\textwidth]{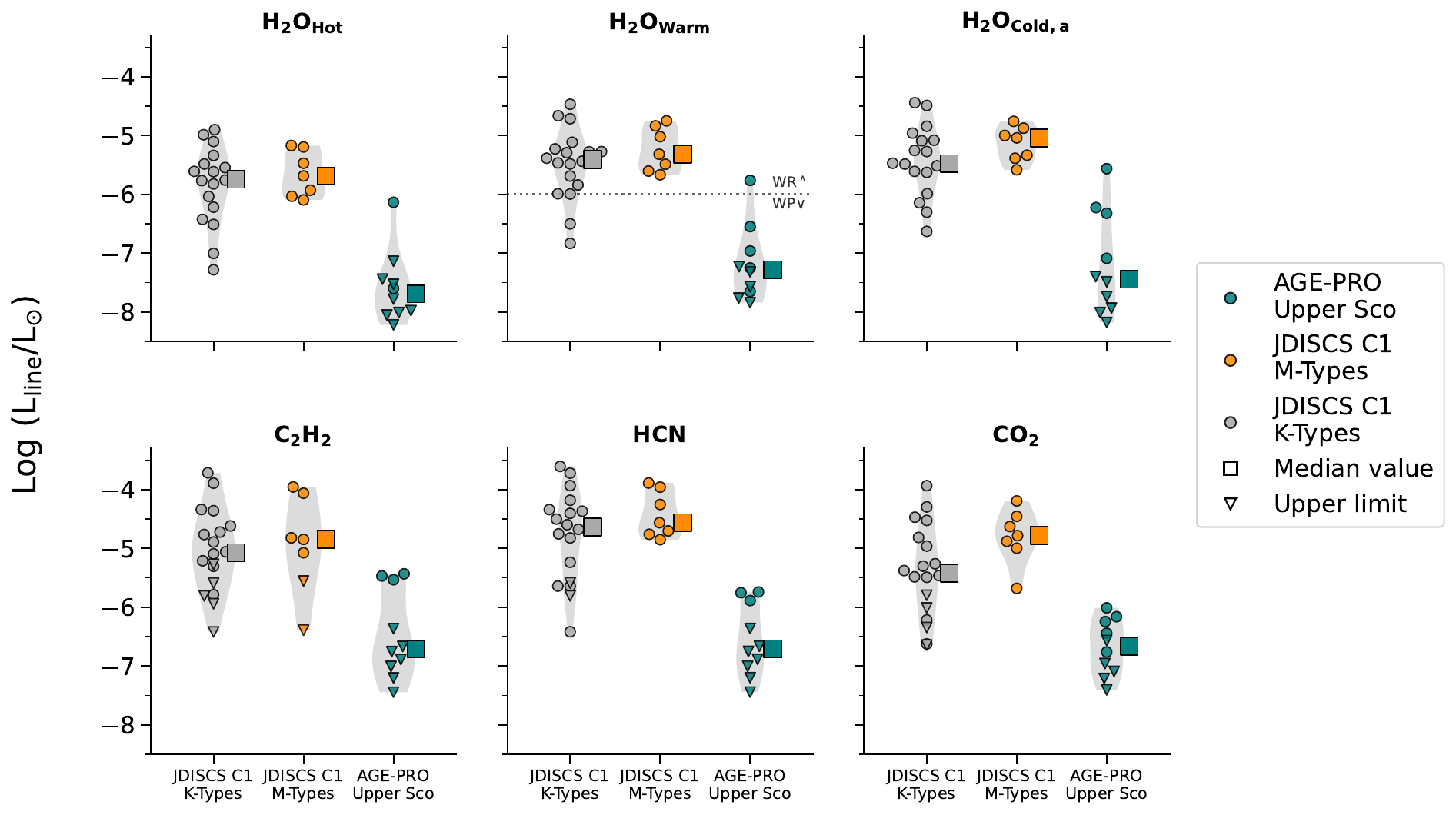}
    \caption{
    Violin plots showing the line luminosity values of the different H$_2$O temperature components, C$_2$H$_2$, HCN, and CO$_2$, sorted by sample and spectral type. Our AGE-PRO Upper Sco sample ($\sim$2-6~Myr) is compared to the younger JDISCS~C1 sample ($\sim$1–3~Myr), where our Upper Sco sample is entirely M-types. This demonstrates the trend of strong water depletion and generally weaker molecular emission in our more evolved sample.
    \label{fig:violin}}
\end{figure*}

Fig.~\ref{fig:violin} shows our integrated line luminosities of H$_2$O temperature components and C$_2$H$_2$, HCN, and CO$_2$ (as defined in Sec.~\ref{subsec:integrations} and reported in Sec.~\ref{subsec:luminosities}), grouped by AGE-PRO Upper Sco and JDISCS~C1 sample.
This reveals a population-level decline in mid-IR molecular line luminosities with age: the integrated line luminosities for the set of detected H$_2$O temperature components and CO$_2$ are systematically lower in Upper Sco than JDISCS~C1 by factors of $\sim$10-1000, consistent with the gas depletion of the AGE-PRO Upper Sco region found in \citet{Trapman2025b} and similar to the results of \citet{Mallaney2026}. This dependence holds for our H$_2$O$\rm_{Cold,a}$ line luminosity as well, even though Zhang et al. (submitted) do not observe this with exclusively younger sources from Ophiuchus.
This is with the exception of USco~5, our one WR source in the sample, which seemingly falls in line with the lower L$_{H_2O_{Warm}}$ end of the JDISCS~C1 sample in H$_2$O emission.

When these line luminosities are instead plotted against accretion luminosity (taken from Empey et al. (accepted) and \citealt{Manara_PPVII,Andrews18b,Alcala17}, respectively) as is done in Fig.~\ref{fig:lacc}, where the accretion luminosity L$\rm_{acc}$ serves as a proxy for age, we find that this result generally holds.
However, we note that this is not especially surprising as stellar luminosity (L$_{\star}$) and L$_{acc}$ decrease with age \citep{Hartmann1998}, and therefore we can expect the individual molecular luminosities to also decrease with time. Indeed, when our calculated line luminosities are normalized by the continuum luminosity at 13 $\upmu$m and the same relation is plotted in Fig.~\ref{fig:lacc_norm}, we can see this dependence largely disappear.

Fig.~\ref{fig:histogram} also demonstrates that our sample is highly water-depleted compared to the younger JDISCS~C1 sample, with 100\% of the M-type in JDISCS~C1 sample qualifying as Water-Rich (WR), compared to only 10\% of our Upper Sco sample (USco~5). Additionally, with nearly $\sim$90\% detections of WR spectra among the JDISCS~C1 K-types, this depletion of water also seems to be present relative to higher mass stars. This is confirmed by \citet{Salyk2026}, who identify 22 of the JDISCS~C1 sample as possessing three distinct temperature components of water vapor, while our sample only includes two disks (USco~2 and 5) that necessitate three temperature components of water to fit the spectra and hot H$_2$O diagnostic line present at ~17.32 $\upmu$m. The greater absence of hot H$_2$O compared to warm H$_2$O in our sample agrees with Xie et al. (in press), who find the strongest correlation between H$_2$O$\rm_{Hot}$ and L$\rm_{acc}$ out of the three H$_2$O temperature components. However, we note that these non-detections could also potentially be due to contamination of the spectra due to photospheric absorption. Nonetheless, this also aligns with our expectations of relatively colder molecular emission with a lower stellar luminosity, which we explore further in Sec.~\ref{subsubsec:cold}.

\subsubsection{More Complex Organic Molecules} \label{subsubsec:complexity}
It is notable that, even though only 30\% of our total sample contains organic molecules, every disk that contains C$_2$H$_2$ and HCN also contains at least one COM (as defined in \citealt{Herbst-vanDishoeck2009}). This is because C$_2$H$_2$ and HCN are the most commonly detected organics among the JDISCS~C1 M-types, with $\sim$85\% and 100\% detection rates, respectively, while only two sources in the older JDISCS~C1 sample show definitive detections of COMs (DoAr~33 from \citet{Colmenares24} and GO~Tau), both of which are K-type stars \citep{Arulanantham25}. This means that the majority of JDISCS~C1 source, and all of its M-types, do not fall into our classification of an Organic-Rich disk, as reflected in Fig.~\ref{fig:histogram}.
This generally agrees with Xie et al. (in press), who find that the detection of COMs such as C$_4$H$_2$ increases with age, potentially indicative of a higher inner-disk C/O with age. All three of our OR disks also exhibit detections of more simple organic molecules/isotopologues ($\mathrm{^{13}CCH_2}$, HC3N, CH$_3$, see Table~\ref{tab:detections}) that are not detected among JDISCS~C1 outside of DoAr~33 and GO~Tau, expanding our agreement with this result. 

\subsubsection{Colder Organic Molecules} \label{subsubsec:cold}

We notably detect a rich array of organic emission lines in three Upper Sco disks comparable in diversity to the most organic‐rich MINDS and JDISCS~C1 spectra (e.g., J160532 from \citealt{Tabone23}; DoAr~33 from \citealt{Colmenares24}), but at markedly colder average excitation temperatures (see Table~\ref{tab:bestfit_params}). Specifically, we find an average C$_2$H$_2$ excitation temperature of $\sim$300 K where detected, compared to an average $\sim$950 K in the relatively older ($\sim$1-3~Myr) JDISCS~C1 M-type population, with a similar $\sim$300 K vs. $\sim$790 K for HCN, respectively \citep{Arulanantham25}. The consistently cold temperatures of the organics in our OR sample lie below the relatively warm organics of the VLMS J160532 \citep{Tabone23,Kanwar2026} and T~Tauri DoAr~33 \citep{Colmenares24} by a few hundred K, and are colder than the VLMS Sz~28 \citep{Kanwar2024b,Kaeufer2024b} and ISO-ChaI~147 \citep{Arabhavi24} by $\sim$100K, aligning with the general trend of these latter two papers that M-dwarfs tend to possess colder organics than K-type stars.
We also find that the carbon-based molecules CO$_2$ and $^{13}$CO$_2$ exhibit similarly low $\lesssim$300 K slab model temperatures (Table~\ref{tab:bestfit_params}), compared to an average of $\sim$630 K in the JDISCS~C1 M-type population where detected, suggesting a similar trend.

We also find that detected $^{13}$CCH$_2$ emission is approximately $\sim$20-50 K colder than that of C$_2$H$_2$ ($^{12}$C$_2$H$_2$) in each of our OR disks; if $^{13}$CCH$_2$ indeed probes deeper layers of the disk at similar due to its lower optical depth, then these cooler excitation temperatures line up with our expectations of cooler gas present near the midplane. 
We additionally find especially low slab model temperatures for HC$_3$N, C$_2$H$_6$, and C$_4$H$_2$ at $<$200 K, which we comment on further in Sec.~\ref{subsec:or}. However, due to the relatively large uncertainties on retrieved temperatures, we recommend that these values be taken with caution.
While lower excitation temperatures are expected from the lower accretion and stellar luminosities in our older sample, we theorize further explanations in Sec.~\ref{subsubsec:destruction}.
The emitting areas shown in Table~\ref{tab:bestfit_params} correspond to average projected emitting radii of $\sim$0.125 au for C$_2$H$_2$ and $\sim$0.175 au for HCN, compared to $\sim$0.7 and $\sim$1.1 au, respectively in the JDISCS~C1 M-type sample \citep{Arulanantham25}. However, due to the degeneracy between log(N) and r$\rm_{emit}$ (log(A)), we cannot make conclusive statements about the implications of these differences.

\begin{figure*}[ht!]
    \centering
    \includegraphics[width=\textwidth]{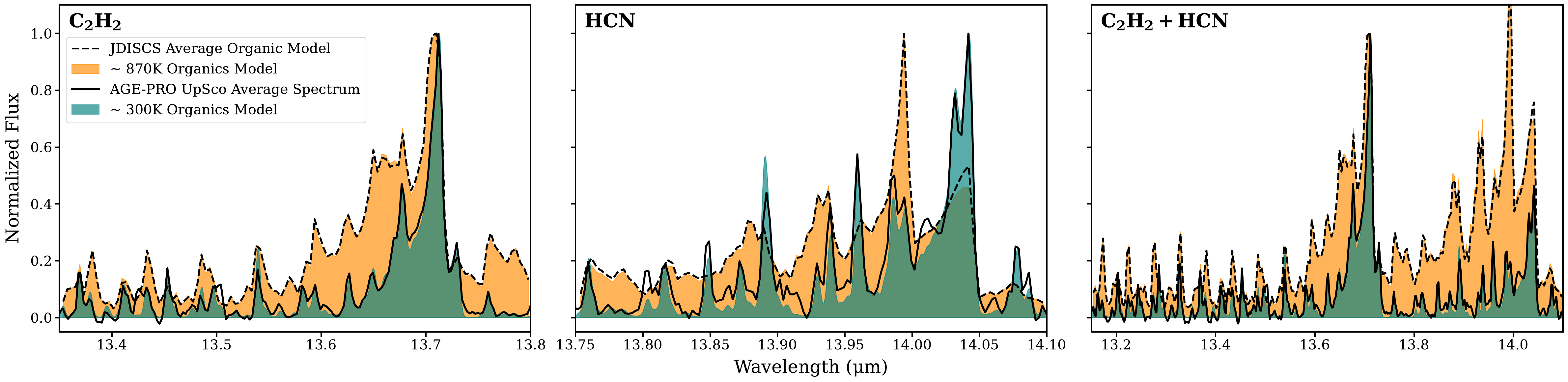}
    \caption{$\mathrm{C_2H_2}$ and HCN cold organic feature shapes in the AGE-PRO Upper Scorpius ($\sim$2-6~Myr) sample compared to the warmer organic features in the JDISCS~C1 (1-3~Myr) sample \citep{Arulanantham25}. The JDISCS~C1 Average Organic Model is computed from only the average $\mathrm{C_2H_2}$ and HCN contributions to the best-fitting models of disks in JDISCS~C1 that demonstrate detections of both species, as warm $\mathrm{H_2O}$ was especially blended with the organic features in the JDISCS~C1 sample. We then fit fixed-temperature models (at the average temperature of C$_2$H$_2$ and HCN combined for each sample) to the average JDISCS~C1 organics model contributions and to the average spectrum of the Organic-Rich (OR) AGE-PRO Upper Sco disks ($\sim870$ K and $\sim$300 K, respectively). This demonstrates that distinctly different temperature components are needed for each. In the left and middle panels, each model and spectrum is normalized to the maximum feature strength of the plotted region. In the right panel, the models and spectra are normalized to the peak of the C$_2$H$_2$, as HCN features tend to be larger than C$_2$H$_2$ in JDISCS~C1 but all C$_2$H$_2$ features in the AGE-PRO Upper Scorpius sample are larger than HCN.
    \label{fig:cold_organics}}
\end{figure*}

We confirm the presence of our cold organics by matching the line widths of the average emission features present in our OR disks to our low-temperature slab models, and contrast that with the analogous average data and models of higher-temperature organics in JDISCS~C1 in Fig.~\ref{fig:cold_organics}. As the Upper Sco features are narrower, indicative of less excitation of upper level energies, and they peak at different central wavelengths, we can more confidently report the low temperature outputs of our slab models. This result aligns with the results of Xie et al. (in press), who find that molecular gas is on average colder in Upper Sco due to an overall lower stellar and accretion luminosity.

\subsubsection{Summary of Evolutionary Trends} \label{subsubsec:trends}
    The comparison between our 2-6~Myr Upper Sco sample and both spectral types of the 1-3~Myr JDISCS~C1 sample reveals three population-level evolutionary trends in inner-disk chemistry:
\begin{enumerate}
    \item \textbf{Declining water emission:} H$_2$O detection rates drop from $>$90\% to 50\%, and line luminosities decrease by 1-2~dex, with only one Upper Sco disk retaining water luminosities comparable to the lower end of the younger sample.

    \item \textbf{Diversifying carbon chemistry:} While simple hydrocarbons (C$_2$H$_2$, HCN) become less frequently detected, the fraction of disks showing \textit{complex} organic inventories (C$_4$H$_2$, HC$_3$N, C$_2$H$_6$, C$_6$H$_6$) increases, and flux ratios of carbon-bearing species to H$_2$O shift upward.

    \item \textbf{Cooling molecular reservoirs:} Hydrocarbon excitation temperatures drop by a factor of $\sim$2-3 between the two samples, from $\sim$700-1000~K to $\lesssim$300~K, suggesting that the molecular emission in evolved disks arises from cooler regions than in their younger counterparts.
\end{enumerate}

We note three important caveats. First, the JDISCS~C1 sample is biased toward disks with bright submillimeter continuum emission (many were selected from the DSHARP survey; \citealt{Andrews18b}), which may favor more massive and structured outer disks than a volume-complete sample would yield. Second, the two samples span overlapping but not identical stellar-mass ranges: our Upper Sco sample extends to lower masses, accretion luminosities, and stellar luminosities than most of the JDISCS~C1 targets. Third, our AGE-PRO Upper Sco sample only consists of 10 disks, which makes classifying trends and "outliers" from JDISCS~C1 and MINDS extremely tentative. Disentangling the effects of age from those of stellar mass will require a larger, mass-matched comparison, which will be enabled by the full AGE-PRO survey across Ophiuchus, Lupus, and Upper Sco (Waggoner et al., in prep).

\section{Discussion} \label{sec:discussion}

The Upper Sco MIRI/MRS survey reveals a diverse chemical landscape in the inner regions of 2-6\,Myr disks. Our sample of 10 systems, selected on the basis to cover the span of millimeter continuum flux within the narrow stellar mass range, reveal four distinct chemical classes: Organic-Rich, CO$_2$-Dominated, Water-Rich, and Molecule-Absent. This diversity is distinctive from previous MIRI surveys of 1-3\,Myr-old T~Tauri disks which show Water-Rich as the dominant type (Sec.~\ref{subsubsec:depletion}). In the discussion section, we examine the physical mechanisms that may cause this diversity and discuss the implications for late-stage planet formation.

\begin{figure*}[ht!]
    \centering
    \includegraphics[width=\textwidth]{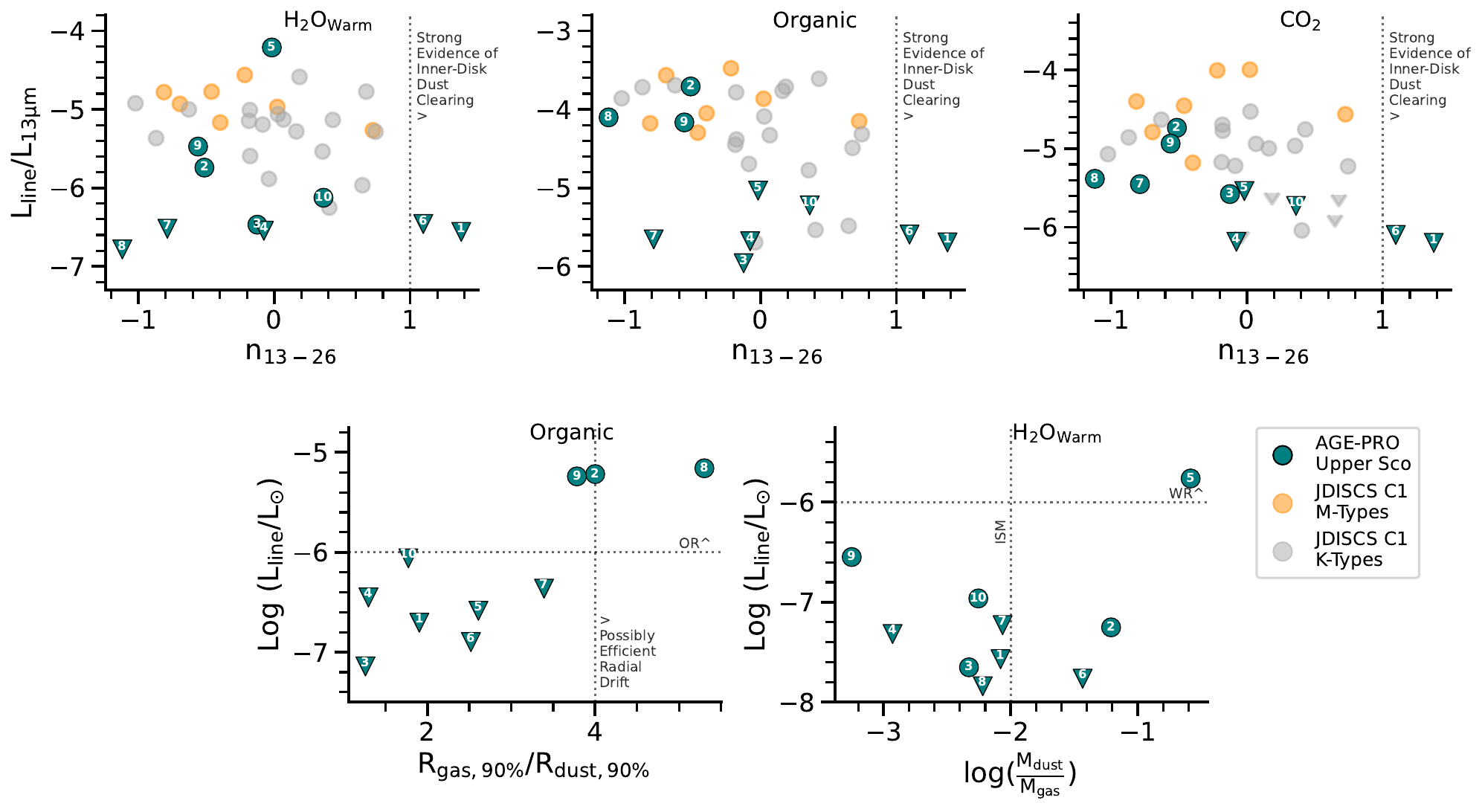}
    \caption{
    The top three panels show several line luminosities (as defined in Sec.~\ref{subsec:integrations}) \textit{normalized} by the 13~$\upmu$m continuum luminosity, plotted as a function of the n$_{13-26}$ spectral index, which shows how quickly the flux changes from 13-26 $\upmu$m, indicative of a possible inner-disk dust cavity, for our Upper Scorpius sample (2-6~Myr) compared to the younger JDISCS~C1 sample ($\sim$1–3~Myr) 
    These show that (excluding our faint emitter USco~4, see Sec.~\ref{subsec:ma}) our two Molecule-Absent (MA) sources, USco~1 and 6, consistently show among the lowest molecular emission across warm H$_2$O, organics, and CO$_2$, both of which possess exceptionally high n$_{13-26}$ indices that are indicative of inner-disk dust clearing.
    \\
    The bottom-left panel displays the \textit{unnormalized} organic line luminosity (see Sec.~\ref{subsec:integrations}) as a function of the $R_{\rm gas,90\%}$/$R_{\rm dust,90\%}$, with a horizontal dotted line indicating the luminosity criterion of our Organic-Rich (OR) disks (Table~\ref{tab:classes}) and a vertical dotted line indicating the threshold for efficient radial drift of pebbles ($\gtrsim$4, \citealt{facchini17,Trapman19,Trapman20_drift}). This shows that each disk in our sample near or within this threshold correspond to all three of our OR sources (USco~2, 8, and 9), as discussed in Sec.~\ref{subsubsec:drift}. \\
    The bottom-right panel displays the \textit{unnormalized} warm H$_2$O line luminosity as a function of the dust-to-gas mass ratio (d2g) in log scale, with a horizontal dotted line indicating the luminosity criterion of a Water-Rich (WR) disk (Table~\ref{tab:classes}) and a vertical dotted line roughly indicating the interstellar medium (ISM) d2g value ($\sim$0.01, \citealt{Bohlin78}). This shows that USco~5, our sole WR disk, possesses an abnormally high d2g ($\sim$0.26), which we discuss possible implications of in Sec.~\ref{subsec:d2g}.
    \label{fig:spectral_trends}}
\end{figure*}

\subsection{Inner-Disk Clearing} \label{subsec:cavities}

\subsubsection{In the Context of Molecule-Absent Spectra} \label{subsubsec:ma_cavities}
Inner-disk cavities, regions of strongly reduced dust and/or gas surface density within the \textit{innermost} few au, provide a natural mechanism to explain our Molecule-Absent spectra. If the warm molecular layer that MIRI probes is physically absent or optically thin, line emission will be suppressed regardless of the volatile content at larger radii \citep{vanderMarel15}.

The mid-infrared spectral index $n_{13-26}$, which measures how steeply the continuum rises from 13 to 26~$\upmu$m, serves as a diagnostic of inner \textit{small dust grain} cavities \citep{Brown2007,Furlan2009}: positive values indicate flux rising toward longer wavelengths, consistent with emission from a wall at the edge of an inner dust cavity rather than from a continuous disk extending to the dust sublimation radius. We use the definition of a ``mm+IR cavity" from \citet{Mallaney2026} to qualify the existence of an \textit{inner-disk} dust cavity: if mm observations (from \citet{Vioque2025} in this case) and the mid-infrared $n_{13-26}$ spectral index \textit{both} show evidence of a cavity, we consider the existence of an inner-disk dust cavity real, while if only one shows evidence, we label it as tentative (see Table~\ref{tab:sources}). Full SEDs of all sources can be found in Fig.~\ref{fig:seds}.

In our sample, the Molecule-Absent disks USco~1 and 6 show the overall most positive $n_{13-26}$ values ($>$1) of both AGE-PRO Upper Sco and JDISCS~C1, as is shown in the top three panels of Fig.~\ref{fig:spectral_trends}. These disks are also have inner-disk dust cavities identified in millimeter continuum images \citep{Vioque2025}. Therefore, an inner-disk cavity can naturally explain the Molecule-Absent spectra of USco~1 and 6, the former of which is illustrated in the fourth panel of Fig.~\ref{fig:cartoon}. We note that the MA disk USco~4 is excluded from this discussion due to its compact, highly-inclined, and faint nature (see Sec.~\ref{subsec:reduction} for more details).

\subsubsection{In the Context of Water-Poor and Water-Absent Spectra} \label{subsubsec:wa_cavities}

Inner-disk clearing also could hypothetically provide an explanation for the remaining Water-Absent disks in our sample, USco~7 and 8 (excluding USco~4, see Sec.~\ref{subsec:reduction}). If a cavity does exist in the CO$_2$-Dominated disk USco~7, the inner-disk gas cavity edge could sit between the H$_2$O and CO$_2$ ice lines, allowing the sublimated CO$_2$ to emit past inner-disk gas cavity's edge while water is absent or frozen onto dust grains in the cleared inner region \citep{oberg11,Grant23,Vlasblom2024}. 
Furthermore, HCN’s ice line has also been proposed to lie a few au outside that of H$_2$O \citep{Bergner2022}; if the inner-disk gas cavity radius similarly lies between that of H$_2$O and HCN, this could yield a combination of both elevated organics/CO$_2$ and water-absence, the scenario seen in USco~8. Depending on the exact inner-disk gas cavity radii relative to the snowline, (or inner-disk \textit{gap} e.g., see \citealt{Kanwar2026}), these explanations can also be extended to the water-poor CD/OR cases of USco~2, 3 and 9.

However, this scenario is tentatively inconsistent with all five of these disks (USco~2, 3, 7, 8, and 9), as their negative $n_{13-26}$ suggests no evidence of an inner-dusk dust cavity. In particular, the Water-Absent USco~7 and 8 exhibit the most negative $n_{13-26}$ index in our sample. 
We note the caveat that MIRI spectra is only sensitive to gas depletion inside a few au of the disk due to the high upper energy levels of most lines.
If the mid-IR dust is decoupled from the gas, the $n_{13-26}$ index is not necessarily representative of the presence of an inner-disk \textit{gas} cavity as well. Similarly, the resolution of ALMA in \citet{Vioque2025} is -much too low to resolve a $\sim$1-2 au cavity needed to reproduce \citet{Vlasblom2024}.
In our sample, USco~9 has an inner gas cavity at 10-20\,au based on CO J=2-1 images \citep{Vioque2025,Trapman2025a} but a negative $n_{13-26}$ index, suggesting the innermost disk is not significantly depleted in smaller dust grains.
While the negative $n_{13-26}$ indices do not definitively rule out the presence of inner-disk dust cavities in these sources, they suggest that the CO$_2$-Dominated and Organic-Rich disks in our sample that are also Water-Poor or Water-Absent may require additional explanations.

\subsection{C/O Evolution: From Water-Rich to Organic-Rich} \label{subsec:evolution}

\begin{figure*}[hb!]
    \centering
    \includegraphics[width=\textwidth]{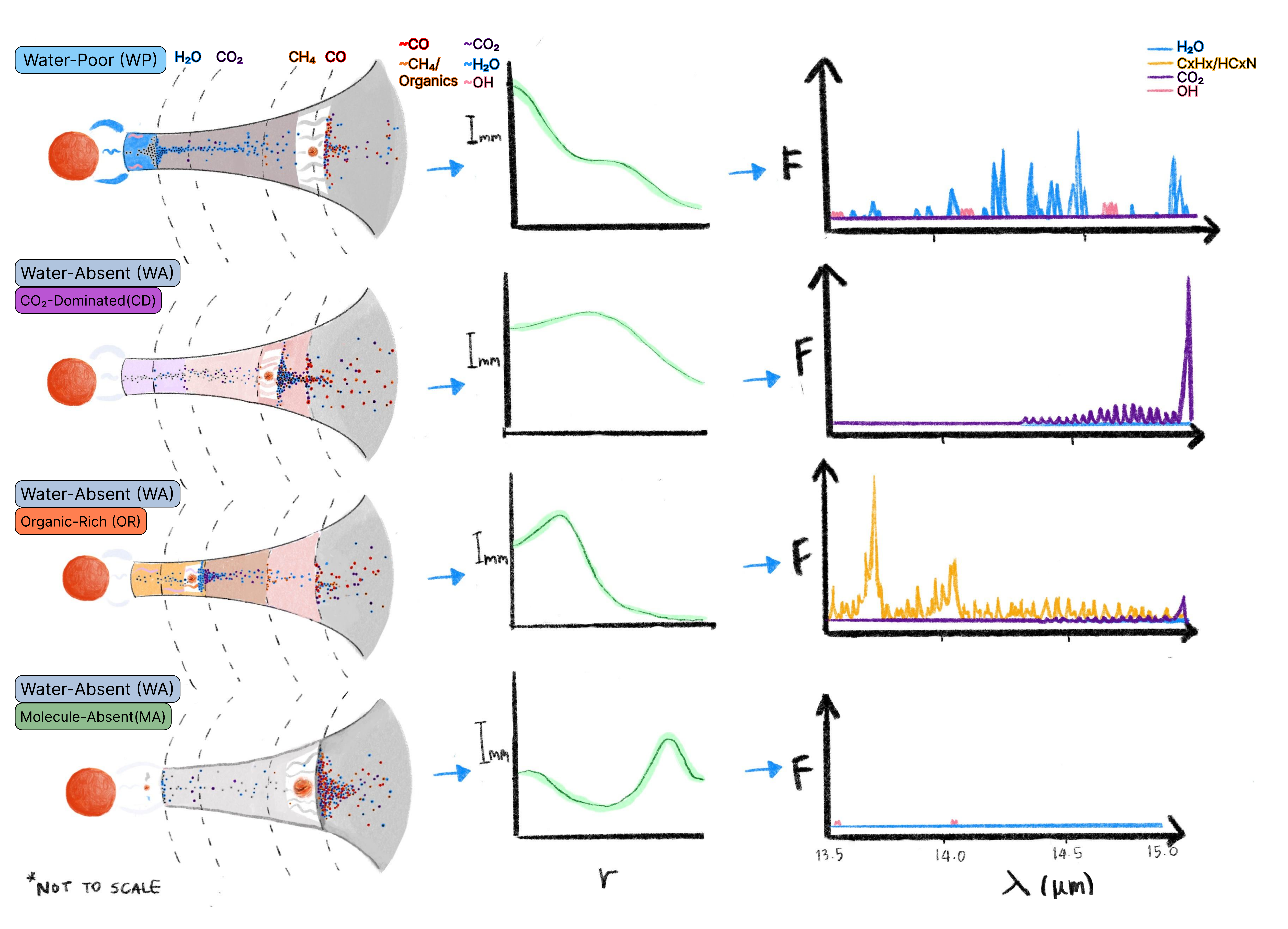}
    \caption{
    Cartoon depicting the different outer-disk dust trapping scenarios outlined in Sec.~\ref{subsec:dust_trapping} for ALMA-resolved disks with distinct substructure from the AGE-PRO Upper Scorpius sample. Radial profiles (central column) of the Water-Poor (WP) disk USco~10, Water-Absent (WA)/CO$_2$-Dominated (CD) disk USco~7, WA/Organic-Rich (OR) disk USco~8, and WA/Molecule-Absent (MA) disk USco~1 are all taken from \citet{Vioque2025}. This cartoon illustrates how multiple physical mechanisms are likely working simultaneously to produce the corresponding diversity in spectra (right column) observe in our sample. This shows how the different radii at which dust traps, if protoplanet induced, can lead to radically different chemical outcomes; gas streams can transport sublimated molecules across protoplanetary gaps, but will halt the inward drift of pebbles if the dust trap exists past the ice line of that molecule, depriving the inner disk of that species as discussed in Sec.~\ref{subsec:factors}. This also portrays the centrally-concentrated continuum emission of our WR and WP disks and the inner-disk dust cavities of our MA disks outlined in Sec.~\ref{subsec:d2g} and \ref{subsubsec:ma_cavities}, respectively.
    \label{fig:cartoon}}
\end{figure*}

The change of C/O elemental ratio of gas of the innermost disk is another potential mechanism that could drive the chemical diversity of our sample.
\citet{Mah23} proposed a two-phase C/O evolution: an early phase where inward pebble drift preferentially delivers H$_2$O-rich ices (making the inner disk sub-stellar in C/O), followed later by the relatively slower inward advection of carbon-rich gas that drives the C/O upward (super-stellar). 
This serves as an intuitive explanation for the much higher fraction of Organic-Rich disks we observe in our evolved sample compared to younger disks (see Fig.~\ref{fig:histogram}): the complex hydrocarbons of these disks are a natural byproduct of the chemical reactions in high C/O regions \citep{Walsh14,Semenov-Wiebe2011,Wei2019,Kanwar2024a,Raul2025}.
This tentatively hints that OR disks are simply a more evolved version of the same Water-Rich disks we see in younger regions (e.g., \citealt{Arulanantham25}).

Furthermore, \citet{Mah23} and \citet{Sellek2025} proposed that VLMS evolve and increase their C/O ratio at a faster rate than T~Tauris, due to faster pebble drift timescale in lower stellar mass systems. Indeed, the OR disks USco~2 and 8 fall within the mass range of a VLMS ($\lesssim 0.3M_{\odot}$), consistent with their relatively carbon-rich inner-disks for the younger age bin of the Upper Sco sample (see Table~\ref{tab:sources}).
The next two lowest-mass stars in our sample (USco~7 and 3, $\sim 0.35-0.4M_{\odot}$) are both of our CO$_2$-Dominated disks, potentially serving as a marker of the chemical-midpoint of this transition from sub-stellar to super-solar C/O, in which processes such as the redox reactions of volatile reprocessing deplete the H$_2$O/CO to produce the CO$_2$ we observe \citep{Furuya-Aikawa2014,Furuya2022,Bosman18}. However, the Water-Rich disk of our sample (USco~5) also falls within the VLMS mass range, and thus requires an alternate explanation (see Sec.~\ref{subsec:d2g}).

\subsubsection{Carbon Grain Destruction and Cold Organics} \label{subsubsec:destruction}
Carbon grain destruction, the thermal/radiative processing of refractory carbonaceous material near ice lines, provides a natural pathway to inject refractory carbon into the gas and drive secondary carbon chemistry with more complex products (e.g., C$_2$H$_6$, C$_4$H$_2$, C$_6$H$_6$) \citep{Kress2010,Lee2010,Siebenmorgen2010,Alata2014,alata15,anderson17,Wei2019,Bosman21_carbon_grains,Colmenares24}.
Applied to our Upper Sco sample, the consistent presence of complex hydrocarbons (C$_4$H$_2$, HC$_3$N) in our Organic-Rich subset (as well as C$_2$H$_6$ \& C$_6$H$_6$ in only USco~2 and 8) points to carbon grain destruction with sufficient irradiation to allow complex hydrocarbon formation but that does not immediately evacuate the newly-liberated carbon.

Our Organic-Rich disks (USco~2, 8, and 9) display robust hydrocarbons at relatively low excitation temperatures ($\lesssim300$~K, see Sec.~\ref{subsubsec:cold}, Table~\ref{tab:bestfit_params}). 
\citet{Colmenares24} posit that low accretion systems such as DoAr~33 could similarly possess low temperature organics as newly-liberated carbon from the soot line can more easily diffuse outward, and as carbon grain sublimation is theoretically irreversible \citep{Li2021}, will not refreeze out and will only get colder. While this is difficult to confirm due to the degeneracy between A and N, this makes intuitive sense as higher accretors would heat up the disk surface layers more and therefore lead to higher temperature organics.
Additionally, if hydrogen commonly reacts directly with these refractory carbon grains into small molecules such as CH$_3$ or CH$_4$ \citep{Raul2025,Kanwar2024a}, these molecules would also form at relatively cold temperatures \citep{Li2021}.
This is consistent with carbon possibly being liberated and chemically processed in cooler, optically-thicker layers or at modest radii where the released species can survive and emit in the mid-IR (e.g., HCN, see \citealt{Bergner2018,Bergner2021,Paneque-Carreno2023}).

\subsubsection{The Possibility of Rapid Pebble Drift} \label{subsubsec:drift}

The timescales of this multi-stage evolutionary picture is dependent on the efficiency of inward pebble drift. \textit{Rapid} pebble drift delivers volatiles to the inner disk early in a disk’s life  \citep{Banzatti20,Kalyaan21,Banzatti23a,Mah23,Trapman19,Trapman20_drift}. The net result is smaller dust radii relative to gas radii (large $R_{\rm gas}/R_{\rm dust}$) and centrally-concentrated continuum emission, where a $R_{\rm gas,90\%}$/$R_{\rm dust,90\%}\gtrsim4$ is often used as a signpost of efficient pebble drift \citep{facchini17,Facchini19,Trapman19,Vlasblom2025}. The expectation is that, assuming an inner-disk gas cavity is not present past the H$_2$O snowline (e.g., \citealt{Vlasblom2024} for which we only find evidence of for USco~1 and 6 in Sec.~\ref{subsec:cavities}), inward pebble drift will deliver water-ice to the inner disk where they sublimate and enrich the inner-disk in water vapor (see e.g., \citealt{Ciesla06,Banzatti20,Banzatti23b,Kalyaan2023}).

The sources in our sample with the highest $R_{\rm gas,90\%}$/$R_{\rm dust,90\%}$ are USco~8 at $\sim5.3$, USco~2 at $\sim$4, USco~9 at $\sim$3.8, and to a lesser extent USco~7, with $\sim$3.4 (see Table~\ref{tab:sources}. calculated from \citealt{Vioque2025} and \citealt{Trapman2025a}). This is notable as the three highest $R_{\rm gas,90\%}$/$R_{\rm dust,90\%}$s in our sample correspond exactly to our three OR disks. This is illustrated in the bottom-left panel of Fig.~\ref{fig:spectral_trends}, which plots L$_{organic}$ as a function of $R_{\rm gas,90\%}$/$R_{\rm dust,90\%}\gtrsim4$ with the threshold for efficient pebble drift and our luminosity criterion of an OR disk labeled as dotted lines.
Delivery of water through efficient radial pebble drift may therefore be a serviceable explanation of the H$_2$O detections seen in USco~2 and 9. 

However, USco~7 and 8, the latter of which has the largest $R_{\rm gas,90\%}$/$R_{\rm dust,90\%}$ of our sample, both possesses markedly Water-Absent (WA) spectra. This either (1) implies that due to the relatively old age of our sample, that most of the gas phase H$_2$O has already accreted onto the central star (2) that some physical mechanism has preferentially delivered carbon-based molecules to the inner-disk without delivering H$_2$O in these sources.

\subsection{The Role of Outer-Disk Dust Traps in Water-Absent Spectra} \label{subsec:dust_trapping}

Pressure traps are another mechanism that can regulate the C/O ratio in the inner disk gas.
If the pressure gradient that usually enables pebble drift reverses (e.g., at the edge of a planet-opened gap, dust ring), a local pressure maximum forms, particles pile up, and inward drift is halted \citep{pinilla12,Lambrechts14,birnstiel12}; this would in turn impede the advection processes mentioned in Sec.~\ref{subsec:evolution}, altering whether the inner disk actually experiences either the oxygen- or carbon-enrichment phases in the first place \citep{Kalyaan21,Kalyaan2023,Mah23,Krijt2025,Luo2026}. We note that \citet{Vioque2025} report USco~6, 7, and 8 as possessing inner-disk dust \textit{cavities}, while we choose to refer to these as dust \textit{traps} due to their extended nature (see \citet{Mallaney2026} for more on cavities and Table~\ref{tab:sources} for our dust ring radii).

Recent modeling suggests that the origin of a gap matters \citep{Lienert2024}: gaps produced by forming planets produce a distinct observable signature from those produced by internal photoevaporation, as planets affect solids and gas differently. In planetary gaps, gas can still flow across the opening through streams, and can therefore still produce a higher inner disk C/O by cutting off icy pebble (e.g., H$_2$O) delivery while admitting carbon-bearing species in the gas-phase due to their farther-out ice lines \citep{Bergner2022,Lienert2024,gorti11,Pinilla17}.

Strikingly, all of the ALMA-resolved, Water-Absent disks in our sample (USco~1, 6, 7, and 8) exactly correspond to the Upper Sco sources with prominent dust rings (peak brightness temperature is not located in the inner disk, see \citealt{Vioque2025}) at large radii.
Furthermore, the distinction of gap-origin indicates that if the ringed structures in USco~7 and 8 (the only two WA disks with strong molecular features) are created due to planetary formation rather than internal photoevaporation, we would still expect to see the carbon-based molecules we observe in these disks as long as water emission is depleted.
As the shouldered ring structures in USco~9 and 10 (see \citealt{Vioque2025}), the only other two disks that display substructure, are much smaller relative to the steep-outward radial profiles of the inner disk, and both display moderate detections of water in the sample, it is plausible that either the shouldered rings were simply not large enough, or formed after the timescale of mass pebble drift and therefore did not prevent inward drift. The substructure scenarios of a ``shouldered" radial structure and dust ring trapping are both illustrated in Fig.~\ref{fig:cartoon}. We also note that the modeling of \citet{Kurtovic2025} find that most disks in the AGE-PRO Upper Sco sample show evidence of either weak or strong dust trapping, therefore perhaps extending this mechanism as an explanation for the generally water-poor nature of our sample (see Sec~\ref{subsubsec:depletion}).

\subsection{High Dust-to-Gas Ratio in the Context of Water-Rich Spectra} \label{subsec:d2g}
USco~5 stands apart as the only WR disk of our sample.
Additionally, despite having the highest \textit{normalized} H$_2$O$_{Warm}$ luminosity ($\mathrm{L_{H_2O_{Warm}}/L_{13\mu m}}$) of both the AGE-PRO Upper Sco and JDISCS~C1 populations by factors of 2-180 (see Fig.~\ref{fig:lacc_norm}), as well as the highest S/N spectrum in our sample, USco~5 surprisingly possesses no detections of C$_2$H$_2$, HCN, or CO$_2$.
This source also stands out by having the largest L$\rm_{acc}$ value of our sample, as well as an abnormally high disk-averaged d2g ($\sim$0.26) compared to ISM values ($\sim$0.01, \citealt{Bohlin78}), largely due to having the lowest M$\rm_{gas}$ of the sample by over an order of magnitude (see Table~\ref{tab:sources}; computation done using values from \citealt{Agurto-Gangas25} and \citealt{Trapman2025a}). The bottom-right panel of Fig.~\ref{fig:spectral_trends} illustrates this, showing the warm water line luminosity (L$\rm_{H_2O_{Warm}}$) as a function of d2g, where the log scale of both axes serves to visually show to which USco~5 is an outlier on both axes. While no trends can be discerned from this panel, the abnormally high d2g could place the source in a regime where H$_2$O dominates carbon-based molecules to this extent.

It has been proposed that a high d2g could lead to ``dust back" reactions that stall or even reverse the net inward advection of gas in the disk, increasing overall disk lifetimes \citep{Dipierro2018,Garate2020}.
It has also been proposed that a high d2g can lead to dust-shielding, which can prevent UV rays from penetrating deep in the gas of the inner-disk and slowing down the photodissociation of water vapor \citep{woitke16}.
Notably, all of the sources in our sample that do exhibit water emission lines (USco~2, 3, 5, 9, and 10) are consistent with either the steep inner disk radial profiles of our sample (USco~3, 9, and 10, the latter of which is depicted in Fig.~\ref{fig:cartoon}) or those that possess a compact ($\lesssim$50 au) gas disk (USco~2, 3, and 5, with the exception of USco~4), suggesting that centrally-concentrated continuum emission can also be indicative of inward pebble delivery of H$_2$O and/or dust-shielding.

\subsection{Factors Controlling Chemical Diversity} \label{subsec:factors}

The discussions above identify several mechanisms that can shape inner-disk chemistry: a high d2g maintaining the inner-disk water vapor reservoir, inner cavities that deplete the inner-disk of gas, C/O evolution driven by different drift or carbon grain processing, and outer dust traps that block water delivery to the inner disk. With both the mm substructure \citep{Vioque2025} and chemical compositions (this work) of part of our sample in view, a natural question is whether these physical properties can at all predict our observed chemical classes.
In order to produce such a diverse range of chemical classes in our single Upper Sco sample, multiple mechanisms are likely operating simultaneously, which we depict in the cartoon in Fig.~\ref{fig:cartoon}. This illustration shows the proposed disk structure, radial profiles from \citet{Vioque2025}, and resulting JWST spectra from our analysis for the centrally-concentrated continuum of the WP source USco~10, the outer disk traps present at different radii in the WA sources USco~7, 8, and 1, as well as the inner-disk clearing of USco~1.

For example, this Figure suggests that a disk with hints of a protoplanetary-induced dust trap past the ice line of CO could partially deprive the inner disk of carbon, especially if this dust trap forms between the transition from water-rich to organic-rich discussed in Sec.~\ref{subsec:evolution}, allowing for the majority of pebbles to reach the inner disk during the former phase and resulting in a centrally-concentrated mm radial profile (as discussed in Sec.~\ref{subsec:d2g}) and a water-dominated spectrum with no observable carbon chemistry, as is all the case with the WP disk USco~10 depicted in the top row of Fig.~\ref{fig:cartoon}. In the two scenarios depicted in the middle two rows, the difference between an outer-disk dust trap outside and inside of the theoretical ice line of CH$_4$/other donors to the carbon chemistry \citep{Kanwar2024a,Raul2025} could possibly lead to the difference between water-absent CD and OR disks; in the former scenario gas streams may permit CO to travel to the inner disk and create observable amounts of CO$_2$ \citep{Furuya-Aikawa2014,Furuya2022,Bosman18} as is seen in USco~7, while in the latter scenario the carbon-carriers travel to the inner disk and create organic-dominated chemistry, as is seen in USco~8. Lastly, the presence of a outer-disk dust trap in combination with a confirmed inner-disk dust cavity as discussed in Sec.~\ref{subsubsec:ma_cavities} can result in the MA spectra we observe in both USco~6 and 1, the latter of which is depicted in the bottom row. However, without incredibly high resolution observations of more disks in both the mm and infrared, these mechanisms will remain largely speculative.

However, due to the diversity in physical parameters we observe within this one sample (see Table~\ref{tab:sources}), the timescales of pebble drift, inner-disk cavity opening, and outer-disk pressure trap formation will be different and not necessarily correlated within our sample with age.
Without the resolution to confirm the presence of inner-disk \textit{gas} cavities (Sec.~\ref{subsubsec:wa_cavities}) or being able to rule out rapid pebble drift as a cause of water accretion onto the central star (Sec.~\ref{subsubsec:drift}), the trend of water-absence in our sample cannot be solely attributed to the correlation with the presence of outer disk traps.
Alternatively, the gas-phase H$_2$O can also be reduced by other mechanisms such as continued grain growth and sequestration of volatiles in large bodies by locking ices in planetesimals and pebbles over long timescales \citep{hogerheijde11_water,Bergin-vanDishoeck2012,Krijt16}. Lastly, the initial chemical conditions of each source can still vary within the Upper Sco cluster, which will also ultimately contribute to the chemical diversity of the sample.

Building on previous works, our analysis of 10 sources in Upper Sco suggest there are at least three primary factors that play a role in regulating the chemical outcome: age, stellar mass (VLMS vs. T~Tauri), and disk structure. However, these are not independent parameters, and with a small sample of 10 disks (only 7 of which are ALMA-resolved), we cannot disentangle the multiparametric effects of each of these on the chemical diversity we observe. This study must be expanded to larger sample of older disks (e.g., Xie et al., in press) in order to break these degeneracies, and more 1.3 mm observations must be taken of WP and WA disks in order to confirm the physical correlations we see here.
Further comparisons to the younger Ophiuchus ($<$1~Myr) and Lupus (1-2~Myr) star-forming regions will be presented in a forthcoming paper (Waggoner et al. in prep).

\subsection{Implications for Late‐Stage Planet Formation} \label{subsec:planets}

The chemical inventory we map in 2-–6~Myr disks has direct consequences for planet assembly and volatile delivery.
The bulk composition of planetesimals and the solid component of planets is set by the solid-phase inventory (ices and refractories) and by dynamical transport (e.g., pebble drift and trapping, \citealt{Liu2020}); hence, elevated gas-phase C/O does not on its own guarantee carbon-rich solids unless solids have also been preferentially enriched in carbon by transport or local processing \citep{Madhusudhan12,booth17,Bergin2024}. Such a distinction matters observationally: an atmosphere that forms by direct accretion of a carbon-rich gas will tend to show signatures of high C/O, whereas a planet assembled primarily from carbon-rich planetesimals entails a more complex combination of bulk composition and atmosphere \citep{Moses13,Hakim2019,Krijit2023,Bergin2023,Lin-Seager2025,Li2026}.  
  
Furthermore, the persistence of organic species at cooler temperatures suggests reservoirs of complex molecules available for incorporation into forming planetesimals during the epoch of gas dispersal. 
If carbon‐rich solids accumulate in the terrestrial region, they may seed early protoplanets with an inventory of prebiotic organics \citep{Madhusudhan12,oberg11}.
Finally, reduced H$_2$O line luminosities hint that water delivery pathways could become increasingly inefficient by $\sim$2–6~Myr, constraining the timescale for volatile-rich planet formation. In combination with dynamical models of pebble drift and photoevaporation \citep{Lambrechts19, Owen2011}, this provides further credence to theories that suggest water‐bearing terrestrial worlds such as Earth must be enriched by mechanisms such as bombardment \citep{Albertsson2014}, suggesting a more complex process that requires more observations to understand. By benchmarking the chemical state of inner disks at the end of their gas‐rich phase, this study provides critical boundary conditions for models of rocky planet composition and water-enrichment.

\section{Conclusions} \label{sec:conclusion}

In this paper, we have presented a JWST/MIRI MRS survey of 10 protoplanetary disks in the Upper Scorpius association, probing the inner few au chemistry of evolved (2-–6~Myr) disks that mark the final gas‐rich phase in the AGE-PRO program. While noting the small sample size and that observations of many more disks will need to be made with JWST and future observatories in order to truly constrain population-level trends, we find:
\begin{enumerate}

\item \textbf{Water vapor luminosity declines with age.}  
Analysis compared to younger disks shows that line luminosities for all three temperature components of H$_2$O decrease with stellar age, with 4/10 showing poor water luminosities and 5/10 disks showing no water detections at all (Sec.~\ref{subsubsec:depletion},~\ref{subsec:comparison}).

\item \textbf{Inner regions of old disks diverge into distinct chemical categories.}
Empirically, our sample spans a range of Water-Rich (WR), Water-Poor (WP), Water-Absent (WA), Organic-Rich (OR), CO$_2$-Dominated (CD), and Molecule-Absent (MA) systems, demonstrating that disks at later ages in a similar environment can exhibit markedly different chemical architectures in the inner few au (Sec.~\ref{subsubsec:water_class},~\ref{subsubsec:special_class},~\ref{subsec:comparison}).

\item \textbf{Organic emission in older disks is distinctly colder than younger disks.} 
Each detection of an organic molecule in our sample (e.g., C$_2$H$_2$, $^{13}$CCH$_2$, HCN, HC$_3$N, C$_2$H$_6$, C$_4$H$_2$, C$_6$H$_6$) displays a noticeably cold excitation temperature ($\lesssim$300 K) compared to younger ($\sim$1–3~Myr) samples ($\sim$600-1000 K).
For the broad Q branches of molecules such as C$_2$H$_2$ and HCN, a distinctly narrower feature strength is indicative of less excitation of the upper energy levels (Sec.~\ref{subsubsec:cold}).

\item \textbf{Water absence is tentatively correlated with outer-disk dust traps.} 
When comparing the physical properties of our disks to the chemistry, several tentative patterns emerge (e.g., the sources with the highest $R_{\rm gas,90\%}$/$R_{\rm dust,90\%}$ are all Organic-Rich disks, our sole Water-Rich disk has an abnormally high dust-to-gas mass ratio). However, the most noticeable is that each of the four ALMA-resolved Water-Absent sources (USco~1, 6, 7, and 8) in our sample displays a prominent outer-disk dust trap, suggesting the importance of dust in controlling the inner-disk chemistry (Sec.~\ref{subsec:dust_trapping},~\ref{subsec:factors}).

\end{enumerate}

\begin{acknowledgments}
This work is based on observations made with the NASA/ESA/CSA James Webb Space Telescope. The data were obtained from the Mikulski Archive for Space Telescopes at the Space Telescope Science Institute, which is operated by the Association of Universities for Research in Astronomy, Inc., under NASA contract NAS 5-03127 for JWST. These observations are associated with JWST GO Cycle 2 program ID 3034 (PI: K. Zhang). Support for E.R. and K.Z. through this program was provided by NASA through a grant from the Space Telescope Science Institute, which is operated by the Association of Universities for Research in Astronomy, Inc., under NASA contract NAS 5-03127.
We thank Blue Rachapradit for their invaluable help with the illustration in Fig.~\ref{fig:cartoon}.
CFM is funded by the European Union (ERC, WANDA, 101039452). Views and opinions expressed are however those of the author(s) only and do not necessarily reflect those of the European Union or the European Research Council Executive Agency. Neither the European Union nor the granting authority can be held responsible for them.
PP acknowledges funding from the UK Research and Innovation (UKRI) under the UK government’s Horizon Europe funding guarantee from ERC (under grant agreement No 101076489).
N.S.B. acknowledges support from NASA Grant No. 80NSSC25K0301 issued through the NNH24ZDA001N-FINESST program.
TK was supported by Science and Technology facilities Council
(STFC) grant no. ST/Y002415/1.
JM acknowledges support from ANID -- Millennium Science Initiative Program -- Center Code NCN2024\_001.
SK acknowledges support from STFC Grant ST/Y002415/1.

\end{acknowledgments}

\begin{contribution}


\end{contribution}

\facilities{
JWST MIRI/MRS \citep{Rieke2015,Wright2015,Wells2015,Argyriou20}
}

\software{
  astropy \citep{Astropy2022},
  emcee \citep{emcee},
  spectools\_ir \citep{Salyk22},
  ctool \citep{Pontoppidan_2026_ctool},
  iSLAT \citep{Jellison24, iSLAT_code},
  matplotlib \citep{Hunter2007},
}

\appendix

\section{Individual Source Notes} \label{sec:comments}
In this section, we provide miscellaneous notes and comments about each source, sorted by Chemotype, as well as various comments broadly applicable to each Chemotype.

\subsection{Notes on Organic-Rich Sources} \label{subsec:or}
Our Organic-Rich (OR) disks (USco~2, 8, and 9) display the richest carbon chemistry, with robust detections of C$_2$H$_2$, HCN, $^{13}$CCH$_2$, HC$_3$N, CO$_2$, C$_4$H$_2$, CH$_3$, C$_2$H$_6$, C$_6$H$_6$, and tentative $^{13}$CO$_2$ and CH$_4$, (see Table~\ref{tab:detections}, Fig.~\ref{fig:contributions_org}, \ref{fig:upsco2_contribution}).
Other OR sources in the literature include J160532 (\citealt{Tabone23}, VLMS), Sz~28 (\citealt{Kanwar2024b}, VLMS), ISO-ChaI 147 (\citealt{Arabhavi24}, VLMS), DoAr~33 (\citealt{Colmenares24}, K-type), and GO~Tau (K-type).
DoAr~33 and GO~Tau are both included in the JDISCS~C1 sample, which show detections of only one COM in their spectra (C$_4$H$_2$), compared to the 1, 2, and 3 COMs that we find in USco~9, 8 and 2, respectively (see Table~\ref{tab:detections}).
However, we note that do not detect any emission of C$_2$H$_4$ or C$_3$H$_4$ in any of our OR spectra, as are found in the VLMS sources Sz~28 and Iso-ChaI~147, broadly agreeing with Xie et al. (in press) that lower mass sources tend to exhibit more complex hydrocarbons.

A key aspect of the organics found in our evolved ($\sim$2-6~Myr) AGE-PRO Upper Sco sample is that they are found at relatively low excitation temperatures (see Sec.~\ref{subsubsec:cold}). As seen in Fig.~\ref{fig:cold_organics}, we note that in JDISCS~C1, on average HCN (emitted at $\sim$820 K) features are more often brighter than C$_2$H$_2$ (emitted at $\sim$920 K) features within the same disk. Conversely, in our Upper Sco OR sources, all three C$_2$H$_2$ features are substantially stronger (see Fig.~\ref{fig:contributions_org}) than HCN (emitted on average at $\sim$290 K vs. $\sim$240 K). However, this does not necessarily indicate a an inverse relationship between emitting temperature and emitting mass as was tested for in Fig.~14 of \citet{Arulanantham25}, where no correlation was found.

We also acknowledge that HC$_3$N and C$_4$H$_2$ emit at low enough temperatures (see Table~\ref{tab:bestfit_params}) in our models that the log(N) and/or log(A) must be increased until their P- and R- branches become discernible in our best fits (see USco~2 and 8 in Fig.~\ref{fig:contributions_org}). While these optically thick P- and R- branches of HC$_3$N and C$_4$H$_2$ are not directly visible in our continuum-subtracted spectra, this could simply be a consequence of our continuum subtraction algorithm over-subtracting this excess emission. This oversubtraction may also be the case for the R branch of C$_2$H$_2$ emission in all three OR disks, as well as the detected and tentative C$_2$H$_6$ and $^{13}$CO$_2$ features we observe in USco~2 and 8, respectively (see Fig.~\ref{fig:contributions_org}) Another possible explanation is that these exceedingly cold molecules are emitting along a column density and temperature gradient, rather than at a single value (see \citealt{Kaeufer24a} for more). This is exemplified by the extremely large log(A) error bars on our C$_4$H$_2$ detections (Table~\ref{tab:bestfit_params}), the best-fitting values of which imply an average emitting radius of $\gtrsim$2.5 au, the temperatures of which ($<100$ K) start to become unphysical (see Sec.~\ref{subsec:or} for more). We also note that the hot CO$_2$ branch at $\sim$16.2 $\upmu$m is not fit to well in USco~2 and 9, implying that CO$_2$ may also be emitting along a column density and temperature range. However, as this feature is also not fit to well in \citet{Vlasblom2025}, we note that this could also be an unidentified molecular feature separate from cold CO$_2$.

\textit{USco~2 (2MASS J16054540-2023088):}
Upper Sco 2 is one of two Organic-Rich disks that also shows detections of C$_2$H$_6$ and C$_6$H$_6$, and the only one which also contains a definitive detection of $^{13}$CO$_2$ and tentative detection of CH$_4$ (see Fig.~\ref{fig:upsco2_contribution}).
USco~2 is also the only other disk in the sample that displays a definitive hot-water component (see Fig.~\ref{fig:contributions_org},~\ref{fig:upsco2_contribution}) outside of our ``Water-Rich" disk USco~5. We note that we include this H$_2$O temperature component from 11.75-18 $\upmu$m in replacement of the warm H$_2$O component included in this wavelength window for the other sources in our sample (see Sec.~\ref{subsec:model}). This is done in order to improve the fit to the spectrum compared with when both components are simultaneously included in this region, in which case most water features are overshot by the model.
However, this hot-water component is at the lower end of our hot-water temperature prior ($\gtrsim$600 K) at $\sim$640 K, indicating that it may not be necessary for a complete fit $>$ 11.75 $\upmu$m.

\textit{USco~8 (2MASS J16221532-2511349):}
USco~8 is the second of the two Organic-Rich disks that exhibits detections C$_2$H$_6$ and C$_6$H$_6$, as well as a tentative detection of $^{13}$CO$_2$ (Fig.~\ref{fig:upsco2_contribution}).
However, we interestingly find that USco~8 is the only Organic-Rich disk with no detections of any water features in the spectrum, qualifying it as a Water-Absent disk (see Sec.~\ref{subsubsec:depletion}).
This source also produces the smallest CO$_2$ signal of this class, also possibly suggesting general oxygen depletion or dust trapping beyond the CO$_2$ ice line (see Sec.~\ref{subsec:dust_trapping} for more details). We qualify the C$_2$H$_6$ and $^{13}$CO$_2$ features in this spectrum as tentative due to the low S/N of these lines (see Fig.~\ref{fig:contributions_org}).

\textit{USco~9 (2MASS J16082324-1930009):}
USco~9 displays the weakest C$_4$H$_2$ and HC$_3$N features of our three Organic-Rich sources (see Fig.~\ref{fig:contributions_org}, Table~\ref{tab:bestfit_params}). This source also notably has non-detections of some of the most complex organic molecules in our sample (C$_2$H$_6$ and C$_6$H$_6$), perhaps suggesting
the destruction of its less stable complex hydrocarbons due to its relatively older age in the sample (see Table~\ref{tab:sources}, Sec.~\ref{subsubsec:destruction} for more discussion). Interestingly, USco~9 qualifies as a highly-inclined object ($>70^{\circ}$) with an inclination of $\sim$75$^{\circ}$, yet displays one of the highest S/N spectra of our sample, therefore being similar to one of the exceptions to the result found in Zhang et al. (submitted) that highly-inclined tend to show weak molecular emission.

\subsection{Notes on \texorpdfstring{CO$_2$}{CO2}-Dominated Sources} \label{subsec:cd}
We report a distinct set (USco~3 and 7) of CO$_2$-Dominated disks (CD) that exhibit distinct CO$_2$ signatures while lacking strong H$_2$O components in our slab fits (see Fig.~\ref{fig:contributions_standard}).
USco~3 and 7 are not the first CO$_2$-Dominated (CD) disks that have been detected within the literature. \citet{pontoppidan10a} first identified several sources with Spitzer that did not have any molecular detections outside of CO$_2$ (similar to USco~7), including GW~Lup and HT~Lup, the latter of which is included in the JDISCS~C1 sample. Here, we do not classify HT~Lup as a CD disk due to only possessing a $\mathrm{L_{CO_2}}/\mathrm{L_{H_2O_{Warm}}}$ of $\sim$4.5 and L$\rm_{organic}$ of $\sim$0.4, both of which fall just short of our criteria of a CD disk (Table~\ref{tab:classes}). However, GW~Lup was confirmed to be a CD disk in \citet{Grant23}, as well as the two disks CX~Tau \citep{Vlasblom2024} and XUE~10 \citep{Frediani2025} from the XUE survey \citep{Ramirez-Tannus2025}. Lastly, the JDISCS~C1 targets Sz~114 \citep{Xie2023} and MY~Lup \citep{Salyk2025} are the two CD disks identified within the JDISCS~C1 sample, as seen in Fig.~\ref{fig:histogram}. While GW~Lup, MY~Lup, and XUE~10 notably exhibit several CO$_2$ isotopologues in their spectra unlike USco~3 and 7, Sz~114 and CX~Tau only show confirmed detections of $^{13}$CO$_2$, showing that CO$_2$ isotopologues are not necessarily an indicator of disks dominated by a CO$_2$ feature.

\textit{USco~7 (2MASS J16202863-2442087 ):}
USco~7 is the most prominent CO$_2$-Dominated disk in our sample, exhibiting CO$_2$ as its only detected molecule with no detected warm/cold water emission at all, qualifying it as a WA disk (see Sec.~\ref{subsubsec:water_class}). We identify multiple tentative cold water features in this source such as at $\sim$21.85 $\upmu$m, but due to the lack of other cold H$_2$O transitions in the spectrum we leave this as tentative. We note that \citet{Miley2025} classify USco~7 as an extremely variable disk, which could have implications for X-ray driven chemistry as was found in IM~Lup \citep{cleeves17}, as well as a tentative close binary system, which could lead to an inner-disk cavity not resolved by ALMA in \citet{Vioque2025}.

\textit{USco~3 (2MASS J16020757-2257467):}
We identify USco~3 as a CO$_2$-Dominated disk, with a sharp CO$_2$ feature and weak warm water emission, falling into the category of WP disks (see Sec.~\ref{subsubsec:water_class}). While the warm H$_2$O diagnostic feature is not present at an S/N$>$3 in this spectrum, we observe the presence of multiple other water emission lines from the $\sim$15.675-17.675 $\upmu$m region and therefore classify this emission as real. We note that this disk displays the coldest CO$_2$ slab model temperature of our entire sample, followed only by the other CD disk (USco~7).

\subsection{Notes on Molecule-Absent Sources} \label{subsec:ma}
We identify a group of Molecule-Absent (MA) disks that show no robust non-H$_2$ molecular detections above our S/N threshold (3). The presence of atomic lines are not included in this criterion (see Table~\ref{tab:atomic_detections}). We note that this definition is distinctly different from the definitions of molecule-poor (MP) disks as defined in \citet{Mallaney2026} and Xie et al. (in press), who define MP sources as those with significantly subluminous emission compared to full disks and as lacking detections of both C$_2$H$_2$ and H$_2$O, respectively. Our Molecule-Absent sources in this sample are USco~1, 4, and 6.

With the exception of an overly-noisy region from $\sim$15-15.6 $\upmu$m seen in all three MA disks (as well as USco~3, 7 and 10; see Fig.~\ref{fig:identifications}), we find that the noise-level of USco~1 and 6 do not exceed 0.5 mJy within the loosely-defined organic region ($\sim$11.75-18 $\upmu$m), and therefore the lack of molecular features in these two disks cannot solely be attributed to a noisy spectra. This result is conspicuous due to the relatively high fraction of Molecule-Absent disks within a single sample of a region, which is higher than that of surveys of younger regions with a larger population (see Sec.~\ref{subsec:comparison}, Fig.~\ref{fig:histogram}).
    
\textit{USco~1 (2MASS J16120668-3010270):}
USco~1 shows tentative detections of asymmetric OH doublets (see Fig.~\ref{fig:identifications} and \ref{fig:upsco1_contribution}) and no detections of H$_2$O, which we note is potentially somewhat similar to the case of \citet{Zannese2024}. We note that this is the only fully-resolved 1.3 mm continuum with ALMA in our entire AGE-PRO sample \citep{Vioque2025}, exhibiting an extremely large dust cavity ($\sim$80 au), agreeing with the large ($>$1) n$_{13-26}$ index we discuss in Sec.~\ref{subsubsec:ma_cavities}. Similarly, this source also shows a tentative gas cavity, where $^{13}$CO and C$^{18}$O show significant depletion in the moment zero maps of this dust cavity, but $^{12}$CO does not (see Fig.~3 of \citealt{Sierra2024}). This is the only source in our sample with a tentative detection of a protoplanet \citep{Sierra2024}, allowing for an interesting potential comparison with the source PDS~70, which also contains several actively forming protoplanets \citep{Keppler2019,Casassus22,Perotti2023,Christiaens2024,Jang2024,Rampinelli2024,Blakely2025}.

\textit{USco~4 (2MASS J16111742-1918285):}
USco~4 stands out in that we find no detections of any molecular features or atomic lines. USco~4 shows noisy features up to 5 mJy within the organic  region ($\sim$11.75-18 $\upmu$m), consistent with the faint emission seen in \citet{Zhang25} likely due to its relatively high inclination of $79^\circ$.
This agrees with the result from Zhang et al. (submitted) that highly-inclined ($>70^{\circ}$) disks tend to show weak molecular emission due to this observational effect, though we note that the OR disk USco~9 has a similar inclination of $75^\circ$ (see Table~\ref{tab:sources}).

\textit{USco~6 (2MASS J16163345-2521505):}
USco~6 shows a tentative detection of CO$_2$ (see Figs.~\ref{fig:identifications},~\ref{fig:contributions_standard}). However, with the lowest line luminosities of CO$_2$ in the sample (Table~\ref{tab:line_luminosities}), we qualify that this feature is most likely not real. \citet{Vioque2025} find a dust ring at $\sim$35 au, as well as a gas cavity, agreeing with the large ($>$1) n$_{13-26}$ index we find for the source, as discussed in Sec.~\ref{subsubsec:ma_cavities}.

\subsection{Notes on Other Sources} \label{subsec:other}
\textit{USco~5 (2MASS J16145026-2332397):}
USco~5 stands out with the highest integrated water luminosity in our sample (Table~\ref{tab:line_luminosities}) for all three temperature components, two distinct OH temperature components, and a distinct lack of hydrocarbon features (see Fig.~\ref{fig:upsco5_contribution1}).
We also find that USco~5 is the only disk in our sample that shows a definitive detection of CO (Fig.~\ref{fig:upsco5_contribution2}).
This stands out due to the near ubiquitous detection rate of C$_2$H$_2$ and HCN in JDISCS~C1 (see Sec.~\ref{subsec:comparison}). The only disk in JDISCS~C1 with definitive detections of water but no organics \textit{and} CO$_2$ is Sz~129, which has much lower normalized warm H$_2$O emission.
As this source has an unusually high dust-to-gas ratio (see Table~\ref{tab:detections}), and sits in our older age bin at $\sim$3.8$_{-1.9}^{+0.9}$~Myr, we explore possible explanations in Sec.~\ref{subsubsec:depletion}.
The best-fitting model for USco~5 is displayed in Fig.~\ref{fig:upsco5_contribution1} and \ref{fig:upsco5_contribution2}.

\textit{USco~10 (2MASS J16090075-1908526):}
USco~10 is the only other disk besides USco~5 to not fall into any particularly defined Chemotype.
This disk firmly falls into the WP category of disks, with a substantial normalized H$_2$O$_{Cold,a}$ luminosity ($\mathrm{L_{H_2O_{Cold,a}}/L_{13\mu m}}$) compared to much of the JDISCS~C1 sample (see Fig.~\ref{fig:lacc_norm}).
However, we find that USco~10 in particular is the only other disk in the sample to contain two distinct OH temperature components, and with the addition of a tentative CO$_2$ detection is indicative of rich oxygen chemistry.

\section{Additional Tables} \label{subsec:additional_tables}

\movetabledown=5.2325cm
\begin{rotatetable}
\begin{deluxetable*}{c|l|c|c!{\color{gray!50}\vrule}c!{\color{gray!50}\vrule}c!{\color{gray!50}\vrule}c|ccc!{\color{gray!50}\vrule}c!{\color{gray!50}\vrule}c}
\tablewidth{0pt}
\tablecaption{Defined line luminosities (in units of $10^{-7}$ L$_\odot$) for each source and molecule, sorted primarily by Water Classification Type (as defined in Table~\ref{tab:classes}). Each wavelength range we define for computation is listed for each molecule, where '-' indicates a derived value from the other luminosities. For molecules marked with a '$^*$', line luminosities were computed using the continuum-subtracted data rather than the model, either because the model didn't fully encompass the feature in the case of the water diagnostic lines \citep{Banzatti25} or models were not run in the case of C$_6$H$_6$ and CH$_3$. Upper limits for the non-detected molecules C$_2$H$_2$, HCN, CO$_2$, and H$_2$O temperature components are marked in \textcolor{gray}{gray}, upper limits for all tentative detections are marked in \textcolor{orange}{orange}, and upper and lower limits for derived quantities where at least one detection is present in the calculation are marked in black. Most of these regions are visualized in Fig.~\ref{fig:identifications}. The four quantities in the bottom of the table correspond to the derived luminosities used to define our different classes of disks (see Sec.~\ref{subsec:integrations}).
\label{tab:line_luminosities}}
\tablehead{
\multicolumn{2}{c}{\textbf{Water Classification}} &
\multicolumn{1}{|c|}{\textbf{WR}} &
\multicolumn{4}{c}{\textbf{WP}} &
\multicolumn{5}{c}{\textbf{WA}} \\
\tableline
\multicolumn{2}{c}{Chemotype} &
\multicolumn{1}{|c|}{} &
\multicolumn{1}{c!{\color{gray!50}\vrule}}{OR} &
\multicolumn{1}{c!{\color{gray!50}\vrule}}{CD} &
\multicolumn{1}{c!{\color{gray!50}\vrule}}{OR} &
\multicolumn{1}{c|}{} &
\multicolumn{1}{c}{} &
\multicolumn{1}{c}{MA} &
\multicolumn{1}{c!{\color{gray!50}\vrule}}{} &
\multicolumn{1}{c!{\color{gray!50}\vrule}}{CD} &
\multicolumn{1}{c}{OR} \\
\tableline
\multicolumn{1}{c|}{Molecule} & \multicolumn{1}{l}{Range (\(\upmu\)m)} &
\multicolumn{1}{|c|}{USco~5} & \multicolumn{1}{c!{\color{gray!50}\vrule}}{USco~2} & \multicolumn{1}{c!{\color{gray!50}\vrule}}{USco~3} &
\multicolumn{1}{c!{\color{gray!50}\vrule}}{USco~9} & \multicolumn{1}{c|}{USco~10} & \multicolumn{1}{c}{USco~1} &
\multicolumn{1}{c}{USco~4} & \multicolumn{1}{c!{\color{gray!50}\vrule}}{USco~6} & \multicolumn{1}{c!{\color{gray!50}\vrule}}{USco~7} &
\multicolumn{1}{c}{USco~8}
}
\startdata
C$_2$H$_2$ & 13.610--13.751 &
\color{gray}$<1.29$ & $29.35 \pm 0.31$ & \color{gray}$<0.36$ &
$33.99 \pm 0.28$ & \color{gray}$<4.33$ & \color{gray}$<0.98$ &
\color{gray}$<1.76$ & \color{gray}$<0.63$ & \color{gray}$<2.15$ &
$36.95 \pm 0.29$ \\
HCN & 13.910--14.065 &
\color{gray}$<1.3$ & $13.0 \pm 0.17$ & \color{gray}$<0.36$ &
$18.2 \pm 0.22$ & \color{gray}$<4.36$ & \color{gray}$<0.99$ &
\color{gray}$<1.77$ & \color{gray}$<0.63$ & \color{gray}$<2.17$ &
$17.65 \pm 0.21$ \\
CO$_2$ & 14.936--15.014 &
\color{gray}$<0.81$ & $5.7 \pm 0.09$ & $1.73 \pm 0.08$ &
$9.77 \pm 0.13$ & \color{orange}$<2.71$ & \color{gray}$<0.62$ &
\color{gray}$<1.1$ & \color{orange}$<0.39$ & $6.91 \pm 0.45$ &
$3.62 \pm 0.11$ \\
C$_2$H$_{\rm 2,tot}$ & 12.0--16.0 &
- & $80.63 \pm 1.0$ & - &
$89.85 \pm 1.16$ & - & - &
- & - & - &
$102.99 \pm 1.14$ \\
HCN$_{\rm tot}$ & 12.0--16.0 &
- & $31.41 \pm 0.66$ & - &
$40.87 \pm 1.06$ & - & - &
- & - & - &
$42.47 \pm 0.97$ \\
CO$_{\rm 2,tot}$ & 12.0--16.0 &
- & $14.36 \pm 0.6$ & $3.62 \pm 0.64$ &
$26.02 \pm 1.04$ & - & - &
- & - & $14.25 \pm 3.82$ &
$7.44 \pm 0.94$ \\
\arrayrulecolor{gray!50}\tblrulevskip\hline\arrayrulecolor{black}\tblrulevskip
H$_2$O$_\mathrm{Hot}^*$ & 17.317--17.330 &
$7.34 \pm 0.07$ & $0.25 \pm 0.01$ & \color{gray}$<0.06$ &
\color{gray}$<0.1$ & \color{gray}$<0.73$ & \color{gray}$<0.17$ &
\color{gray}$<0.3$ & \color{gray}$<0.11$ & \color{gray}$<0.36$ &
\color{gray}$<0.09$ \\
H$_2$O$_\mathrm{Warm}^*$ & 17.490--17.520 &
$17.29 \pm 0.16$ & $0.56 \pm 0.02$ & $0.22 \pm 0.03$ &
$2.83 \pm 0.04$ & $1.09 \pm 0.08$ & \color{gray}$<0.27$ &
\color{gray}$<0.48$ & \color{gray}$<0.17$ & \color{gray}$<0.59$ &
\color{gray}$<0.14$ \\
H$_2$O$_{\mathrm{Cold,a}}^*$ & 23.805--23.830 &
$27.35 \pm 0.28$ & $0.82 \pm 0.07$ & \color{orange}$<0.07$ &
$4.77 \pm 0.18$ & $5.96 \pm 0.23$ & \color{gray}$<0.18$ &
\color{gray}$<0.33$ & \color{gray}$<0.12$ & \color{orange}$<0.4$ &
\color{gray}$<0.1$ \\
H$_2$O$_{\mathrm{Cold,b}}^*$ & 23.880--23.910 &
$25.52 \pm 0.27$ & $1.36 \pm 0.1$ & \color{gray}$<0.08$ &
$4.91 \pm 0.2$ & $5.65 \pm 0.26$ & \color{gray}$<0.21$ &
\color{gray}$<0.37$ & \color{gray}$<0.13$ & \color{gray}$<0.46$ &
\color{gray}$<0.11$ \\
\arrayrulecolor{gray!50}\tblrulevskip\hline\arrayrulecolor{black}\tblrulevskip
OH$_\mathrm{Hot}$ & 20.035--20.065 &
$4.91 \pm 0.14$ & $0.29 \pm 0.03$ & - &
$1.78 \pm 0.06$ & $2.09 \pm 0.44$ & \color{orange}$<0.30$ &
- & - & - & - \\
OH$_\mathrm{Cold}$ & 24.595--24.630 &
$5.67 \pm 0.1$ & - & - &
- & $3.19 \pm 0.29$ & - &
- & - & - &
- \\
C$_2$H$_6$ & 11.25--12.75 &
- & $10.33 \pm 0.39$ & - &
- & - & - &
- & - & - &
\color{orange}$<2.07$ \\
C$_6$H$_6^*$ & 14.825--14.85 &
- & $1.7 \pm 0.02$ & - &
- & - & - &
- & - & - &
$3.03 \pm 0.04$ \\
HC$_3$N & 15.050--15.090 &
- & $3.55 \pm 0.06$ & - &
$3.39 \pm 0.09$ & - & - &
- & - & - &
$8.59 \pm 0.09$ \\
$^{13}$CO$_2$ & 15.393--15.430 &
- & $0.55 \pm 0.06$ & - &
- & - & - &
- & - & - &
\color{orange}$<0.27$ \\
C$_4$H$_2$ & 15.908--15.938 &
- & $1.6 \pm 0.04$ & - &
$0.92 \pm 0.06$ & - & - &
- & - & - &
$1.95 \pm 0.06$ \\
CH$_3^*$ & 16.485--16.510 &
- & $1.07 \pm 0.02$ & - &
$0.98 \pm 0.02$ & - & - &
- & - & - &
$1.09 \pm 0.03$ \\
\arrayrulecolor{black!100}\tblrulevskip\hline\arrayrulecolor{black}\tblrulevskip
L$_\mathrm{organic}$ &  &
\color{gray}$<2.6$ & $60.61 \pm 0.73$ & \color{gray}$<0.72$ &
$57.49 \pm 0.45$ & \color{gray}$<8.69$ & \color{gray}$<1.97$ &
\color{gray}$<3.53$ & \color{gray}$<1.26$ & \color{gray}$<4.32$ &
$69.26 \pm 0.52$ \\
L$_{\mathrm{CO2}}$/L$_{\mathrm{H2O_{Warm}}}$ &  &
\color{gray}$<0.05$ & $10.17$ & $7.86$ &
$3.45$ & \color{gray}$<2.49$ & - &
- & - & $>11.71$ &
$>25.86$ \\
L$_{\mathrm{CO2}}$/L$_{\mathrm{organic}}$ &  &
- & $0.09$ & $>2.41$ &
$0.17$ & - & - &
- & - & $>1.6$ &
$0.05$ \\
\enddata
\end{deluxetable*}
\end{rotatetable}
\clearpage

\clearpage
\startlongtable
\begin{deluxetable*}{lcccccccccc}
    \tabletypesize{\footnotesize}
    \tablewidth{0pt}
    \tablecaption{Best-fitting model parameters for each source. Best-fitting temperatures are rounded to the nearest 10 K, log N to the nearest 0.1 cm$^{-2)}$, log A to the nearest 0.01 au$^2$, and log M to the nearest 0.01 M$_\Earth$. We report upper limits for all non-detections of the three H$_2$O temperature components, C$_2$H$_2$, HCN, and CO$_2$. Tentative detections (upper-limits included) are marked in orange, while non-detection upper limits are marked in gray. Non-LTE emission (e.g., hot and cold OH, CH$_4$, and CO), as well as unphysically cold ($\lesssim$150 K) values (e.g., C$_2$H$_6$, C$_4$H$_2$, see Sec.~\ref{subsec:or} for more) should be taken with caution due to the LTE emission assumption of our slab models (see Sec.~\ref{subsec:model}).
    \label{tab:bestfit_params}}
    \tablehead{
      \colhead{} & \colhead{USco~1} & \colhead{USco~2} & \colhead{USco~3} & \colhead{USco~4} & \colhead{USco~5} & \colhead{USco~6} & \colhead{USco~7} & \colhead{USco~8} & \colhead{USco~9} & \colhead{USco~10}
    }
    \startdata
        \multicolumn{11}{l}{\textbf{H$_2$O} \textbf{(Hot)}} \\ 
        T (K) & - & $620_{-10}^{+30}$ & - & - & $890_{-0}^{+0}$ & - & - & - & - & - \\ 
        log N (cm$^{-2}$) & - & $18.7_{-0.2}^{+0.1}$ & - & - & $18.3_{-0.0}^{+0.0}$ & - & - & - & - & - \\ 
        log A (au$^2$) & - & $-2.17_{-0.01}^{+0.03}$ & - & - & $-1.34_{-0.00}^{+0.00}$ & - & - & - & - & - \\ 
        log M (M\_Earth) & \color{gray}$<-10.24$ & $-7.88_{-0.03}^{+0.02}$ & \color{gray}$<-10.54$ & \color{gray}$<-10.21$ & $-6.96_{-0.00}^{+0.00}$ & \color{gray}$<-10.34$ & \color{gray}$<-10.25$ & \color{gray}$<-10.46$ & \color{gray}$<-10.38$ & \color{gray}$<-10.25$ \\ 
        \midrule
        \multicolumn{11}{l}{\textbf{H$_2$O} \textbf{(Warm)}} \\ 
        T (K) & - & $350_{-10}^{+0}$ & $320_{-40}^{+160}$ & - & $500_{-0}^{+0}$ & - & - & - & $440_{-0}^{+0}$ & $520_{-40}^{+10}$ \\ 
        log N (cm$^{-2}$) & - & $17.2_{-0.7}^{+0.2}$ & $19.3_{-2.1}^{+0.4}$ & - & $17.7_{-0.0}^{+0.0}$ & - & - & - & $18.3_{-0.0}^{+0.0}$ & $17.8_{-0.0}^{+0.0}$ \\ 
        log A (au$^2$) & - & $-1.13_{-0.12}^{+0.13}$ & $-1.70_{-0.16}^{+0.16}$ & - & $-0.10_{-0.01}^{+0.00}$ & - & - & - & $-0.87_{-0.00}^{+0.01}$ & $-1.33_{-0.03}^{+0.17}$ \\ 
        log M (M\_Earth) & \color{gray}$<-10.23$ & $-7.88_{-0.30}^{+0.06}$ & $-6.40_{-0.40}^{+0.15}$ & \color{gray}$<-10.22$ & $-6.40_{-0.00}^{+0.00}$ & \color{gray}$<-10.39$ & \color{gray}$<-10.26$ & \color{gray}$<-10.48$ & $-6.55_{-0.00}^{+0.00}$ & $-7.39_{-0.02}^{+0.03}$ \\ 
        \midrule
        \multicolumn{11}{l}{\textbf{H$_2$O} \textbf{(Cold)}} \\ 
        T (K) & - & - & - & - & $200_{-0}^{+10}$ & - & \color{orange}$240_{-60}^{+100}$ & - & $270_{-10}^{+10}$ & $230_{-20}^{+0}$ \\ 
        log N (cm$^{-2}$) & - & - & - & - & $15.0_{-0.1}^{+0.1}$ & - & \color{orange}$19.9_{-2.1}^{+1.2}$ & - & $17.0_{-0.2}^{+0.7}$ & $17.3_{-0.8}^{+0.0}$ \\ 
        log A (au$^2$) & - & - & - & - & $2.97_{-0.27}^{+0.20}$ & - & \color{orange}$-1.09_{-1.23}^{+0.70}$ & - & $0.14_{-0.28}^{+0.10}$ & $0.17_{-0.04}^{+0.74}$ \\ 
        log M (M\_Earth) & \color{gray}$<-10.23$ & \color{gray}$<-10.70$ & \color{gray}$<-10.57$ & \color{gray}$<-10.24$ & $-5.98_{-0.03}^{+0.01}$ & \color{gray}$<-10.41$ & \color{orange}$-5.10_{-0.28}^{+0.39}$ & \color{gray}$<-10.48$ & $-7.07_{-0.04}^{+0.08}$ & $-6.44_{-0.01}^{+0.01}$\\ 
        \midrule
        \multicolumn{11}{l}{\textbf{OH} \textbf{(Hot)}} \\
        T (K) & \color{orange}$3470_{-1440}^{+420}$ & $1230_{-80}^{+340}$ & - & - & $3170_{-10}^{+10}$ & - & - & - & $1050_{-60}^{+10}$ & $1210_{-20}^{+30}$ \\
        log N (cm$^{-2}$) & \color{orange}$15.9_{-1.4}^{+0.8}$ & $14.7_{-1.1}^{+2.4}$ & - & - & $12.1_{-0.1}^{+0.3}$ & - & - & - & $12.5_{-0.5}^{+0.7}$ & $12.5_{-0.5}^{+0.7}$ \\
        log A (au$^2$) & \color{orange}$-2.83_{-0.17}^{+1.28}$ & $-0.94_{-1.59}^{+1.51}$ & - & - & $2.17_{-0.35}^{+0.12}$ & - & - & - & $2.23_{-0.68}^{+0.58}$ & $2.09_{-0.73}^{+0.58}$ \\
        log M (M\_Earth) & \color{orange}$-10.85_{-0.09}^{+0.60}$ & $-10.23_{-0.36}^{+0.10}$ & - & - & $-9.67_{-0.00}^{+0.00}$ & - & - & - & $-9.19_{-0.01}^{+0.09}$ & $-9.43_{-0.01}^{+0.01}$ \\
        \midrule
        \multicolumn{11}{l}{\textbf{OH} \textbf{(Cold)}} \\
        T (K) & - & - & - & - & $490_{-0}^{+0}$ & - & - & - & - & $120_{-0}^{+0}$ \\
        log N (cm$^{-2}$) & - & - & - & - & $17.3_{-0.0}^{+0.0}$ & - & - & - & - & $23.5_{-0.3}^{+0.3}$ \\
        log A (au$^2$) & - & - & - & - & $-0.04_{-0.01}^{+0.01}$ & - & - & - & - & $0.66_{-0.06}^{+0.07}$ \\
        log M (M\_Earth) & - & - & - & - & $-6.73_{-0.01}^{+0.01}$ & - & - & - & - & $-1.68_{-0.01}^{+0.00}$ \\
        \midrule
        \multicolumn{11}{l}{\textbf{C$_2$H$_6$}} \\ 
        T (K) & - & $100_{-80}^{+30}$ & - & - & - & - & - & \color{orange}$100_{-80}^{+20}$ & - & - \\ 
        log N (cm$^{-2}$) & - & $18.4_{-0.4}^{+0.6}$ & - & - & - & - & - & \color{orange}$18.6_{-0.3}^{+0.6}$ & - & - \\ 
        log A (au$^2$) & - & $1.04_{-0.67}^{+1.96}$ & - & - & - & - & - & \color{orange}$0.90_{-0.51}^{+1.75}$ & - & - \\ 
        log M (M\_Earth) & - & $-4.50_{-0.25}^{+0.03}$ & - & - & - & - & - & \color{orange}$-4.40_{-0.16}^{+0.06}$ & - & - \\ 
        \midrule
        \multicolumn{11}{l}{\textbf{C$_2$H$_2$}} \\ 
        T (K) & - & $240_{-10}^{+20}$ & - & - & - & - & - & $190_{-10}^{+0}$ & $290_{-10}^{+0}$ & - \\ 
        log N (cm$^{-2}$) & - & $17.4_{-0.2}^{+0.1}$ & - & - & - & - & - & $17.5_{-0.8}^{+0.3}$ & $16.9_{-0.0}^{+0.0}$ & - \\ 
        log A (au$^2$) & - & $-0.88_{-0.10}^{+0.05}$ & - & - & - & - & - & $-1.00_{-0.65}^{+0.22}$ & $-1.05_{-0.02}^{+0.07}$ & - \\ 
        log M (M\_Earth) & \color{gray}$<-9.35$ & $-7.46_{-0.11}^{+0.04}$ & \color{gray}$<-9.55$ & \color{gray}$<-9.24$ & \color{gray}$<-8.52$ & \color{gray}$<-9.59$ & \color{gray}$<-9.15$ & $-6.45_{-0.04}^{+0.07}$ & $-8.08_{-0.02}^{+0.07}$ & \color{gray}$<-9.27$ \\ 
        \midrule
        \multicolumn{11}{l}{\textbf{$^{13}$CCH$_2$}} \\ 
        T (K) & - & $170_{-30}^{+20}$ & - & - & - & - & - & $240_{-10}^{+20}$ & $120_{-10}^{+30}$ & - \\ 
        log N (cm$^{-2}$) & - & $17.6_{-0.3}^{+0.6}$ & - & - & - & - & - & $19.0_{-0.2}^{+1.1}$ & $16.0_{-0.6}^{+0.1}$ & - \\ 
        log A (au$^2$) & - & $-0.58_{-0.27}^{+0.63}$ & - & - & - & - & - & $-1.44_{-0.11}^{+0.12}$ & $0.90_{-0.86}^{+0.32}$ & - \\ 
        log M (M\_Earth) & - & $-6.89_{-0.25}^{+0.36}$ & - & - & - & - & - & $-6.41_{-0.10}^{+0.28}$ & $-7.08_{-0.14}^{+0.16}$ & - \\ 
        \midrule
        \multicolumn{11}{l}{\textbf{HCN}} \\ 
        T (K) & - & $310_{-10}^{+60}$ & - & - & - & - & - & $260_{-20}^{+10}$ & $300_{-0}^{+0}$ & - \\ 
        log N (cm$^{-2}$) & - & $17.0_{-0.3}^{+0.1}$ & - & - & - & - & - & $17.1_{-0.1}^{+0.1}$ & $16.6_{-0.6}^{+0.1}$ & - \\ 
        log A (au$^2$) & - & $-1.31_{-0.16}^{+0.03}$ & - & - & - & - & - & $-0.92_{-0.03}^{+0.10}$ & $-0.95_{-0.01}^{+0.01}$ & - \\ 
        log M (M\_Earth) & \color{gray}$<-10.40$ & $-8.25_{-0.20}^{+0.04}$ & \color{gray}$<-10.60$ & \color{gray}$<-10.29$ & \color{gray}$<-9.59$ & \color{gray}$<-10.64$ & \color{gray}$<-10.20$ & $-7.77_{-0.03}^{+0.12}$ & $-8.26_{-0.02}^{+0.02}$ & \color{gray}$<-10.32$ \\ 
        \midrule
        \multicolumn{11}{l}{\textbf{CO$_2$}} \\ 
        T (K) & - & $480_{-30}^{+10}$ & $190_{-60}^{+100}$ & - & - & \color{orange}$130_{-110}^{+130}$ & $210_{-10}^{+20}$ & $260_{-40}^{+10}$ & $330_{-30}^{+10}$ & \color{orange}$240_{-170}^{+170}$ \\ 
        log N (cm$^{-2}$) & - & $16.0_{-2.0}^{+0.8}$ & $16.0_{-1.4}^{+0.1}$ & - & - & \color{orange}$13.8_{-1.8}^{+3.3}$ & $15.8_{-0.0}^{+0.0}$ & $13.7_{-1.6}^{+1.1}$ & $17.0_{-0.1}^{+0.2}$ & \color{orange}$17.0_{-2.4}^{+3.7}$ \\ 
        log A (au$^2$) & - & $-1.20_{-0.70}^{+1.97}$ & $-0.62_{-0.48}^{+0.62}$ & - & - & \color{orange}$-1.98_{-2.56}^{+1.02}$ & $-0.03_{-0.02}^{+0.04}$ & $1.36_{-1.11}^{+1.60}$ & $-1.49_{-0.02}^{+0.19}$ & \color{orange}$-1.56_{-0.47}^{+1.01}$ \\ 
        log M (M\_Earth) & \color{gray}$<-9.57$ & $-9.33_{-0.01}^{+0.06}$ & $-8.54_{-0.57}^{+0.50}$ & \color{gray}$<-9.46$ & \color{gray}$<-8.76$ & \color{orange}$-8.19_{-0.87}^{+1.69}$ & $-8.20_{-0.03}^{+0.04}$ & $-8.89_{-0.02}^{+0.18}$ & $-8.41_{-0.04}^{+0.14}$ & \color{orange}$-8.52_{-0.68}^{+1.82}$ \\ 
        \midrule
        \multicolumn{11}{l}{\textbf{HC$_3$N}} \\ 
        T (K) & - & $180_{-10}^{+10}$ & - & - & - & - & - & $160_{-0}^{+0}$ & $170_{-10}^{+0}$ & - \\ 
        log N (cm$^{-2}$) & - & $15.0_{-0.6}^{+1.1}$ & - & - & - & - & - & $16.5_{-0.1}^{+0.1}$ & $13.8_{-0.8}^{+0.9}$ & - \\ 
        log A (au$^2$) & - & $0.37_{-0.92}^{+0.61}$ & - & - & - & - & - & $-0.27_{-0.12}^{+0.49}$ & $1.67_{-0.91}^{+0.82}$ & - \\ 
        log M (M\_Earth) & - & $-8.58_{-0.04}^{+0.04}$ & - & - & - & - & - & $-7.70_{-0.10}^{+0.01}$ & $-8.44_{-0.03}^{+0.11}$ & - \\ 
        \midrule
        \multicolumn{11}{l}{\textbf{$^{13}$CO$_2$}} \\ 
        T (K) & - & $240_{-60}^{+180}$ & - & - & - & - & - & \color{orange}$90_{-60}^{+60}$ & - & - \\ 
        log N (cm$^{-2}$) & - & $14.9_{-0.9}^{+7.1}$ & - & - & - & - & - & \color{orange}$21.0_{-4.1}^{+1.0}$ & - & - \\ 
        log A (au$^2$) & - & $-0.24_{-1.76}^{+1.35}$ & - & - & - & - & - & \color{orange}$0.57_{-1.74}^{+1.43}$ & - & - \\ 
        log M (M\_Earth) & - & $-9.35_{-1.43}^{+0.89}$ & - & - & - & - & - & \color{orange}$-2.38_{-0.64}^{+0.25}$ & - & - \\ 
        \midrule
        \multicolumn{11}{l}{\textbf{C$_4$H$_2$}} \\ 
        T (K) & - & $120_{-10}^{+40}$ & - & - & - & - & - & $70_{-50}^{+10}$ & $170_{-20}^{+20}$ & - \\ 
        log N (cm$^{-2}$) & - & $16.0_{-0.7}^{+0.2}$ & - & - & - & - & - & $17.1_{-0.3}^{+0.3}$ & $14.6_{-1.4}^{+1.1}$ & - \\ 
        log A (au$^2$) & - & $0.06_{-0.23}^{+0.16}$ & - & - & - & - & - & $1.92_{-0.73}^{+0.35}$ & $0.30_{-1.86}^{+1.05}$ & - \\ 
        log M (M\_Earth) & - & $-7.89_{-0.07}^{+0.08}$ & - & - & - & - & - & $-4.92_{-0.26}^{+0.08}$ & $-9.05_{-0.15}^{+0.13}$ & - \\ 
        \midrule
        \multicolumn{11}{l}{\textbf{CO} \textbf{(1-0)}} \\ 
        T (K) & - & - & - & - & $1530_{-160}^{+130}$ & - & - & - & - & - \\ 
        log N (cm$^{-2}$) & - & - & - & - & $18.13_{-0.12}^{+0.17}$ & - & - & - & - & - \\ 
        log A (au$^2$) & - & - & - & - & $-1.65_{-0.05}^{+0.07}$ & - & - & - & - & - \\ 
        log M (M\_Earth) & - &- & - & - & $-7.47_{-0.07}^{+0.11}$ & - & - & - & - & - \\
        \midrule
        \multicolumn{11}{l}{\textbf{CO} \textbf{(3-2),(2-1)}} \\ 
        T (K) & - & - & - & - & $1600_{-140}^{+200}$ & - & - & - & - & - \\ 
        log N (cm$^{-2}$) & - & - & - & - & $14.29_{-0.82}^{+1.78}$ & - & - & - & - & - \\ 
        log A (au$^2$) & - & - & - & - & $1.33_{-1.76}^{+0.67}$ & - & - & - & - & - \\ 
        log M (M\_Earth) & - &- & - & - & $-8.33_{-0.30}^{+0.09}$ & - & - & - & - & - \\
        \midrule
        \multicolumn{11}{l}{\textbf{CH$_4$}} \\ 
        T (K) & - & \color{orange}$890_{-40}^{+30}$ & - & - & - & - & - & - & - & - \\ 
        log N (cm$^{-2}$) & - & \color{orange}$16.3_{-1.5}^{+0.2}$ & - & - & - & - & - & - & - & - \\ 
        log A (au$^2$) & - & \color{orange}$-1.83_{-0.17}^{+1.51}$ & - & - &  & - & - & - & - & - \\ 
        log M (M\_Earth) & - & \color{orange}$-9.47_{-0.01}^{+0.01}$ & - & - &  & - & - & - & - & - \\
    \enddata
\end{deluxetable*}

\begin{deluxetable*}{cccccc}
\tablecaption{Log of JWST Observations for the AGE-PRO Upper Sco Sources \label{tab:obs_log_uppsco}
\label{tab:observations}}
\tablehead{
\colhead{Source} & \colhead{2MASS ID} & \colhead{Program} & \colhead{Date} & \colhead{Visit ID} & \colhead{Exposure (s)}}
\startdata
USco~1 & J16120668-3010270 & 3034 & 2024-07-27 & 03034019001 & 1365 \\
USco~2 & J16054540-2023088 & 3034 & 2024-07-21 & 03034020001 & 2031 \\
USco~3 & J16020757-2257467 & 3034 & 2024-07-16 & 03034021001 & 1520 \\
USco~4 & J16111742-1918285 & 3034 & 2024-07-18 & 03034022001 & 1021 \\
USco~5 & J16145026-2332397 & 3034 & 2024-07-27 & 03034023001 & 1520 \\
USco~6 & J16163345-2521505 & 3034 & 2024-07-27 & 03034024001 & 1520 \\
USco~7 & J16202863-2442087 & 3034 & 2024-08-01 & 03034025001 & 1520 \\
USco~8 & J16221532-2511349 & 3034 & 2024-08-01 & 03034026001 & 1520 \\
USco~9 & J16082324-1930009 & 3034 & 2024-07-20 & 03034027001 & 1354 \\
USco~10 & J16090075-1908526 & 3034 & 2024-07-21 & 03034028001 & 1021 \\
\enddata
\end{deluxetable*}


\begin{table*}[h]
    \centering
    \caption{Atomic line detections in each source, done in a similar style to Table~\ref{tab:detections}. The colored boxes indicate the detections (green), non-detections (grey), and tentative detections (orange) of each corresponding atomic transition.
    \label{tab:atomic_detections}}
    \begin{tabular}{lcccccccccc}
      \hline
       & USco~1 & USco~2 & USco~3 & USco~4 & USco~5 & USco~6 & USco~7 & USco~8 & USco~9 & USco~10 \\
      \hline
    Ni II (6.64~$\upmu$m)      & \mybox{} & \mybox{} & \mybox{} & \mybox{} & \mybox{} & \mybox{} & \mybox{} & \mybox{} & \mybox{} & \mybox{} \\
    Ar II (6.99~$\upmu$m)      & \mybox{} & \mybox{} & \mybox{} & \mybox{} & \mybox{} & \myboxxx{} & \mybox{} & \mybox{} & \myboxxx{} & \myboxxx{} \\
    H I (9–6) (5.91~$\upmu$m)  & \mybox{} & \mybox{} & \mybox{} & \mybox{} & \mybox{} & \mybox{} & \mybox{} & \mybox{} & \mybox{} & \mybox{} \\
    H I (6–5) (7.46~$\upmu$m)  & \myboxxx{} & \mybox{} & \mybox{} & \mybox{} & \myboxxx{} & \mybox{} & \mybox{} & \mybox{} & \myboxx{} & \mybox{} \\
    H I (8–6) (7.50~$\upmu$m)  & \myboxxx{} & \mybox{} & \mybox{} & \mybox{} & \mybox{} & \mybox{} & \mybox{} & \mybox{} & \mybox{} & \mybox{} \\
    H I (10–7) (8.76~$\upmu$m) & \mybox{} & \mybox{} & \mybox{} & \mybox{} & \mybox{} & \mybox{} & \mybox{} & \mybox{} & \mybox{} & \mybox{} \\
    H I (13–8) (9.39~$\upmu$m) & \myboxxx{} & \mybox{} & \mybox{} & \mybox{} & \mybox{} & \mybox{} & \mybox{} & \mybox{} & \mybox{} & \mybox{} \\
    H I (9–7) (11.30~$\upmu$m)& \myboxxx{} & \mybox{} & \mybox{} & \mybox{} & \mybox{} & \mybox{} & \mybox{} & \mybox{} & \mybox{} & \mybox{} \\
    H I (7–6) (12.37~$\upmu$m)& \myboxxx{} & \mybox{} & \mybox{} & \mybox{} & \mybox{} & \mybox{} & \mybox{} & \myboxx{} & \myboxx{} & \myboxx{} \\
    H I (11–8) (12.39~$\upmu$m)& \myboxxx{} & \mybox{} & \mybox{} & \mybox{} & \mybox{} & \mybox{} & \mybox{} & \mybox{} & \mybox{} & \mybox{} \\
    H I (10–8) (16.21~$\upmu$m)& \myboxxx{} & \mybox{} & \mybox{} & \mybox{} & \mybox{} & \mybox{} & \mybox{} & \mybox{} & \myboxx{} & \mybox{} \\
    H I (8–7) (19.06~$\upmu$m)& \mybox{} & \mybox{} & \mybox{} & \mybox{} & \mybox{} & \mybox{} & \mybox{} & \mybox{} & \mybox{} & \mybox{} \\
    Ne II (12.81~$\upmu$m)     & \myboxxx{} & \myboxxx{} & \myboxxx{} & \mybox{} & \mybox{} & \mybox{} & \mybox{} & \myboxxx{} & \myboxxx{} & \myboxxx{} \\
    Ne III (15.56~$\upmu$m)    & \myboxxx{} & \mybox{} & \mybox{} & \myboxx{} & \mybox{} & \mybox{} & \mybox{} & \myboxxx{} & \myboxxx{} & \myboxxx{} \\
    Fe II (14.98~$\upmu$m)     & \mybox{} & \mybox{} & \mybox{} & \mybox{} & \mybox{} & \mybox{} & \mybox{} & \mybox{} & \mybox{} & \mybox{} \\
    Fe II (17.94~$\upmu$m)     & \mybox{} & \mybox{} & \mybox{} & \mybox{} & \mybox{} & \mybox{} & \mybox{} & \mybox{} & \mybox{} & \mybox{} \\
    S I (25.25~$\upmu$m)       & \mybox{} & \mybox{} & \mybox{} & \mybox{} & \mybox{} & \mybox{} & \myboxxx{} & \mybox{} & \mybox{} & \mybox{} \\
      \hline
    \end{tabular}%
\end{table*}

\section{Additional Figures} \label{subsec:additional_figures}

\begin{figure*}[ht!]
    \centering
    \includegraphics[width=.875\textwidth]{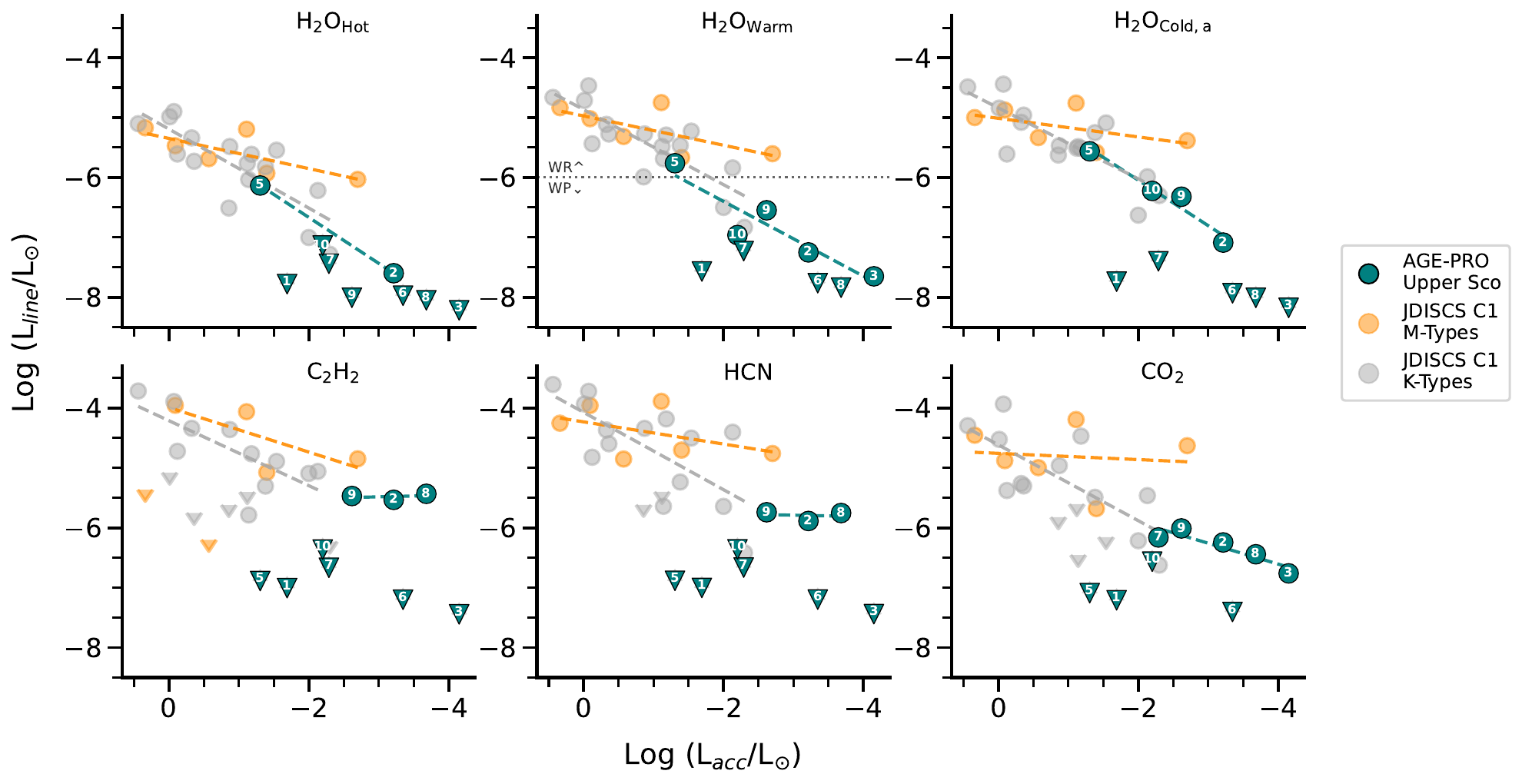}
    \caption{
    Trend plots showing the line luminosity values of the different H$_2$O temperature components, C$_2$H$_2$, HCN, and CO$_2$ as a function of accretion luminosity serving as a proxy of age (where L$\rm_{acc}$ decreases with time; note inverse x-axis). Our AGE-PRO Upper Sco sample ($\sim$2-6~Myr) is compared to the younger JDISCS~C1 sample ($\sim$1–3~Myr) , separated by spectral type, where our Upper Sco sample is entirely M-types. This demonstrates the trend of strong water depletion and generally weaker molecular emission in our more evolved sample, similar to Fig.~\ref{fig:violin}.
    \label{fig:lacc}}
\end{figure*}

\begin{figure*}[ht!]
    \centering
    \includegraphics[width=.9\textwidth]{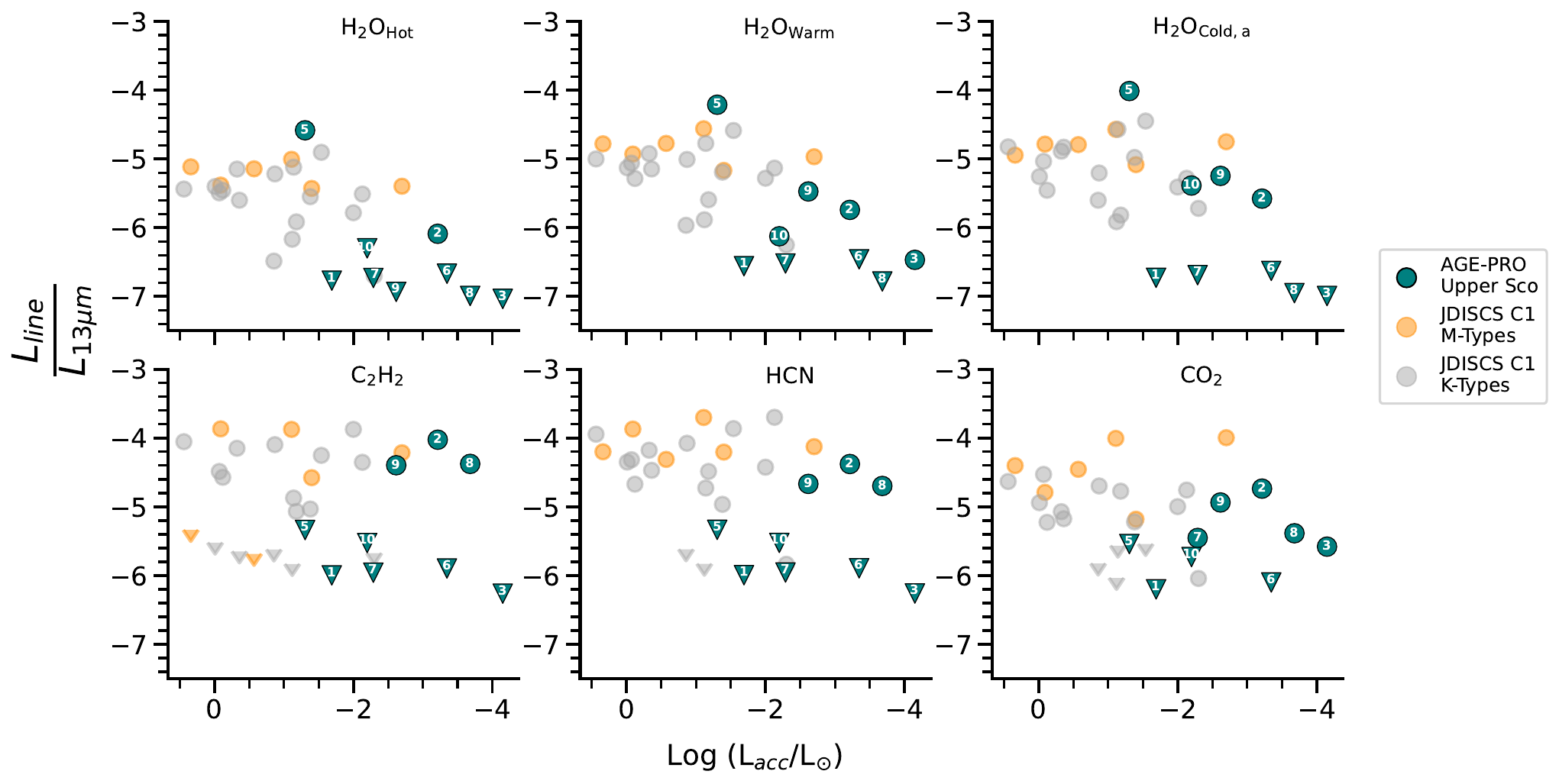}
    \caption{
    Trend plots showing our computed line luminosity values in the same manner as Fig.~\ref{fig:lacc}, except after the line luminosities have been normalized by the 13 $\upmu$m continuum luminosity. This demonstrates the unusually high water abundance of USco~5 and relatively high fraction of Water-Absent and Molecule-Absent spectra in our sample. 
    \label{fig:lacc_norm}}
\end{figure*}

\begin{figure*}[ht!]
    \centering
    \includegraphics[width=\textwidth]{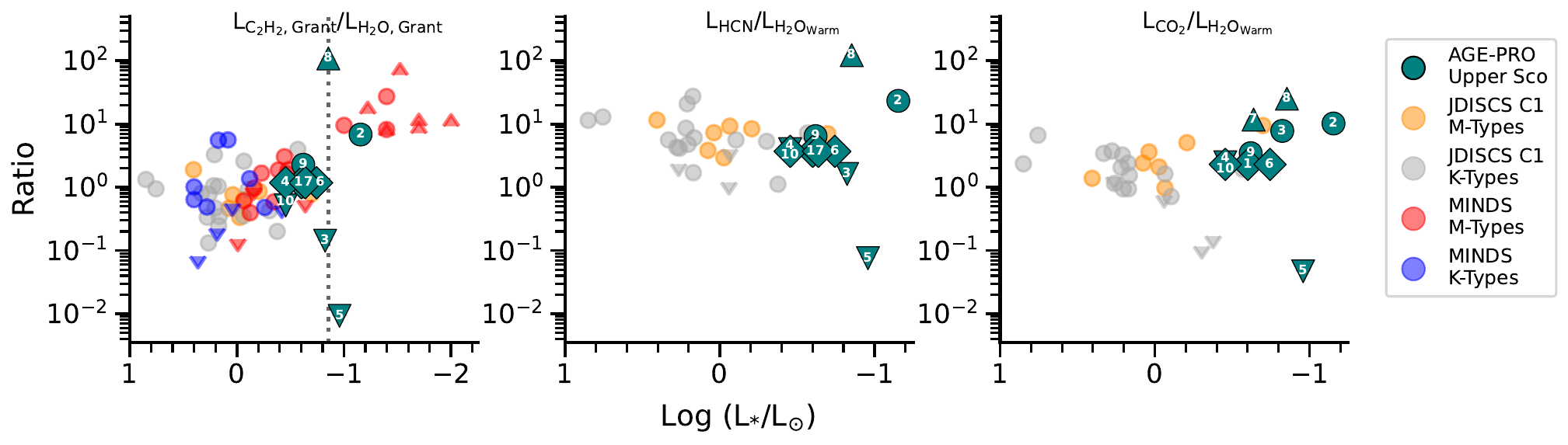}
    \caption{Plot showing the ratio of the line-luminosity of several carbon-based molecules to H$_2$O as a function of stellar luminosity (note inverse x-axis), done in a similar manner to \citet{Grant2025}. We compare this to the JDISCS~C1 (1-3~Myr) \citep{Arulanantham25} and MINDS (left panel, \citealt{Henning24}) samples separated by spectral type. All AGE-PRO Upper Sco disks are M-types. A rough boundary is plotted in the leftmost panel loosely approximating the border between T~Tauri and a VLMS as $\mathrm{\frac{L}{L_{\odot}} \approx 0.23\left(\frac{M_{VLMS}}{M_{\odot}}\right)^{2.3}}$ \citep{Henry1993}, where $\mathrm{M_{VLMS} \approx 0.3M_{\odot}}$. While all 26 M- and K-types in the JDISCS~C1 sample matches the T~Tauris of the MINDS sample, the AGE-PRO Upper Sco sample exhibits multiple outliers to the groupings found in \citet{Grant2025} at this boundary.
    All Upper Sco error bars are smaller than data points.
    \label{fig:flux ratio}}
\end{figure*}

\begin{figure*}[ht!]
    \centering
    \includegraphics[width=\textwidth]{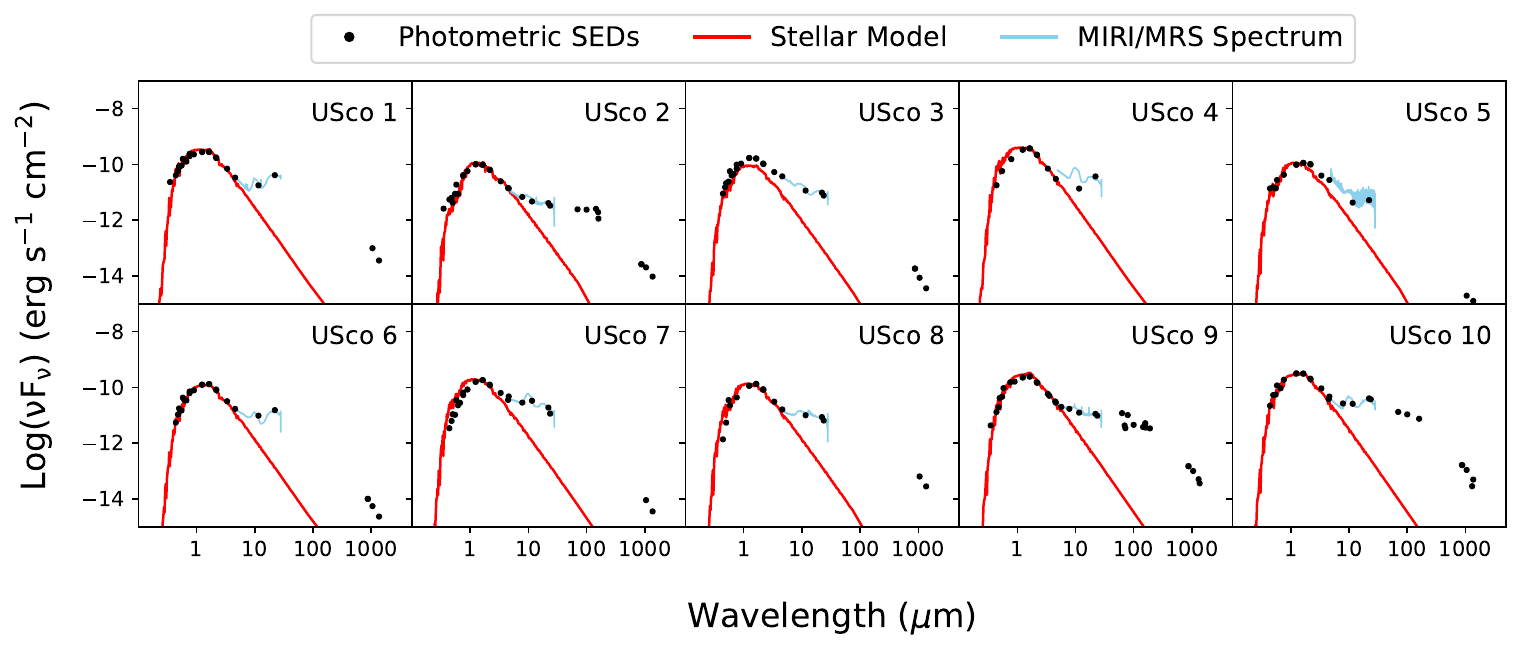}
    \caption{SEDs of the AGE-PRO Upper Sco sample, done in a similar manner to Figure 9 of \citet{Zhang25}, where the red curve shows the PHOENIX stellar photosphere models, the blue curves show the MIRI/MRS spectra from this work, and the SED photometry is taken from \citet{RuizRodriguez2025,AGEPRO_III_Lupus,Agurto-Gangas25}
    \label{fig:seds}}
\end{figure*}

\begin{figure*}[ht!]
    \centering
    \includegraphics[width=\textwidth]{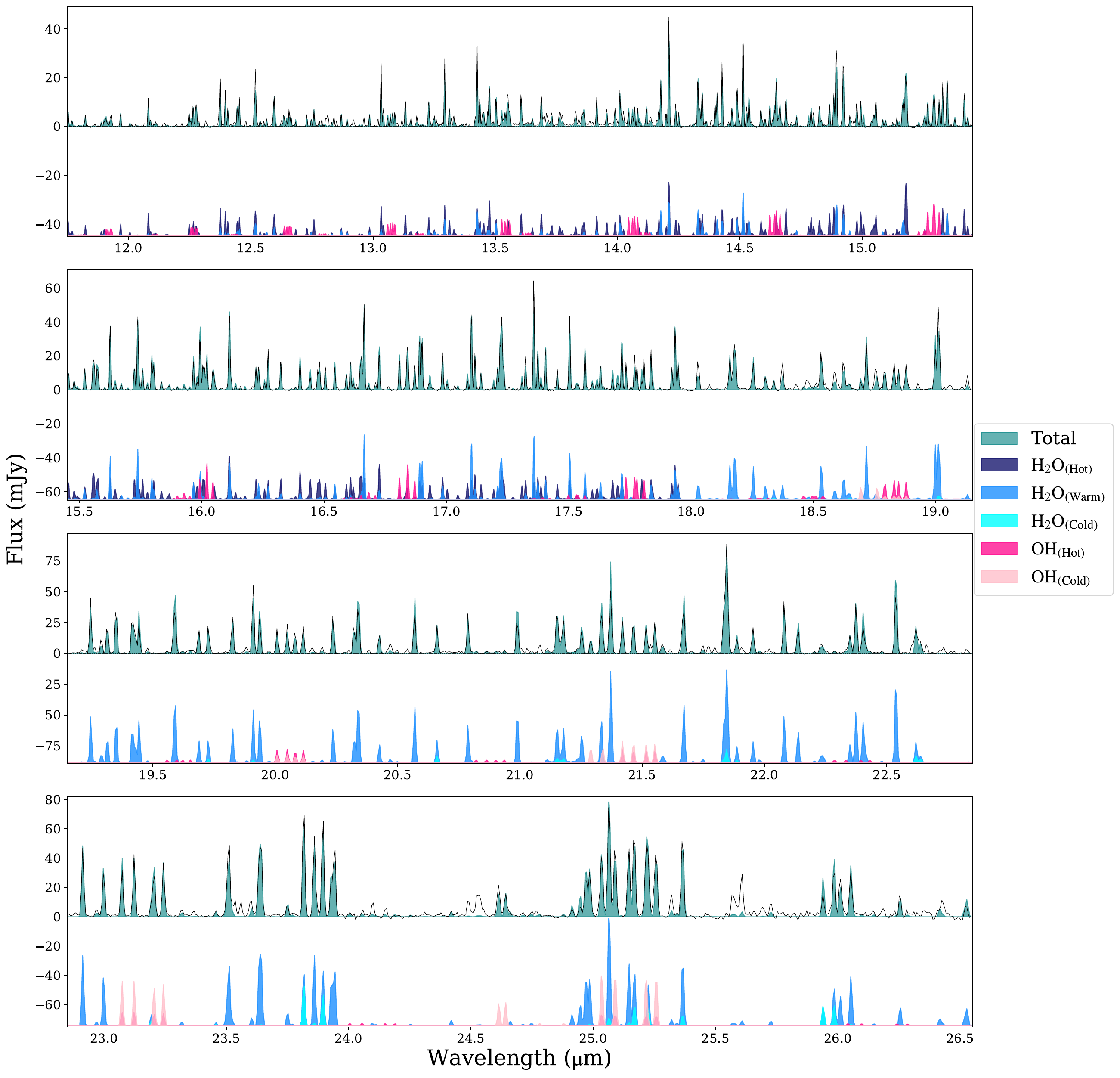}
    \caption{Total best-fitting slab model for our sole Water-Rich (WR) disk USco~5 from $\sim$11.75-26.55 $\upmu$m. See Fig.~\ref{fig:upsco5_contribution2} for best-fitting slab model of low-wavelength region, and see Sec.~\ref{subsec:other} for comments.
    \label{fig:upsco5_contribution1}}
\end{figure*}

\begin{figure*}[ht!]
    \centering
    \includegraphics[width=\textwidth]{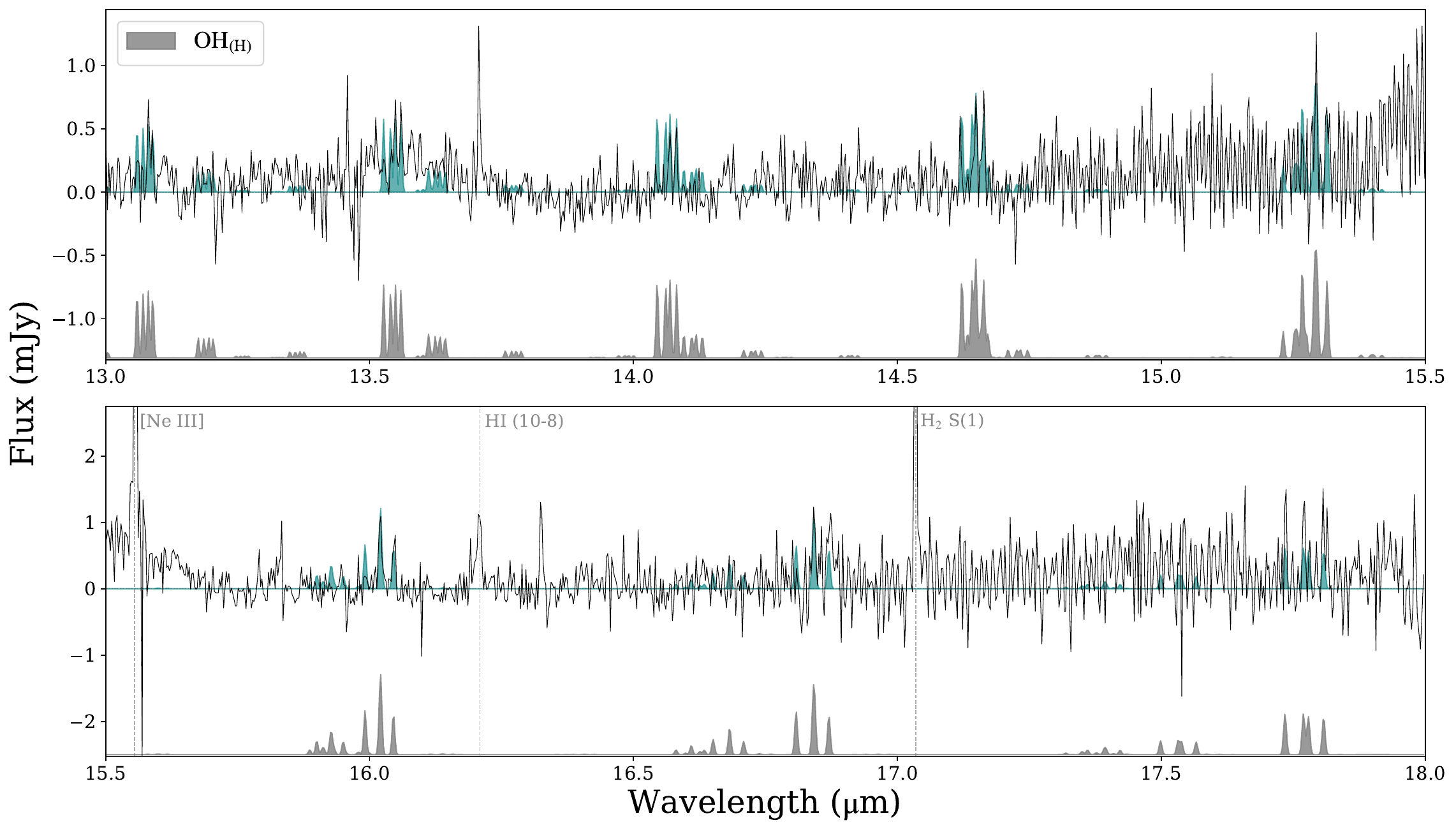}
    \caption{Total best-fitting slab model for the Molecule-Absent disk USco~1 (WA, MA) showing tentative OH detections, possibly due to the presence of asymmetric OH doublets in the spectrum. See Sec.~\ref{subsec:ma} for comments.
    \label{fig:upsco1_contribution}}
\end{figure*}

\begin{figure*}[ht!]
    \centering
    \includegraphics[width=\textwidth]{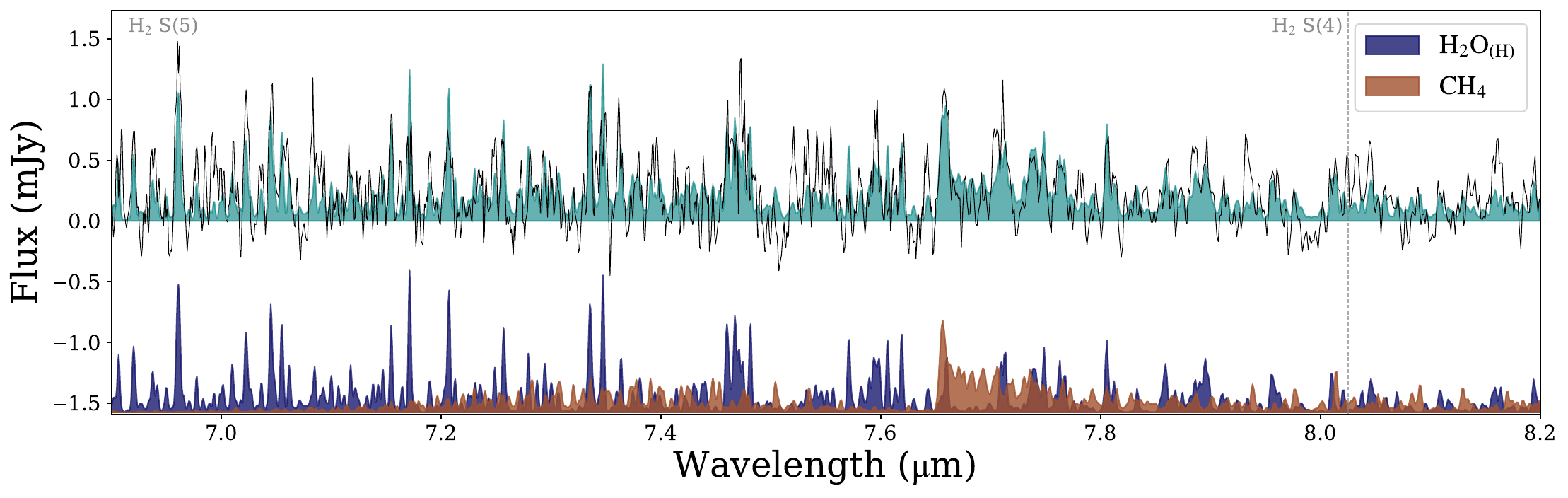}
    \caption{Total best-fitting slab model for lower wavelength regions of USco~2 with a tentative CH$_4$ detection. We note that water emission at these wavelengths is typically not in LTE \citep{Banzatti25}, and therefore the models in this region should be taken with caution, as our methodology employs LTE slab modeling \citep{Salyk22}, which typically overestimates these features.
    \label{fig:upsco2_contribution}}
\end{figure*}

\begin{figure*}[ht!]
    \centering
    \includegraphics[width=\textwidth]{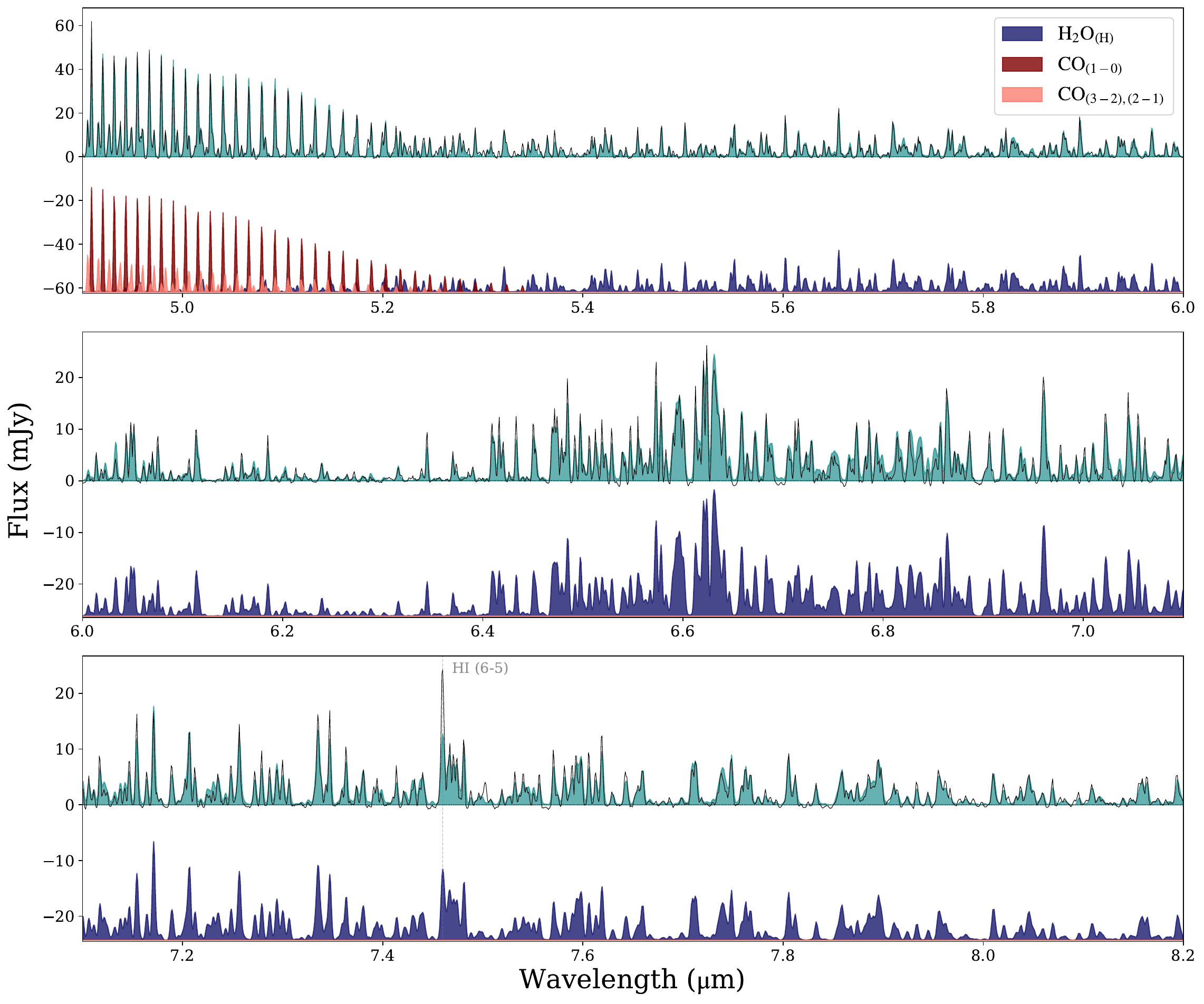}
    \caption{Total best-fitting slab model for the Water-Rich (WR) disk USco~5 showing lower wavelength regions with non-LTE water. Due to non-LTE emission, CO was modeled with two separate components, one for the 1-0 and one for the 3-2 and 2-1 vibrational transitions. The 1-0 transitions correspond to a (1000-1400K) component and the 3-2 and 2-1 transitions correspond to a warmer component (1400-1800K). This was done by fitting only the 1-0 lines, subtracting that fit from the data, and then modeling the 3-2 and 2-1 transitions, the details of which can be found in Waggoner et al. (in prep). See Fig.~\ref{fig:upsco5_contribution1} for best-fitting slab model of longer-wavelength regions and Sec.~\ref{subsec:other} for comments.
    \label{fig:upsco5_contribution2}}
\end{figure*}



\bibliography{references}{}
\bibliographystyle{aasjournalv7}

\end{document}